\documentclass[preprint,journal]{vgtc}
\usepackage[utf8]{inputenc}
\pdfoutput=1

\ifpdf
  \pdfoutput=1\relax                   
  \usepackage{graphicx}                
  \DeclareGraphicsExtensions{.pdf,.png,.jpg,.jpeg} 
\else
  \usepackage{graphicx}                
  \DeclareGraphicsExtensions{.eps}     
\fi%

\graphicspath{{figures/}{pictures/}{images/}{./}} 

\usepackage{amsmath}
\usepackage{amssymb}
\usepackage{microtype}                 
\PassOptionsToPackage{warn}{textcomp}  
\usepackage{textcomp}                  
\usepackage{mathptmx}                  
\usepackage{times}                     
\usepackage{cite}                      
\usepackage{array}
\usepackage{tikz}
\usepackage{bm}
\usepackage{color}
\usepackage[percent]{overpic}
\usepackage{wrapfig}

\definecolor{orangeNils}{rgb}{0,0,0}
\newcommand{\nils}[1]{{\color{orangeNils}#1}}
\definecolor{blueSebastian}{rgb}{0.2,0.3,1.0}

\definecolor{purpleRachel}{rgb}{0.8,0.3,0.9}

\usetikzlibrary{shapes,arrows,decorations.pathreplacing}

\onlineid{1091}

\vgtccategory{Research}
\vgtcpapertype{algorithm}

\title{Volumetric Isosurface Rendering \\with Deep Learning-Based Super-Resolution}

\author{Sebastian Weiss, Mengyu Chu, Nils Thuerey, Rüdiger Westermann}
\authorfooter{
\item[] All authors are with Technical University of Munich.
\item Sebastian Weiss: sebastian13.weiss@tum.de.
\item Mengyu Chu: mengyu.chu@tum.de.
\item Nils Thuerey: nils.thuerey@tum.de.
\item Rüdiger Westermann: westermann@tum.de.
}

\abstract{Rendering an accurate image of an isosurface in a volumetric field typically requires large numbers of data samples. Reducing the number of required samples lies at the core of research in volume rendering. With the advent of deep learning networks, a number of architectures have been proposed recently to infer missing samples in multi-dimensional fields, for applications such as image super-resolution and scan completion. In this paper, we investigate the use of such architectures for learning the upscaling of a low-resolution sampling of an isosurface to a higher resolution, with high fidelity reconstruction of spatial detail and shading. We introduce a fully convolutional neural network, to learn a latent representation generating a smooth, edge-aware normal field and ambient occlusions from a low-resolution normal and depth field. By adding a frame-to-frame motion loss into the learning stage, the upscaling can consider temporal variations and achieves improved frame-to-frame coherence. We demonstrate the quality of the network for isosurfaces which were never seen during training, and discuss 
remote and in-situ visualization as well as focus$+$context visualization as potential applications.%
} 


\teaser{
  \vspace{-6pt}
  \centering
  \begin{subfigure}[t]{0.3\textwidth}
      \begin{overpic}[width=\textwidth]{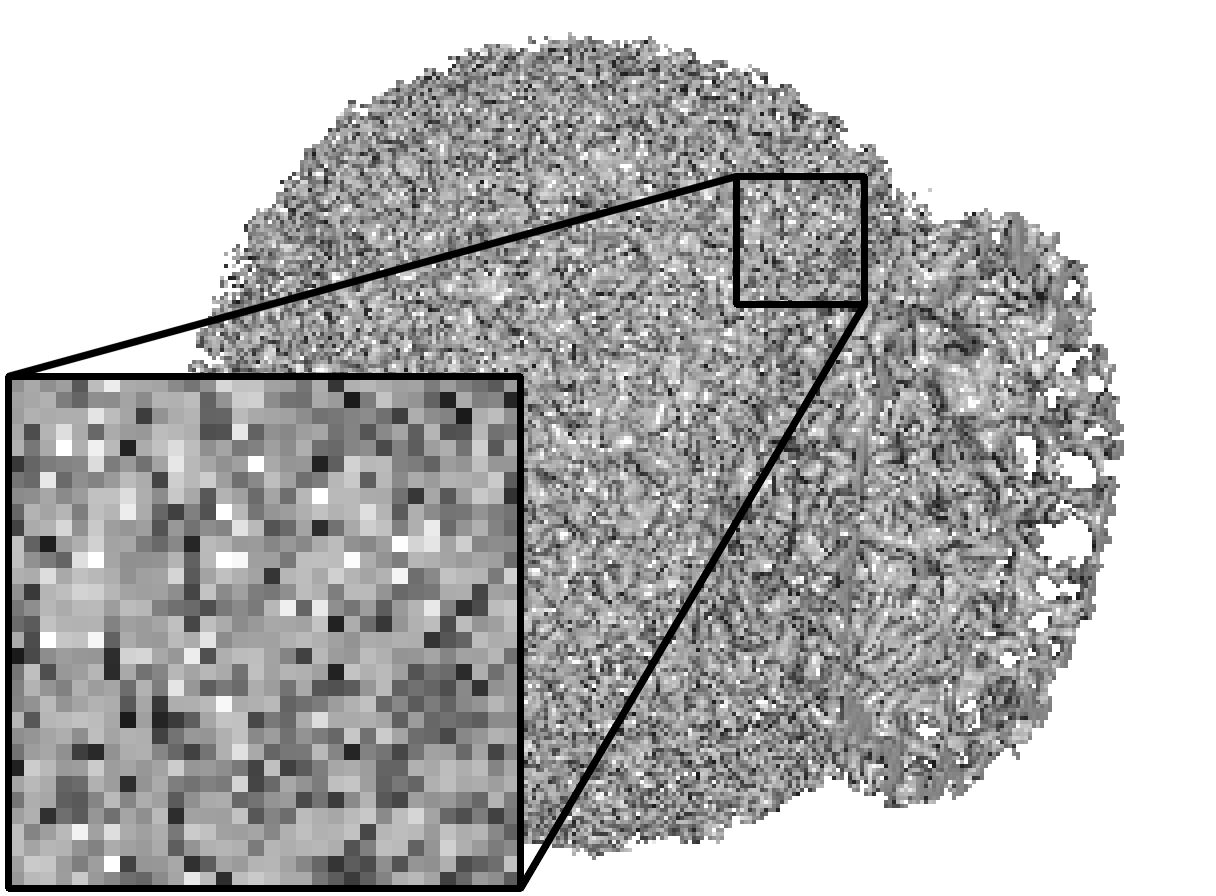}
         \put(80,0){(a)}  
      \end{overpic}
  \end{subfigure}
  \begin{subfigure}[t]{0.3\textwidth}
      \begin{overpic}[width=\textwidth]{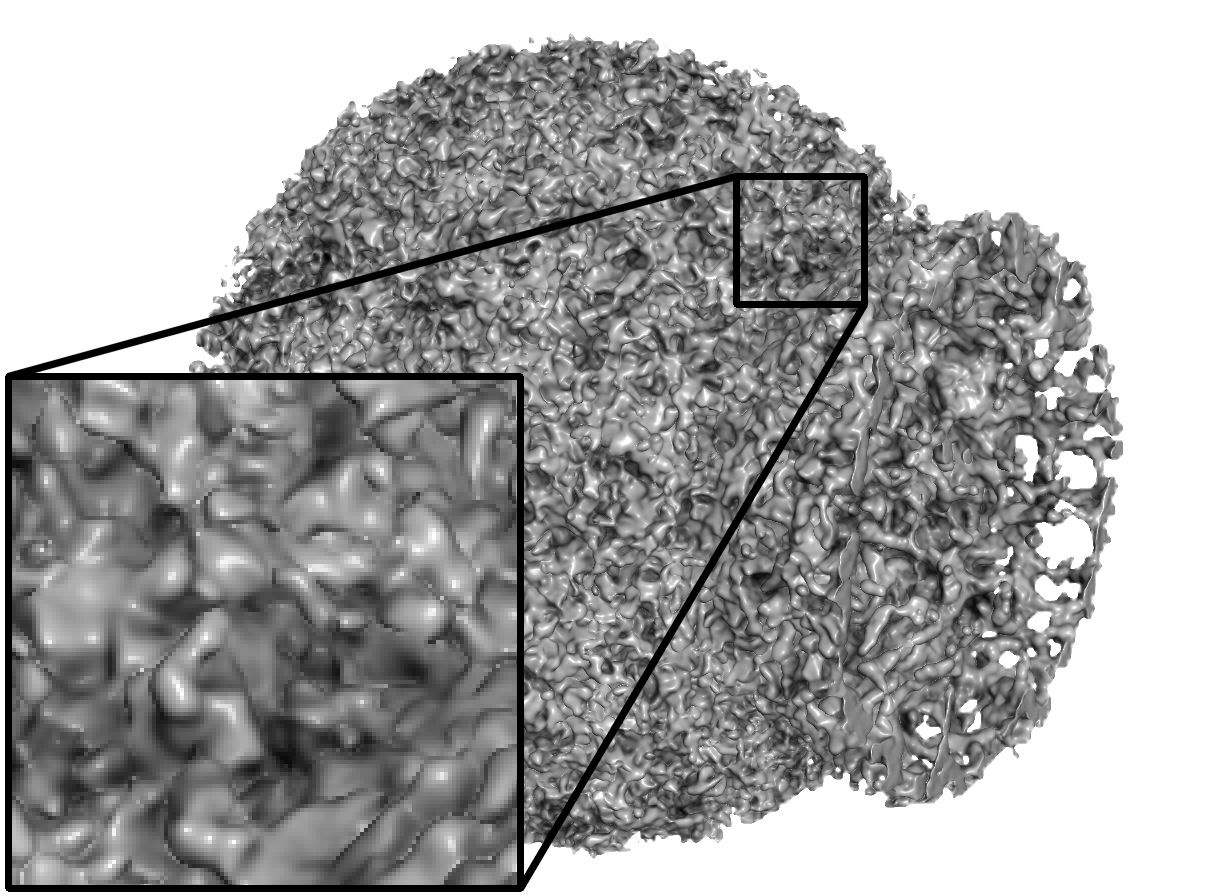}
         \put(80,0){(b)}  
      \end{overpic}
  \end{subfigure}
  \begin{subfigure}[t]{0.3\textwidth}
      \begin{overpic}[width=\textwidth]{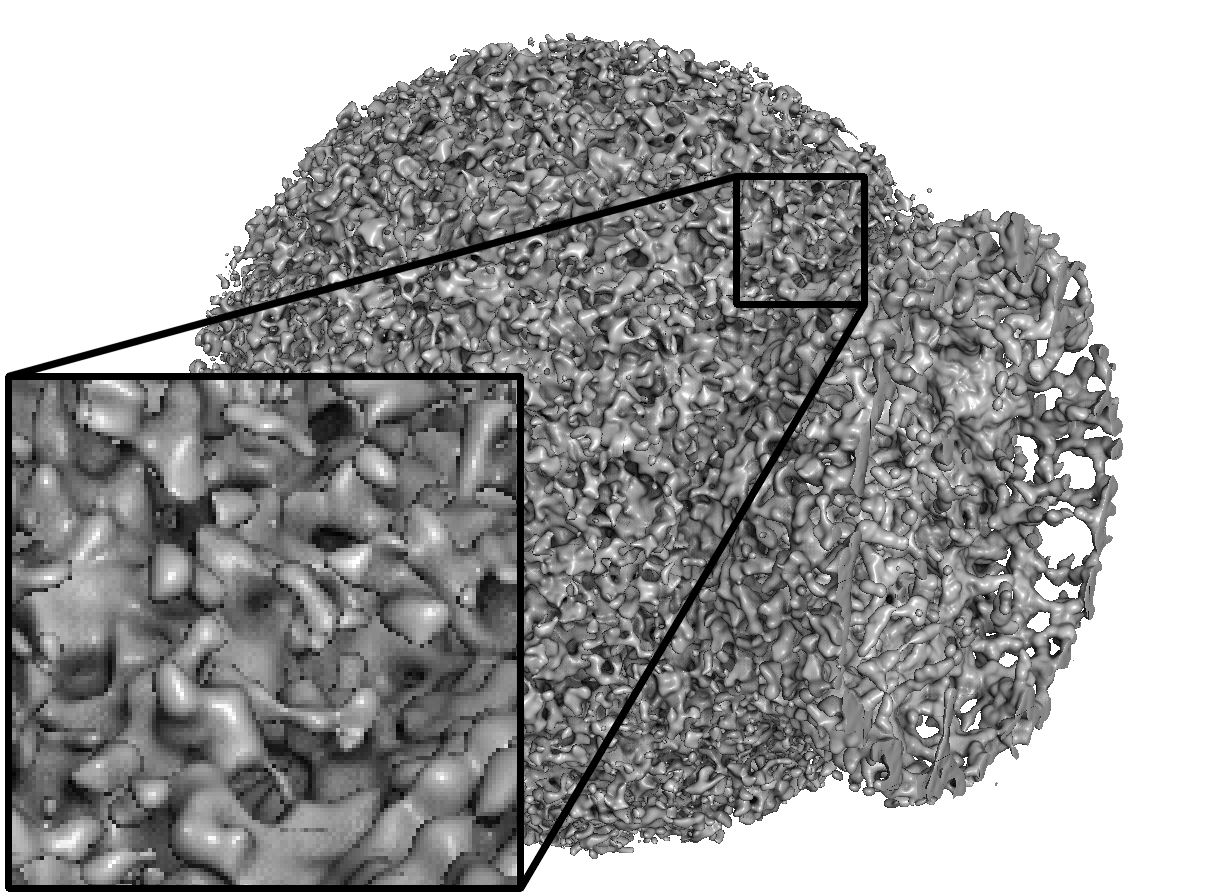}
         \put(80,0){(c)}  
      \end{overpic}
  \end{subfigure}
  \caption{Our super-resolution network can upscale (a) an input sampling of isosurface normals and depths at low resolution (i.e., 320x240), to (b) a high-resolution normal and depth map (i.e., 1280x960) with ambient occlusion. For ease of interpretation, only the shaded output is shown. (c) The ground truth is rendered at 1280x960. Samples are from a $1024^3$ grid, ground truth renders at 0.16 and 18.6 secs w/ and w/o ambient occlusion, super-resolution takes 0.07 sec}
	\label{fig:teaser}
}

\vgtcinsertpkg
\usepackage{subcaption}

\newcommand{\R}{\mathbb{R}}
\newcommand{\LR}{\textit{LR}}
\newcommand{\HR}{\textit{est}}
\newcommand{\GT}{\textit{GT}}
\newcommand{\AO}{{A\!O}}
\newcommand{\etal}{~\textit{et~al.}}

\begin{document}


\firstsection{Introduction}

\maketitle

Over the last decades, much of the research related to volumetric ray-casting for isosurface rendering has been devoted to the investigation of efficient search structures, i.e., data structures that enable to reduce the number of data samples required to determine where a view ray intersects the surface. Despite the high degree of sophistication of these acceleration structures, for large volumes---and in particular for volumes with heterogeneous material distribution---the traversal of these data structures becomes increasingly costly. Since the workload of ray-casting is linear in the number of pixels, frame rates can drop significantly when isosurfaces in large volumes are rendered on high-resolution display systems.

This effect is greatly intensified if global illumination effects are considered. One of the most important global illumination effects for isosurfaces is ambient occlusion. Ambient occlusion (AO) estimates for every surface point the attenuation of ambient light from the surrounding, and uses this information to enhance cavities and locations closely surrounded by other surface parts. Ambient occlusion is simulated by testing along many secondary rays per surface point whether the isosurface is hit. These tests can be greatly accelerated by using a ray-casting acceleration structure, yet in general the number of data samples that need to be accessed increases so strongly that interactive frame rates cannot be maintained.  

In this work, we investigate a novel approach to further reduce the number of samples in isosurface ray-casting with ambient occlusion. This strategy is orthogonal to the use of an acceleration structure, meaning that it works in tandem with this structure to even more aggressively reduce the number of samples. In our investigations, we address the following questions which are paramount in volumetric isosurface rendering: First, we shed light on the question whether an accurate high-resolution image of the surface---a super-resolution image---can be inferred from only the unshaded surface points at a far lower image resolution. Second, we aim at investigating whether ambient occlusions can be inferred from the surface geometry without the need to explicitly compute occlusions on that geometry. 

At first, it seems questionable whether the aforementioned goals can be achieved. From a signal theoretical point of view, it can be argued that new structures---beyond what can be predicted by classical up-scaling filters like bi-linear or -cubic interpolation---cannot be inferred without any further assumptions about their occurrence. On the other hand, recent works in deep learning have demonstrated that such assumptions can be learned by an artificial neural network from corresponding pairs of low- and high-resolution color images. Learned assumptions can then be transferred to a new low-resolution input, to generate a high-resolution variant that adheres to 
\nils{the data seen at traininig time}.
Learning-based image and video super-resolution have achieved remarkable results over the last years, by training networks using low- and high-resolution color images of the same scene~\cite{dong2016image, kappeler2016video}. The proposed recurrent neural network architectures train on high-resolution color images and low-resolution copies that have been generated from these high-resolution images via filtering. Even though similar in spirit to classical super-resolution techniques, it is worth noting that we strive for a conceptually very different approach in this work: Instead of using color images and down-scaled ground truth images for training, we aim at incorporating 3D scene information in the form of per-frame depth and normal fields into the training and reconstruction process. It is our goal to let the network learn the relations between the isosurface geometry sampled at a low and a high resolution, and to infer on the relations between the geometry and shading.

\begin{figure*}[ht]
    \centering
    
    ~ ~
    \begin{subfigure}[b]{0.23\textwidth}
        \begin{overpic}[width=\textwidth]{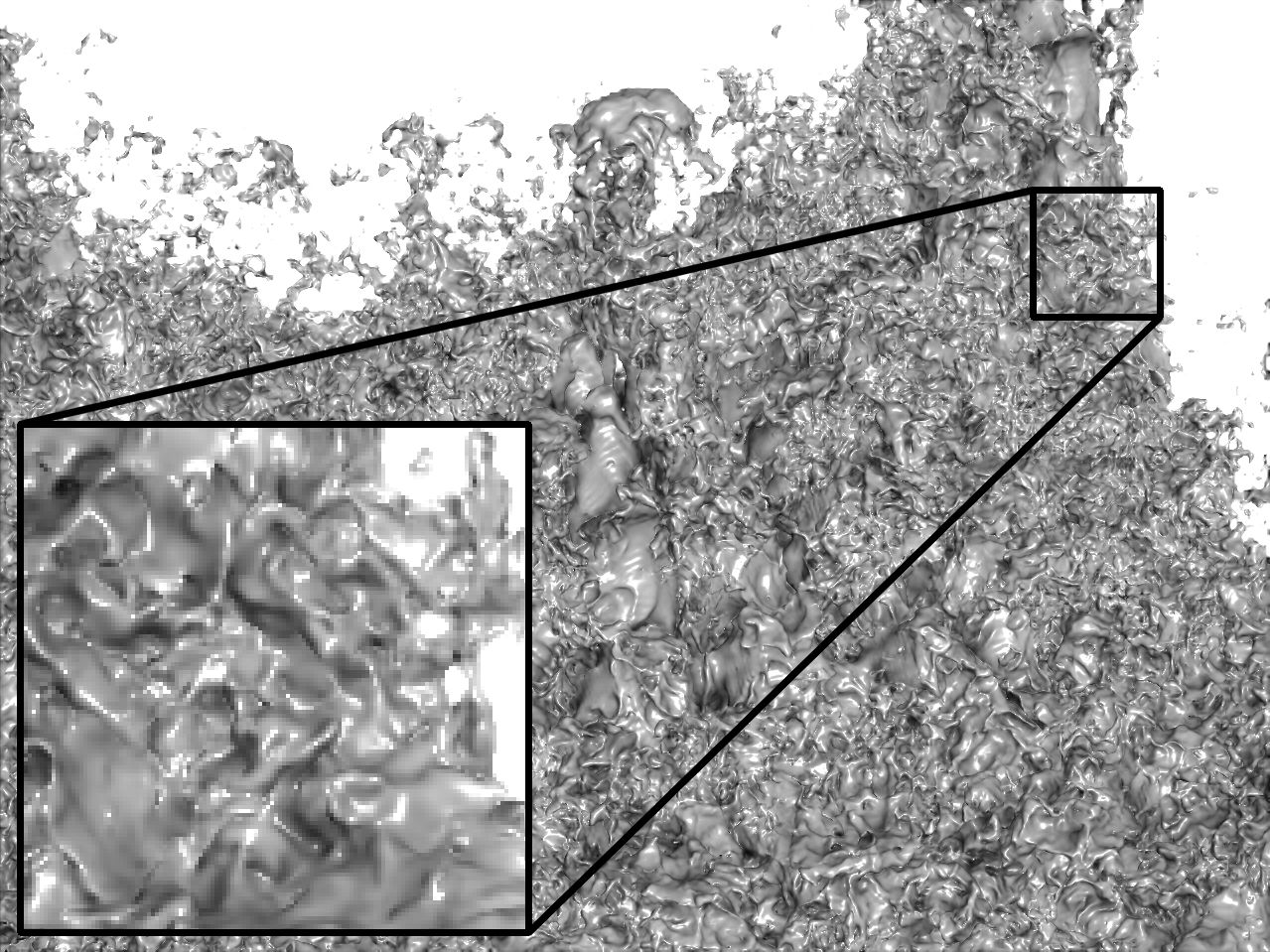}
           \put(-10,5){(a)}  
        \end{overpic}
    \end{subfigure}
    ~ ~
    \begin{subfigure}[b]{0.23\textwidth}
        \begin{overpic}[width=\textwidth]{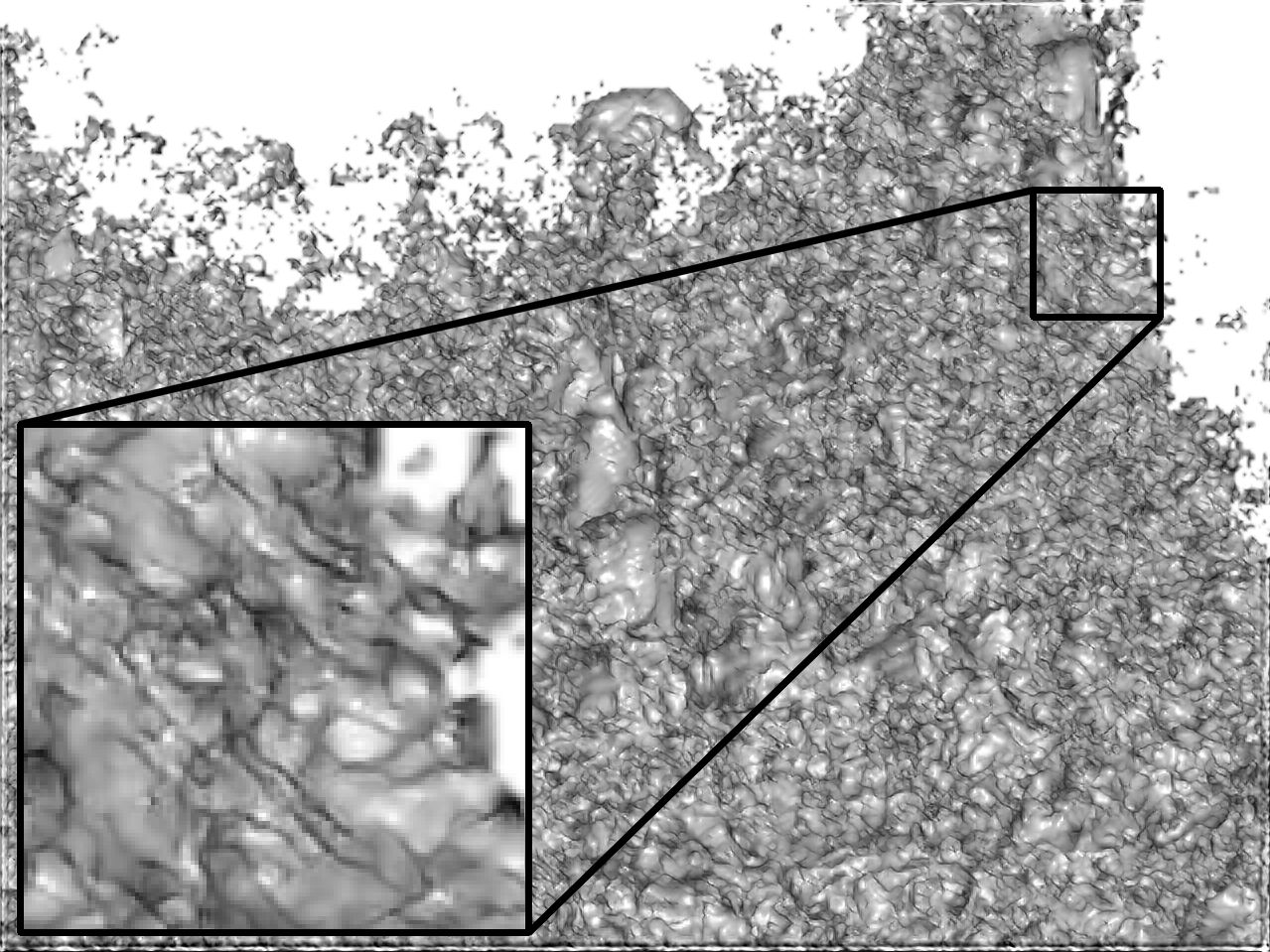}
           \put(-10,5){(b)}  
        \end{overpic}
    \end{subfigure}
    ~ ~
    \begin{subfigure}[b]{0.23\textwidth}
        \begin{overpic}[width=\textwidth]{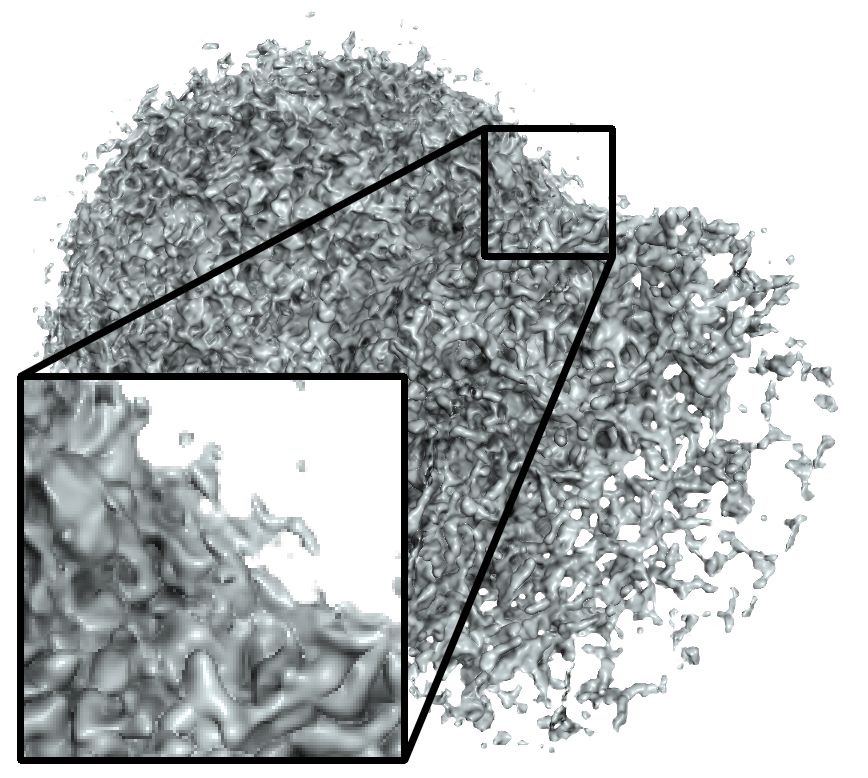}
           \put(-10,5){(c)}  
        \end{overpic}
    \end{subfigure}
    ~ ~
    \begin{subfigure}[b]{0.23\textwidth}
        \begin{overpic}[width=\textwidth]{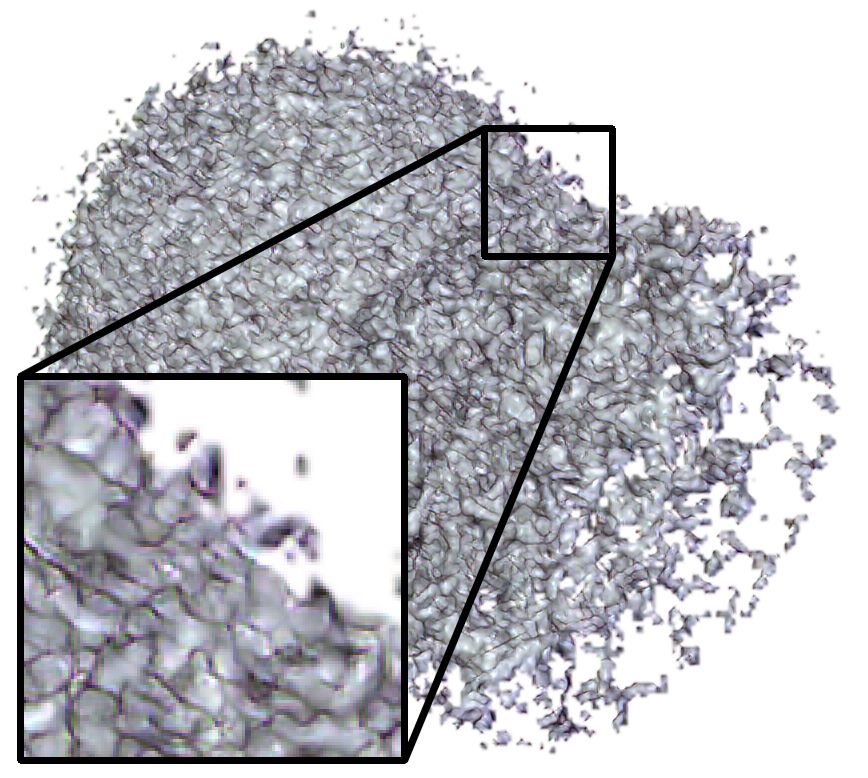}
        \put(-10,5){(d)}  
        \end{overpic}
    \end{subfigure}
    \caption{Super-resolution on depth and normal maps with screen-space shading (a,c) leads to superior reconstruction quality compared to super-resolution on color images (b,d). These images exhibit color bleeding and shifts, while our screen-space shading approach successfully prevents these artifacts. (b) was converted to gray-scale to emphasize geometric differences.}
    \label{fig:ColorErrors}
\end{figure*}

\subsection{Contribution}

Building upon prior works in deep learning and network architectures for image and video super-resolution, we present an artificial neural network that learns to upscale a sampled representation of geometric properties of an isosurface at low resolution to a higher resolution. We introduce a fully convolutional neural network, using the FRVSR-Net~\cite{sajjadi2018frame} as a basis, to learn a latent representation that can generate a smooth, edge-aware depth and normal field, as well as ambient occlusions, from a low-resolution depth and normal field. To support user navigation in the exploration of complex isosurfaces, we integrate a loss function into the training pass that penalizes frame-to-frame inconsistencies and achieves improved temporal coherence in the reconstruction step. 

We use the network architecture to infer images of isosurfaces with four times the resolution of the input images. \autoref{fig:teaser} demonstrates the reconstruction quality of the upscaling process. Since the network is designed to learn high-resolution ambient occlusions from low-resolution depth and normal images, computations of ambient occlusions at runtime are entirely avoided. Compared to volumetric ray-casting at full resolution, the number of samples from the volumetric field can be reduced drastically about a factor of 16 (without ambient occlusions) and 16 x $N_{AO}$ (with ambient occlusions and $N_{AO}$ samples per surface point). A performance gain of one to two orders of magnitude compared to full resolution ray-casting is achieved.

Our specific contributions are:
\begin{itemize}
\item We show that it is beneficial to train the network based on depth and normal images instead of color. Our results indicate that this training process results in an improved learning of geometric surface properties, and avoids that the network implicitly learns an internal color representation (see \autoref{fig:ColorErrors}). 
\item Instead of letting the network learn to infer AO in the high-resolution output from AO in the low-resolution inputs, it is solely learned from low-resolution depth and normal maps. Corresponding training data with AOs at high resolution is provided at training time. At runtime, AO then never needs to be simulated when rendering the low-resolution input images, but is instead inferred by the network.    
\item We show that training and evaluating the network with a differentiable screen-space shading step greatly enhances the visual quality when compared to super-resolution on RGB-images only. Our networks learn to anticipate the influence of shading, and produce geometry information such that the desired final color content is obtained.
\item To let the network learn to maintain frame-to-frame coherence, we additionally add a motion loss for the generated image content. In this way, the network achieves improved reconstruction quality and becomes well suited for interactive exploration tasks.
\end{itemize}

We analyse the quality of the network for isosurfaces that were never shown to the network during the training pass. The quality of upscaling seems to indicate that not a specific isosurface is learned, but rather that the network is able to infer the geometric properties of isocontours in scalar fields. On the other hand, it is not possible in general to quantify the reconstruction error that can be introduced by the network. In an upcoming evaluation we will thoroughly study the error and investigate whether it impacts the fidelity by which certain features are reconstructed. We also outline applications where the error is either acceptable, e.g., during interactive navigation, or is even intentionally integrated, for example in focus$+$context visualization.

\section{Related Work}

Our approach works in combination with established acceleration techniques for volumetric ray-casting of isosurfaces, and builds upon recent developments in image and video super-resolution via artificial neural networks to further reduce the number of data access operations.

\paragraph{Volumetric Ray-Casting of Isosurfaces}

Over the last decades, considerable effort has been put into the development of acceleration techniques for isosurface ray-casting in 3D scalar fields. Direct volume ray-casting of isosurfaces was proposed by Levoy ~\cite{levoy:1989:DSV}. Classical fixed-step ray-casting traverses the volume along a ray using equidistant steps in the order of the voxel size. Acceleration structures for isosurface ray-casting encode larger areas were the surface cannot occur, and ray-casting uses this information to skip these areas with few steps. One of the most often used acceleration structure is the min-max pyramid ~\cite{Danskin:1992:FAV:147130.147155}, a tree data structure that stores at every interior node the interval of data values in the corresponding part of the volume. Pyramidal data structures are at the core of most volumetric ray-casting techniques to effectively reduce the number of data samples that need to be accessed during ray traversal.

Because the min-max pyramid is independent of the selected iso-value, it can be used to loop through the pencil of iso-surfaces without any further adaptations. For pre-selected isosurfaces, bounding cells or simple geometries were introduced to restrict ray-traversal to a surface's interior ~\cite{Sobierajski:1994:VRT:197938.197949,Wan:1999:HPP:319351.319435}. Adaptive step-size control according to pre-computed distance information aimed at accelerating first-hit determination~\cite{sramek:sr:1994}. Recently, SparseLeap~\cite{Beyer2017:SparseLeap} introduced pyramidal occupancy histograms to generate geometric structures representing non-empty regions. They are then rasterized into per-pixel fragment lists to obtain those segments that need to be traversed.

Significant performance improvements have been achieved by approaches which exploit high memory bandwidth and texture mapping hardware on GPUs for sampling and interpolation in 3D scalar fields ~\cite{Kruger:2003:ATG:1081432.1081482,Hadwiger2005RealTimeRA}. For isosurface ray-casting, frame to frame depth buffer coherence on the GPU was employed to speed up first-hit determination~\cite{Klein05exploitingframe-to-frame,Braley:2009:GAI:1666778.1666820}. A number of approaches have shown the efficiency of GPU volume ray-casting when paired with compact isosurface representations, brick-based or octree subdivision, and out-of-core strategies for handling data sets too large to be stored on the GPU ~\cite{Gobbetti2008,treib:giga:2012,PE:VMV:VMV12:047-054,fogal2013}. For a thorough overview of GPU approaches for large-scale volume rendering, let us refer to the report by Beyer\etal~\cite{doi:10.1111/cgf.12605}.

Related to isosurface rendering is the simulation of realistic surface shading effects. Ambient occlusion estimates for every surface point the integral of the visibility function over the hemisphere~\cite{zhukov:AO:1998}. Ambient occlusion can greatly improve isosurface visualization, by enhancing the perception of small surface details. A number of approximations for ambient occlusion simulation in volumetric data sets have been proposed, for instance, local and moment-based approximations of occluding voxels ~\cite{Penner:2008:IAO:2386410.2386420,Hernell:2007:EAE:2386501.2386503} or pre-computed visibility information ~\cite{ropinski:AO:2008}. The survey by Ropinski\etal~\cite{Jnsson2014ASO} provides a thorough overview of the use of global illumination in volume visualization. Even though very efficient screen-space approximations of ambient occlusion exist~\cite{Mittring:2007:FNG:1281500.1281671,Bavoil:2008:IHA:1401032.1401061}, we decided to consider ray-traced ambient occlusion in object-space to achieve high quality.

\paragraph{Deep Learning of Super-Resolution and Shading}

For super-resolution of natural images, deep learning based methods have progressed rapidly 
since the very first method~\cite{dong2016image} 
surpassed traditional techniques in terms of peak signal-to-noise ratio (PSNR). 
Regarding network architectures,
Kim et al. introduced a very deep network~\cite{kim2016accurate}, 
Lai et al. designed the Laplacian pyramid network~\cite{lai2017deep}, 
and advanced network structures have been applied, such as the ResNet~\cite{he2016deep, ledig2017photo} and DenseNet~\cite{huang2017densely,tong2017image} architectures.
Regarding loss formulations, realistic high-frequency detail is significantly 
improved by using adversarial and perceptual losses based on pretrained  networks~\cite{ledig2017photo,sajjadi2017enhancenet}. 
Compared to single-image methods, video super-resolution tasks 
introduce the time dimensions, and as such require
temporal coherence and consistent image content
across multiple frames. While many methods use multiple low-resolution frames~\cite{tao2017spmc, liu2017robust, jo2018deep},
the FRVSR-Net~\cite{sajjadi2018frame} reuses the previously generated high-resolution image
to achieves better temporal coherence. By using a spatio-temporal discriminator, the TecoGAN~\cite{chu2018temporally} network produced results with spatial detail without sacrificing temporal coherence.
Overall, motion compensation represents a critical component when taking into account 
multiple input frames.
Methods either use explicit motion estimation and rely on its accuracy~\cite{liao2015video,Xie2018,sajjadi2018frame,chu2018temporally},
or spend extra efforts implicitly such as detail fusion~\cite{tao2017spmc} and dynamic upsampling~\cite{jo2018deep}. In our setting, we can instead leverage the computation of reliable screen-space motions via raytracing.

In a different scenario, neural networks were trained to infer images from a noisy input generated via path-tracing with low number of paths, of the same resolution as the target, but with significantly reduced variance in the color samples ~\cite{cnn-denoise,mara17towards}. Deep shading~\cite{Nalbach:2017:DSC:3128450.3128458} utilized a neural network to infer shading from rendered images, targeting attributes like position, normals, and reflections for color images of the same resolution. None of these techniques used neural networks for upscaling as we do, yet they are related in that they use additional parameter buffers to improve reconstruction quality of global illumination.

\section{Isosurface Learning}\label{sec:Method}

Our method consists of a pre-process in which an artifical neural network is trained, and the upscaling process which receives a new low-resolution isosurface image and uses the trained network to perform the upscaling of this image. Our network is designed to perform 4x upscaling, i.e. from input images of size $H \times W$ to output images of size $4H \times 4W$. Note, however, that other upscaling factors can be realized simply be adapting the architecture to these factors and re-training the network. 

The network is trained on unshaded surface points. It receives the low-resolution input image in the form of a normal and depth map for a selected view, as well as corresponding high-resolution maps with an additional AO map that is generated for that view. A low- and high-resolution binary mask indicate those pixels where the surface is hit. Once a new low-resolution input image is upscaled, i.e., high-resolution normal and AO maps are reconstructed, screen-space shading is computed and added to AO in a post-process to generate the final color.

The network, given many low- and high-resolution pairs of input maps for different isosurfaces and views, internally builds a so-called latent representation that aims at mapping the low-resolution inputs to their high-resolution counterparts. A loss function is used to penalize differences between the high-resolution learned and ground-truth variants. 
%
The networks we use are trained with collections of randomly sampled views from a small number of exemplary datasets from which meaningful ground-truth renderings of isosurfaces were generated as training data. 
%
We will analyze the behaviour of differently trained networks on test data with isosurfaces the models have never seen during training in \autoref{sec:evaluation}.


\subsection{Input Data}\label{sec:Method:Input}

Both the low- and high-resolution input and ground truth maps are generated via volumetric ray-casting. 
AO in the high-resolution image is simulated by spawning 512 additional secondary rays per surface point, and testing each of them for an intersection with the surface. 
Since we aim at supporting temporally coherent super-resolution, all images have a time subscript $t$, starting with $t=1$ at the first frame.

The following low-resolution input maps of size $H \times W$ are used in the training step: 
\begin{itemize}
    \item $M_t^\LR \in [-1,+1]^{H \times W}$: The binary input mask that specifies for every pixel whether the isosurface is hit (mask=1) or not (mask=-1). Internally, the network learns continuous values, and uses these values to smoothly blend the final color over the background.
    \item $N_t^\LR \in [-1,+1]^{3 \times H \times W}$: The normal map with the normal vectors in screen-space.
    \item $D_t^\LR \in [0,1]^{H \times W}$: The depth map, in which 0 indicates that no hit was found.
\end{itemize}
Thus, the low-resolution input to the network can be written as $I_t^\LR := \{M_t^\LR, N_t^\LR, D_t^\LR\} \in \R^{5 \times H \times W}$. We subsequently call this the low-resolution input image. 

\begin{wrapfigure}{hR}{0.50\linewidth} 
	\vspace{-12pt} \centering
	\includegraphics[width=0.99\linewidth]{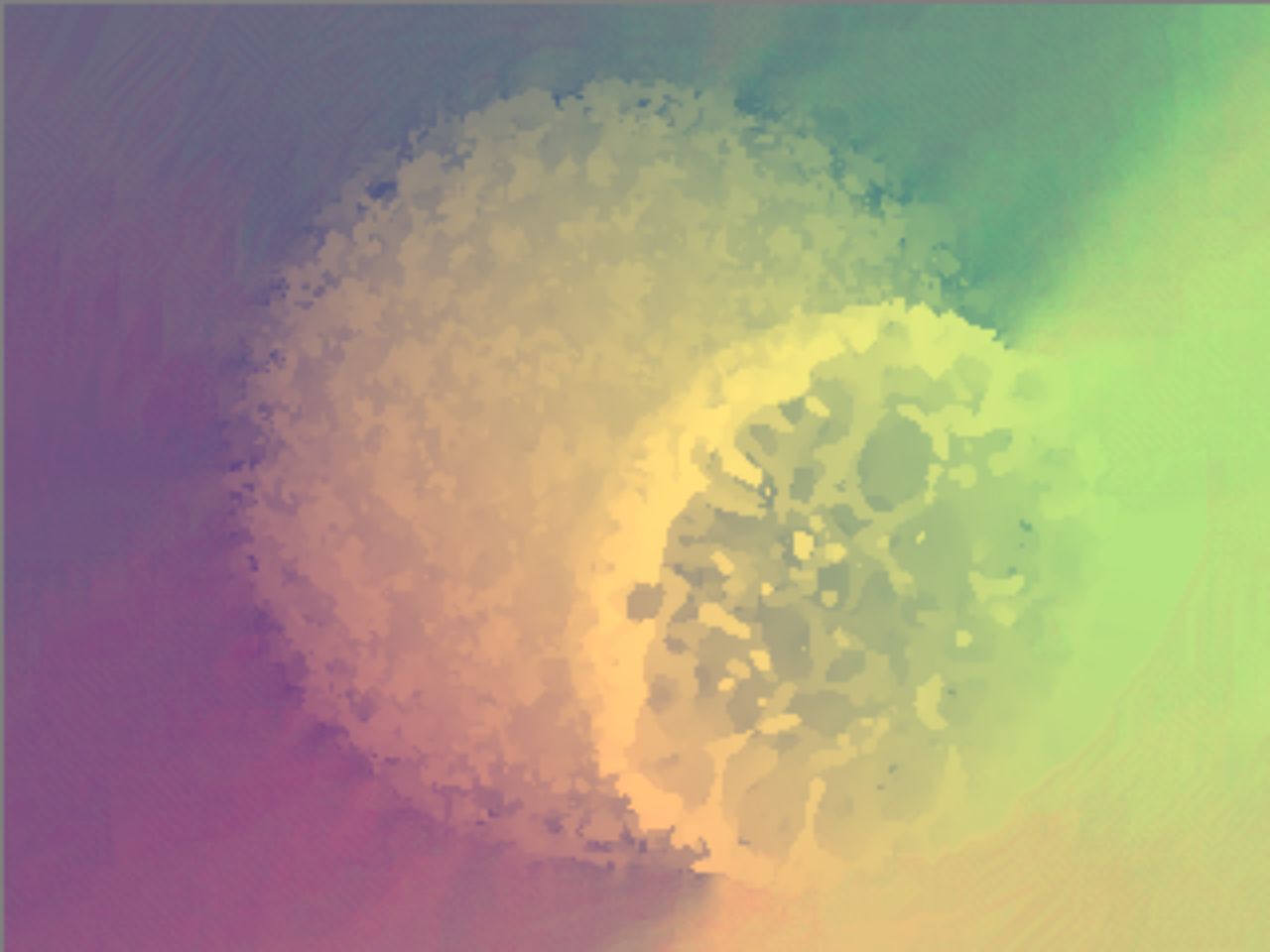}
	\caption{Dense screen-space flow from rotational movement. 2D displacement vectors are color coded.}
	\label{fig:flow}\vspace{-18pt} 
\end{wrapfigure}
Additionally, we generate the following map during ray-casting:
\begin{itemize}
\item $F_t^\LR \in [-1,+1]^{2 \times H \times W}$: A map of 2D displacement vectors, indicating the screen-space flow from the previous view to the current view.
\end{itemize}

\begin{figure}[t]
    \centering
    \resizebox{\linewidth}{60pt}
    {\begin{tikzpicture}[scale=0.7,
	every text node part/.style={align=center},
	every node/.style={minimum size=1.5cm,inner sep=0pt}]

\node[draw,rectangle,fill=blue!20] (Ft) at (-3,-3) {\LARGE $F_t^\LR$};
\node[draw,rectangle,fill=blue!20] (It) at (6,-3) {\LARGE $I_t^\LR$};

\node[draw,diamond,fill=yellow!30] (up) at (0,-3) {Up-\\scaling};
\node[draw,diamond,fill=yellow!30] (warp) at (0, -6) {Warp};
\node[draw,diamond,fill=yellow!30] (flatten) at (3, -6) {Flatten};
\node[draw,diamond,fill=yellow!30] (shading) at (11, -4) {Shading};

\node[draw,rectangle,rounded corners=.2cm,fill=red!40] (net) at (6,-6) {SRNet};

\node[draw,rectangle,fill=green!30] (Otm) at (-3,-6) {\LARGE $O_{t-1}^\HR$};
\node[draw,rectangle,fill=green!30] (Ot) at (9,-6) {\LARGE $O_t^\HR$};
\node[draw,rectangle,fill=green!30] (Ct) at (13.5,-4) {\LARGE $C_t^\HR$};

\node[minimum size=0] (dot1) at (-5,-6) {\LARGE $\cdots$};
\node[minimum size=0] (dot2) at (11,-6) {\LARGE $\cdots$};

\draw[very thick,->] (Ft) -- (up);
\draw[very thick,->] (up) -- (warp);
\draw[very thick,->] (dot1) -- (Otm);
\draw[very thick,->] (Otm) -- (warp);
\draw[very thick,->] (warp) -- (flatten) node[midway,above=-10pt] {\Large $\tilde{O}_{t-1}^\HR$};
\draw[very thick,->] (flatten) -- (net) node[near start,above=-10pt] {\Large $O_f$};
\draw[very thick,->] (It) -- (net);
\draw[very thick,->] (net) -- (Ot);
\draw[very thick,->] (Ot) -- (dot2);
\draw[very thick,->] (Ot) -- (shading);
\draw[very thick,->] (shading) -- (Ct);

\end{tikzpicture}}
    \caption{Schematic illustration of the processing steps. Blue: low-resolution inputs, green: high-resolution outputs, yellow: fixed processing steps, red: trained network.}
    
    \label{fig:Pipeline}
\end{figure}

The screen-space flow is used to align the previous high-resolution results with the current low-resolution input maps. Under the assumption of temporal coherence, the network can then minimize for the deviation of the currently predicted high-resolution map from the temporally extrapolated previous one in the training process. 
To compute the screen-space flow, assume that the current ray hits the isosurface at world position $x_t$, in the low-resolution image. Let $M_t$ and $M_{t-1}$ be the current and previous model-view-projection matrix, respectively. Then, we can project the point $x_t$ into the screen-space corresponding to the current and the previous view, giving $x'_t, x'_{t-1}$. The flow is then computed as $f_t := x'_t - x'_{t-1}$, indicating how to displace the previous mask, depth and normal maps at time $t-1$ to align them with the frame at time $t$. 
Since the described method provides the displacement vectors only at locations in the low-resolution input image where the isosurface is hit in the current frame, 
we use a Navier-Stokes-based image inpainting \cite{bertalmio2001navier} via OpenCV \cite{opencv_library} to obtain a dense displacement field (see \autoref{fig:flow}).
The inpainting algorithm fills the empty areas in such a way that the resulting flow is as incompressible as possible.


For aligning the previous maps, we first upscale the current flow field via bi-linear interpolation to high-resolution. In a semi-lagrangian fashion, we generate new high-resolution maps where every pixel in the upscaled maps retrieves the value in the corresponding high-resolution map from the previous frame, by using the inverse flow vector to determine the target location.  


The high-resolution input data, which is used as ground truth in the training process, is comprised of the same maps as the low-resolution input, plus an AO map $\AO_t^\GT \in [0,1]^{4H \times 4W}$. Here, a value of one indicates no occlusion and a value of zero total occlusion.
Thus, the ground truth image can be written as $O_t^\GT := \{M_t^\GT, N_t^\GT, D_t^\GT, \AO_t^\GT\} \in \R^{6 \times 4H \times 4W}$.
Once the network is trained with $I_t^\LR$ and $O_t^\GT$, it can be used to predict a new high-resolution output image
$O_t^\HR := \{M_t^\HR, N_t^\HR, D_t^\HR, \AO_t^\HR\} \in \R^{6 \times 4H \times 4W}$ from a given low-resolution image and the high-resolution output of the previous frame.

\subsection{Super-Resolution Surface Prediction}
\label{sec:Method:Pipeline}

Once the network has been trained, new low-resolution images are processed by the stages illustrated in \autoref{fig:Pipeline} to predict high-resolution output maps. For the inference step, we build upon  the frame-recurrent neural network architecture of Sajjadi\etal\cite{sajjadi2018frame}.
At the current timestep $t$, the network is given the input $I_t^\LR$ and the previous high-resolution prediction $O_{t-1}^\HR$, warped using the image-space flow, for temporal coherence. It produces the current prediction $O_t^\HR$, and after a post-processing step also the final color $C_t^\HR \in [0,1]^{3 \times 4H \times 4W}$.

\paragraph{1. Upscaling and Warping:}\label{sec:Method:Pipeline:Warping}
After upscaling the screen-space flow $F_t^\LR$, it is used as described to warp all previous estimated maps $O_{t-1}^\HR$, leading to $\tilde{O}_{t-1}^\HR \in \R^{6 \times 4H \times 4H}$.

\paragraph{2. Flattening:}\label{sec:Method:Pipeline:Flattening}
Next, the warped previous maps $\tilde{O}_{t-1}^\HR$ are flattened into the low resolution by applying a space-to-depth transformation \cite{sajjadi2018frame}
\begin{equation}
    S_s : \R^{6 \times 4H \times 4W} \rightarrow \R^{4^2 6 \times H \times W} .
\end{equation}
I.e., every 4x4 block of the high-resolution image is mapped to a single pixel in the low-resolution image. The channels of these $4x4=16$ pixels are concatenated, resulting in a new low-resolution image with 16-times the number of channels: $O_f$.

\paragraph{3. Super-Resolution:}\label{sec:Method:Pipeline:Superres}
The super-resolution network then receives the current low resolution input $I_t^\LR$ (5 channels) and the flattened, warped prediction from the previous frame $O_f$ (i.e., 16*6 channels).
The network then estimates the six channels of the output $O_t^\HR$, the high-resolution mask, normal, depth, and ambient occlusion.

\paragraph{4. Shading:}\label{sec:Method:Pipeline:Shading}
To generate a color image, we apply screen-space Phong shading with ambient occlusion as a post-processing step,
\begin{equation}
    C_\textit{rgb} = \text{Phong}(c_a, c_d, c_s c_m, N_t^\HR)*\AO_t^\HR ,
\end{equation}
with the ambient color $c_a$, diffuse color $c_d$, specular color $c_s$ and material color $c_m$ as parameters.

The network also produces a high-resolution mask $M_t^\HR$ as output. While the input mask $M_t^\LR$ was comprised only of values -1 (outside) and +1 (inside), $M_t^\HR$ can take on any values. Hence, $M_t^\HR$ is clamped first to $[-1,+1]$ and then rescaled to $[0,1]$, leading to $M'_t{}^\HR$. This map then acts exactly like an alpha-channel and allows the network to smooth out edges, i.e., 
\begin{equation}
    C_t^\HR = \text{lerp}(c_\text{bg}, C_\textit{rgb}, M'_t{}^\HR) .
\end{equation}

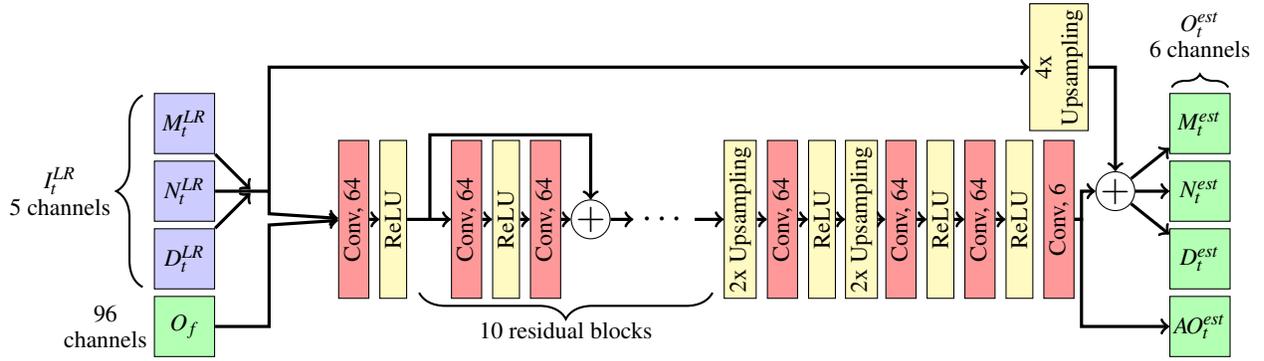
\begin{figure*}
    \centering
    \tikzset{ArchitectureIO/.style={draw,rectangle,minimum size=0.8cm,inner sep=0pt}}
\tikzset{LearnableLayer/.style={draw,rectangle,inner sep=2pt,fill=red!40,rotate=90,minimum width=2.1cm}}
\tikzset{FixedLayer/.style={draw,rectangle,inner sep=2pt,fill=yellow!30,rotate=90,minimum width=2.1cm}}
\tikzset{Connection/.style={very thick, ->}}

\begin{tikzpicture}[scale=0.75,
	every text node part/.style={align=center}]

\node[ArchitectureIO,fill=blue!20] (Min) at (0,0) {$M_t^{LR}$};
\node[ArchitectureIO,fill=blue!20] (Nin) at (0,-1.2) {$N_t^{LR}$};
\node[ArchitectureIO,fill=blue!20] (Din) at (0,-2.4) {$D_t^{LR}$};
\node[inner sep=0pt] (Iin) at (1.2,-1.2) {};
\draw [decorate,decoration={brace,amplitude=10pt},xshift=-4pt,yshift=0pt,thick]
	(-0.6,-2.9) -- (-0.6,0.5) node [black,midway,xshift=-1.1cm] 
	{$I_t^{LR}$\\5 channels};

\node[ArchitectureIO,fill=green!30] (Of) at (0,-3.6) {$O_f$};
\node at (-1.4,-3.6) {96\\channels};

\node[LearnableLayer] (C1) at (3,-1.7) {Conv, 64};
\node[FixedLayer] (R1) at (3.7,-1.7) {ReLU};

\node[LearnableLayer] (C2) at (5,-1.7) {Conv, 64};
\node[FixedLayer] (R2) at (5.7,-1.7) {ReLU};
\node[LearnableLayer] (C3) at (6.4,-1.7) {Conv, 64};
\node[circle,draw,inner sep=0pt] (A1) at (7.2,-1.7) {\LARGE $+$};
\node[minimum size=0] (dot1) at (8.5,-1.7) {\LARGE $\cdots$};
\draw [decorate,decoration={brace,amplitude=10pt},xshift=-4pt,yshift=0pt,thick]
	(9.5,-3) -- (4.3,-3) node [black,midway,yshift=-0.5cm] 
	{10 residual blocks};

\node[FixedLayer] (U1) at (9.85,-1.7) {2x Upsampling};
\node[LearnableLayer] (C4) at (10.6,-1.7) {Conv, 64};
\node[FixedLayer] (R3) at (11.3,-1.7) {ReLU};
\node[FixedLayer] (U2) at (12,-1.7) {2x Upsampling};
\node[LearnableLayer] (C5) at (12.7,-1.7) {Conv, 64};
\node[FixedLayer] (R4) at (13.4,-1.7) {ReLU};
\node[LearnableLayer] (C6) at (14.1,-1.7) {Conv, 64};
\node[FixedLayer] (R5) at (14.8,-1.7) {ReLU};
\node[LearnableLayer] (C7) at (15.5,-1.7) {Conv, 6};
\node[circle,draw,inner sep=0pt] (A2) at (16.5,-1.2) {\LARGE $+$};
\node[FixedLayer,minimum width=1cm] (U3) at (15.5,1) {4x\\Upsampling};

\node[ArchitectureIO,fill=green!30] (Mo) at (18,0) {$M_t^{est}$};
\node[ArchitectureIO,fill=green!30] (No) at (18,-1.2) {$N_t^{est}$};
\node[ArchitectureIO,fill=green!30] (Do) at (18,-2.4) {$D_t^{est}$};
\node[ArchitectureIO,fill=green!30] (AOo) at (18,-3.6) {${A\!O}_t^{est}$};
\draw [decorate,decoration={brace,amplitude=5pt},yshift=+4pt,xshift=0pt,thick]
	(17.5,0.5) -- (18.5,0.5) node [black,midway,yshift=+0.7cm] 
	{$O_t^{est}$\\6 channels};

\draw[Connection] (Min) -- (Iin);
\draw[Connection] (Nin) -- (1.5,-1.2) -- (1.5,-1.6) -- (C1);
\draw[Connection] (Din) -- (Iin);
\draw[Connection] (Of) -- (1.5,-3.6) -- (1.5,-1.8) -- (C1);

\draw[Connection] (C1) -- (R1);
\draw[Connection] (R1) -- (C2);
\draw[Connection] (C2) -- (R2);
\draw[Connection] (R2) -- (C3);
\draw[Connection] (C3) -- (A1);
\draw[Connection] (4.35,-1.7) -- (4.35,-0.2) -- (7.2,-0.2) -- (A1);
\draw[Connection] (A1) -- (dot1);

\draw[Connection] (dot1) -- (U1);
\draw[Connection] (U1) -- (C4);
\draw[Connection] (C4) -- (R3);
\draw[Connection] (R3) -- (U2);
\draw[Connection] (U2) -- (C5);
\draw[Connection] (C5) -- (R4);
\draw[Connection] (R4) -- (C6);
\draw[Connection] (C6) -- (R5);
\draw[Connection] (C7) -- (15.9,-1.7) -- (15.9,-1.2) -- (A2);
\draw[Connection] (C7) -- (15.9,-1.7) -- (15.9,-3.6) -- (AOo);
\draw[Connection] (1.5,-1.2) -- (1.5,1) -- (U3);
\draw[Connection] (U3) -- (16.5,1) -- (A2);
\draw[Connection] (A2) -- (Mo);
\draw[Connection] (A2) -- (No);
\draw[Connection] (A2) -- (Do);

\end{tikzpicture}
    \caption{Network architecture for the SRNet. 
    Within the network, $\oplus$ indicates component-wise addition of the residual. All convolutions use 3x3 kernels with stride 1. Bilinear interpolation was used for the upsampling layers.}
    \label{fig:Architecture}
\end{figure*}

\subsection{Loss Functions}\label{sec:Method:Loss}

In the following, we describe the loss functions we have included into the network. The loss functions are used by the network during training to calculate the model error in the optimization process. They allow for controlling the importance of certain features and, thus, the fidelity by which these features can be predicted by the network. The losses we describe are commonly used in artificial neural networks, yet in our case they are applied separately to different channels of the predicted and ground truth images. The total loss function we used for training the network is a weighted sum of the losses functions below. In \autoref{sec:evaluation}, we analyze the effects of different loss functions on prediction quality.


\paragraph{1. Spatial loss}\label{sec:Method:Loss:Simple}
As a baseline, we employ losses with regular vector norms, i.e. $L_1$ or $L_2$, on the different outputs of the network.
Let $X$ be either the mask $M$, the normal $N$, the ambient occlusion $\AO$ or the shaded output $C$.
Then the L1 and L2 losses are given by:
\begin{align}
    \mathcal{L}_{X,l1} = || X_t^\HR - X_t^\GT ||_1 \nonumber ,~ ~
    \mathcal{L}_{X,l2} = || X_t^\HR - X_t^\GT ||_2^2 .
\end{align}

\paragraph{2. Perceptual loss}\label{sec:Method:Loss:Perc}
Perceptual losses, as proposed by Gatys\etal~\cite{gatys2016image}, Dosovitskiy~and~Brox\cite{NIPS2016_6158} and Johnson\etal\cite{johnson2016perceptual}, have been widely adopted to guide learning tasks towards detailed outputs instead of smoothed mean values.
The idea is that two images are similar if they have similar activations in the latent space of a pre-trained network.
Let $\phi$ be the function that extracts the layer activations when feeding the image into the feature network.
Then the distance is computed by
\begin{equation}
    \mathcal{L}_{X,P} = ||\phi(X_t^\HR) - \phi(X_t^\GT) ||_2^2 .
\end{equation}
As feature network $\phi$, the pretrained VGG-19 network~\cite{VGG} is used. We used all convolution layers in all spatial dimensions as features, with weights scaled so that each layer has the same average activation when evaluated over all input images.

Since the VGG network was trained to recognize objects in color images in the space $[0,1]^3$, the shaded output $C$ can be directly used.
This perceptual loss on the color space can be backpropagated to the network outputs, i.e. normals and ambient occlusions, with the help of the differentiable Phong shading. 
This shading is part of the loss function, and is implemented such that gradients can flow from the loss evaluation into the weight update of the neural network during training.
Hence, with our architecture the network receives a gradient so that it can learn how the output, e.g., the generated normals, should be modified such that the shaded color matches the look of the target image.
When applying the perceptual loss on other entries, the input has to be transformed first. The normal map is rescaled from scale $[-1,+1]^3$ to $[0,1]^3$, and depth and masking maps are converted to grayscale RGB images.
We did not use additional texture or style loss terms \cite{gatys2016image,sajjadi2017enhancenet}, since these introduce artificial details and roughness in the image which is not desired in smooth isosurface renderings.

\paragraph{3. Temporal loss}\label{sec:Method:Loss:Temporal}
All previous loss functions worked only on the current image. To strengthen the temporal coherence and reduce flickering, we employ a temporal $L_2$ loss \cite{chen2017coherent}. We penalize differences between the current high-resolution image $O_t^\HR$ and the previous, warped high-resolution image $O_w$ with
\begin{equation}
    \mathcal{L}_{X,\text{temp}} = || X_t^\HR - \tilde{X}_{t-1}^\HR ||_2^2,
\end{equation}
where $X$ can be $M$, $N$, $\AO$ or $C$.

In the literature, more sophistic approaches to improve the temporal coherence are available, e.g. using temporal discriminators \cite{chu2018temporally}. These architectures give impressive result, but are quite hard to train. We found that already with the proposed simple temporal loss, good temporal coherence can be achieved in our current application. We refer the readers to the accompanying video for a sequence of a reconstruction over time.

\paragraph{4. Loss masking}\label{sec:Method:Loss:Restriction}
During screen-space shading in the post-process (see \autoref{sec:Method:Pipeline:Shading}), the output color is modulated with the mask indicating hits with surface points. Pixels where the mask is -1 are set to the background color. Hence, the normal and ambient occlusion values produced by the network in these areas are irrelevant for the final color.

To reflect this in the loss function, loss terms that do not act on the mask (i.e. normals, ambient occlusions, colors) are itself modulated with the mask so that areas that are masked out don't contribute to the loss.
We found this to be a crucial step that simplifies the network's task: In empty regions, the ground truth images are filled with default values in the non-mask channels, while with loss masking, the network does not have to match these values.

\paragraph{5. Adversarial Training}\label{sec:Method:Loss:Adversarial}
Lastly, we also employed an adversarial loss as inspired 
by Chu\etal~\cite{chu2018temporally}. 
In adversarial training, a discriminator network is trained parallel to the super-resolution network generator. The discriminator receives ground truth images and predicted images, and is trained to classify whether the input was ground truth or not. This discriminator is then used in the loss function of the generator network.
For further details of adversarial training and GANs let us refer to, e.g., Goodfellow\etal~\cite{goodfellow2014generative}.

In more detail, for evaluating the predicted images, the discriminator is provided with
\begin{itemize}
    \item the high-resolution output $O_t^\HR$, and optionally the color $C_t^\HR$,
    \item the input image $I_t^\LR$ as a conditional input to learn and penalize the mismatching between input and output,
    \item the previous frames $I_{t-1}^\LR, O_{t-1}^\HR$, and optionally $C_{t-1}^\HR$ to learn to penalize for  temporal coherence.
\end{itemize}
To evaluate the discriminator score of the ground truth images, the predicted images $\circ^\HR$ are replaced by $\circ^\GT$.

As a loss function we use the binary cross entropy loss. Formally, let $z$ be the input over all timesteps and $G(z)$ the generated results, i.e. the application of our super-resolution prediction on all timesteps.
Let $D(x)$ be the discriminator that takes the high-resolution outputs as input and produces a single scalar score.
Then the discriminator is trained to distinguish fake from real data by minimizing
\begin{equation}
    \mathcal{L}_{\text{GAN},D} = - \log(D(x))-\log(1-D(G(z))).
\end{equation}
The generator is trained to minimize
\begin{equation}
    \mathcal{L}_{\text{GAN},G} = -\log(D(G(z)))
\end{equation}

When using the adversarial loss, we found that it is important to use a network that is pre-trained with different loss terms, e.g. L2 or perceptual, as a starting point. Otherwise the discriminator becomes too good too quickly so that no gradients are available for the generator. 

\section{Learning Methodology}

In this chapter, we describe our used network architecture, as well as the training and inference steps in more detail.

\subsection{Network Architecture}\label{sec:Method:Architecture}

Our network architecture employs a modified frame-recurrent neural network consisting of a series of residual blocks \cite{sajjadi2018frame}. A visual overview is given in \autoref{fig:Architecture}.
The generator network starts with one convolution layer to reduce the 101 input channels (5 from $I_t^\LR$, $6*4^2$ from $O_f$) into 64 channels. Next, 10 residual blocks are used, each of which contains 2 convolutional layers. 
These are followed by two upscaling blocks (2x bilinear upscaling, a convolution and a ReLU) arriving at a 4x resolution, still with 64 channels. In a final step, two convolutions process these channels to reduce the latent feature space to
the desired 6 output channels. The network is fully convolutional, and all layers use 3x3 kernels with stride 1.

The network as a whole is a residual network, i.e. it only learns changes to the input. As shown by previous work \cite{he2016deep}, this improves the generalizing capabilities of the network, as it can focus on generating the residual content. Hence, the 5 channels of the input are bilinearly upsampled and added to the first five channels of the output, producing $M_t^\HR, N_t^\HR$ and $D_t^\HR$.
The only exception is $\AO_t^\HR$, which is inferred from scratch, as there is no low-resolution input for this map.


\subsection{Training and Inference}\label{sec:Method:Training}

The overall loss for training the super-resolution network is a weighted sum of the loss terms described in \autoref{sec:Method:Loss}:
\begin{eqnarray}
\mathcal{L}_{\text{full}} &=& 
    \lambda_{M} \mathcal{L}_{M,l1}+ \lambda_{\AO} \mathcal{L}_{\AO,l1}
    +\lambda_{N} \mathcal{L}_{N,l1} \nonumber \\
&&  +\lambda_{G} \mathcal{L}_{\text{GAN},G} 
    +\lambda_{N,P} \mathcal{L}_{N,P}+\lambda_{C,P} \mathcal{L}_{C,P}  \\
&&  +\lambda_{\AO\text{temp}} \mathcal{L}_{\AO,\text{temp}}
    +\lambda_{N\text{temp}} \mathcal{L}_{N,\text{temp}}
    +\lambda_{C\text{temp}} \mathcal{L}_{C,\text{temp}} \nonumber.
\end{eqnarray}
For a comparison of three networks with different weights on the loss functions, see \autoref{sec:evaluation},
the corresponding details of the $\lambda$ weights are given in \autoref{tab:NetworkLosses}.
%
\begin{table}[b]
    \centering
    \begin{tabular}{r|p{5.2cm}|l}
        Network & Losses & PSNR \\ \hline
        GAN 1 & $\mathcal{L}_{M,l1}+\mathcal{L}_{\AO,l1}+0.1\mathcal{L}_{C,P}+\mathcal{L}_{\text{GAN},G}$ & 19.095 \\
        GAN 2 & $\mathcal{L}_{M,l1}+\mathcal{L}_{\AO,l1}+0.1\mathcal{L}_{N,P}+\mathcal{L}_{\text{GAN},G}$ \newline No AO in the shading for the GAN & 17.175 \\
        L1-normal & $\mathcal{L}_{M,l1}+\mathcal{L}_{\AO,l1}+10\mathcal{L}_{N,l1}+0.1\mathcal{L}_{C,\text{temp}}$ & 21.659 
    \end{tabular}
    \caption{Loss weights for three selected networks that are used to generate the result images. The PSNR of our methods are much higher than the ones of nearest-neighbor and bilinear interpolations which are 14.282 and 14.045 respectively.}
    \label{tab:NetworkLosses}
\end{table}

\begin{figure}
    \vspace{-6pt}
    \centering
    \includegraphics[width=0.24\linewidth]{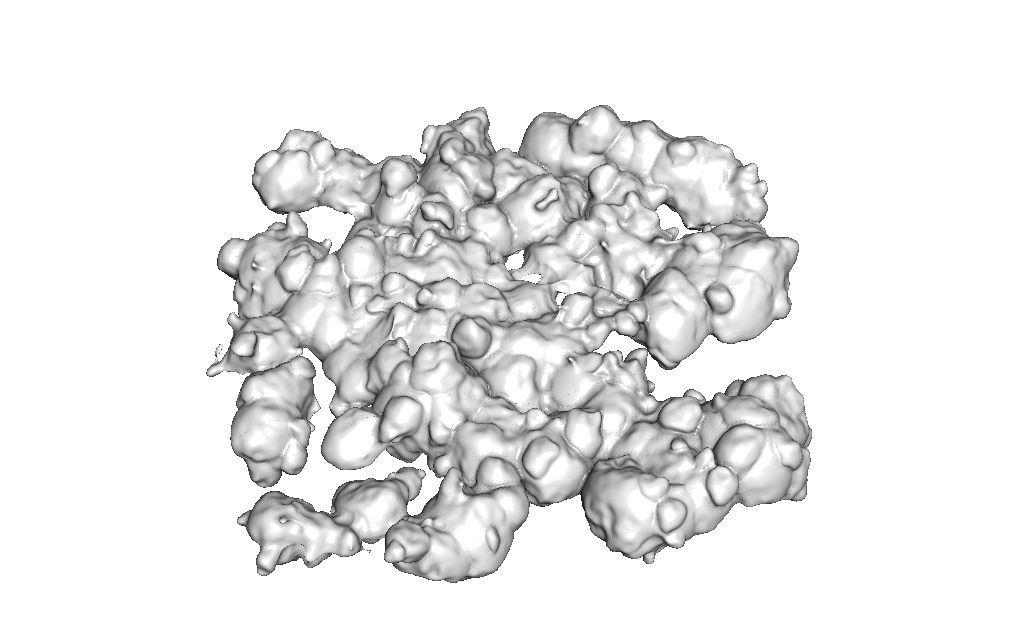}
    \includegraphics[width=0.24\linewidth]{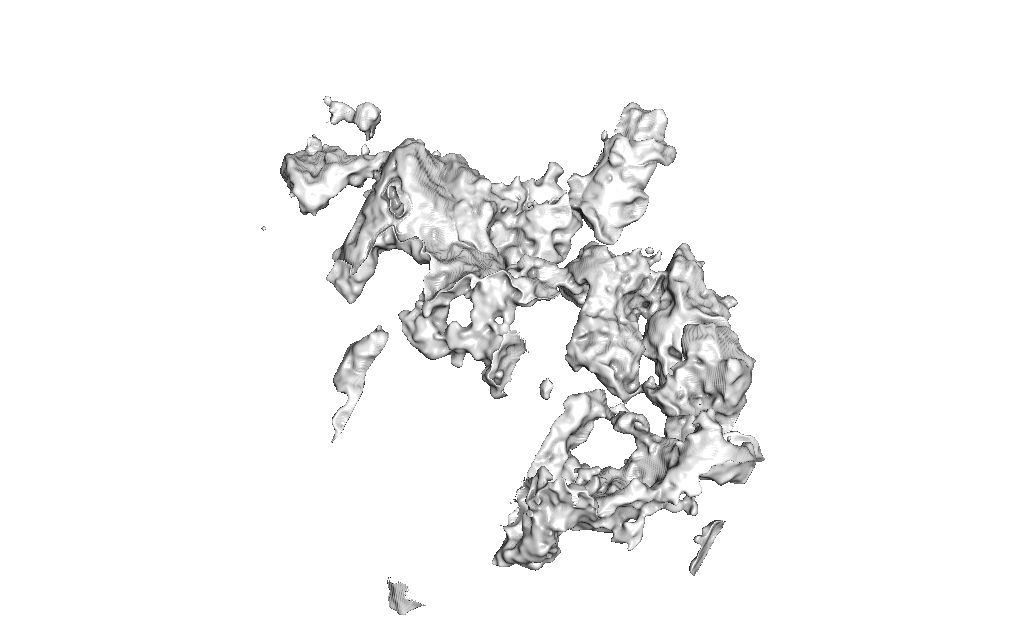}
    \includegraphics[width=0.24\linewidth]{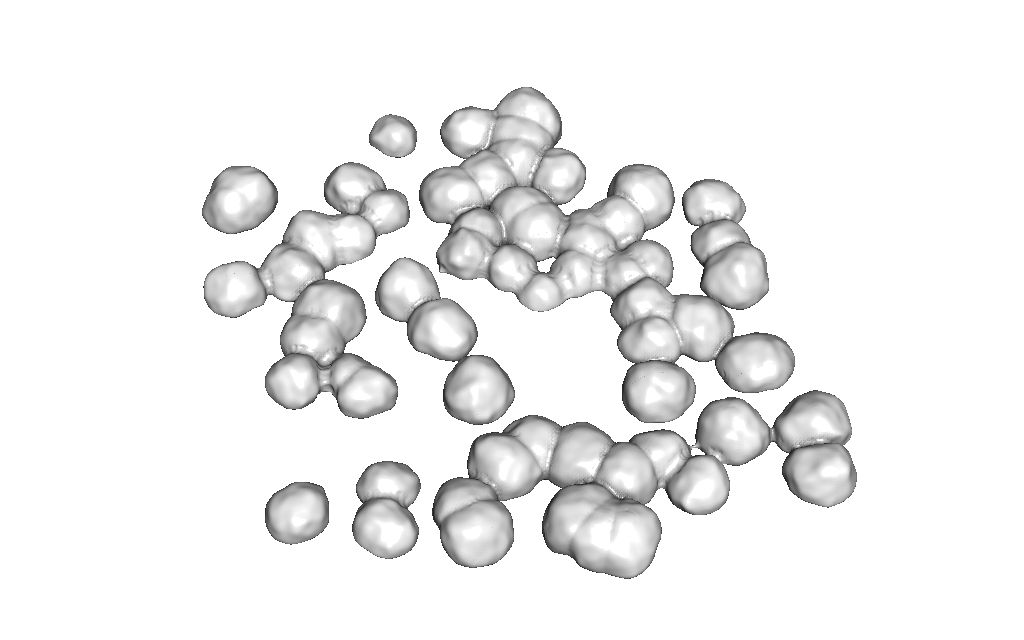}
    \includegraphics[width=0.24\linewidth]{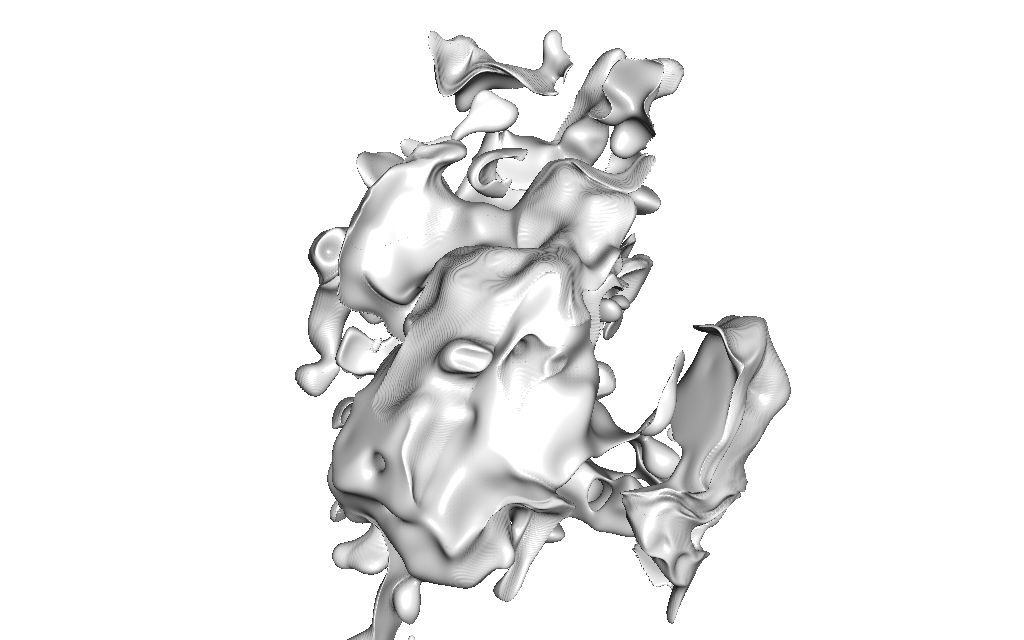}\\
    \includegraphics[width=0.24\linewidth]{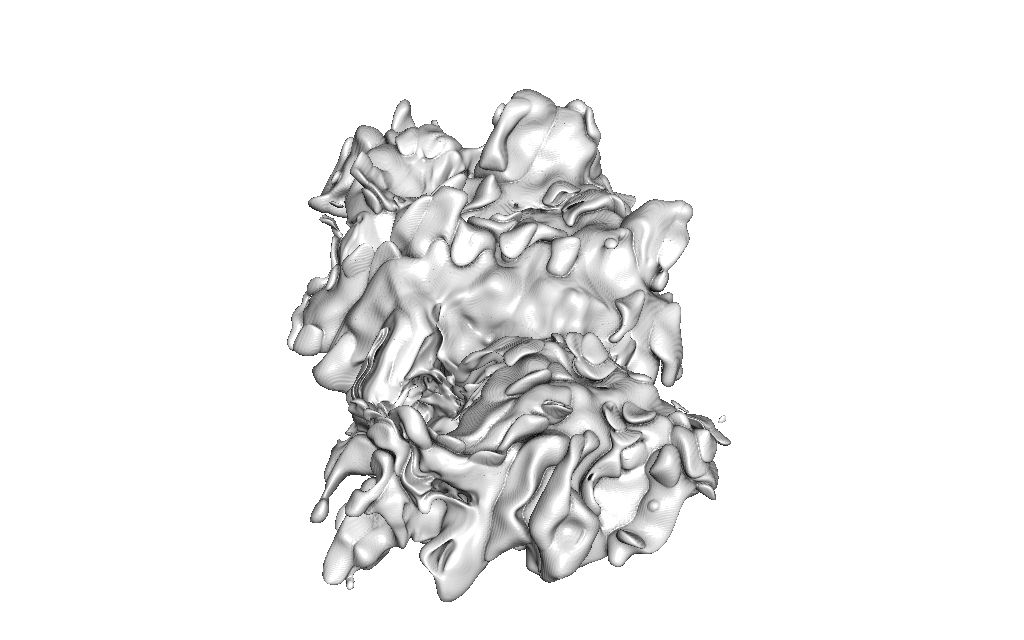}
    \includegraphics[width=0.24\linewidth]{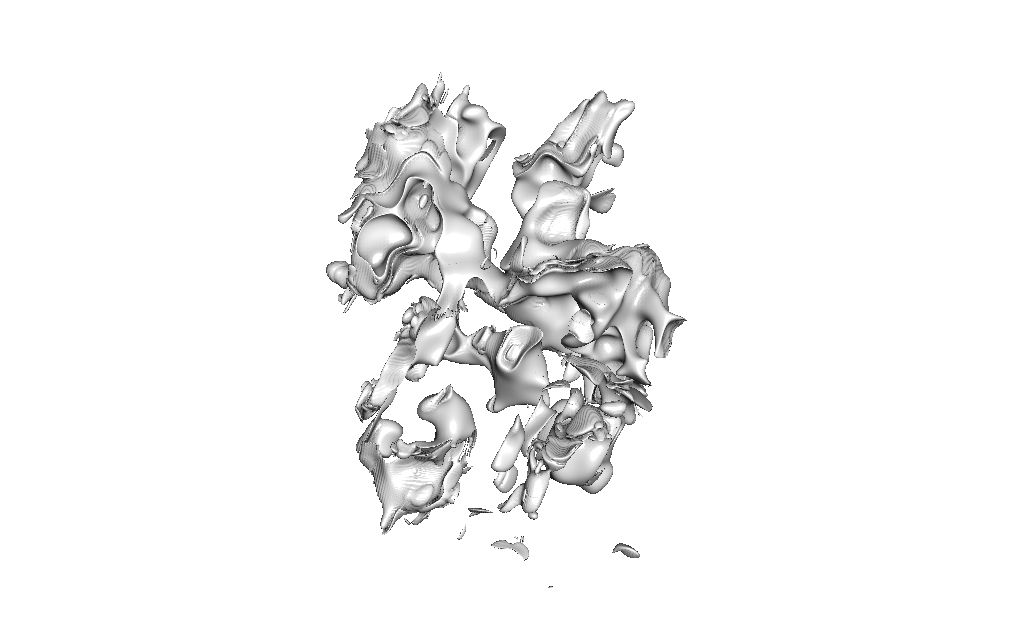}
    \includegraphics[width=0.24\linewidth]{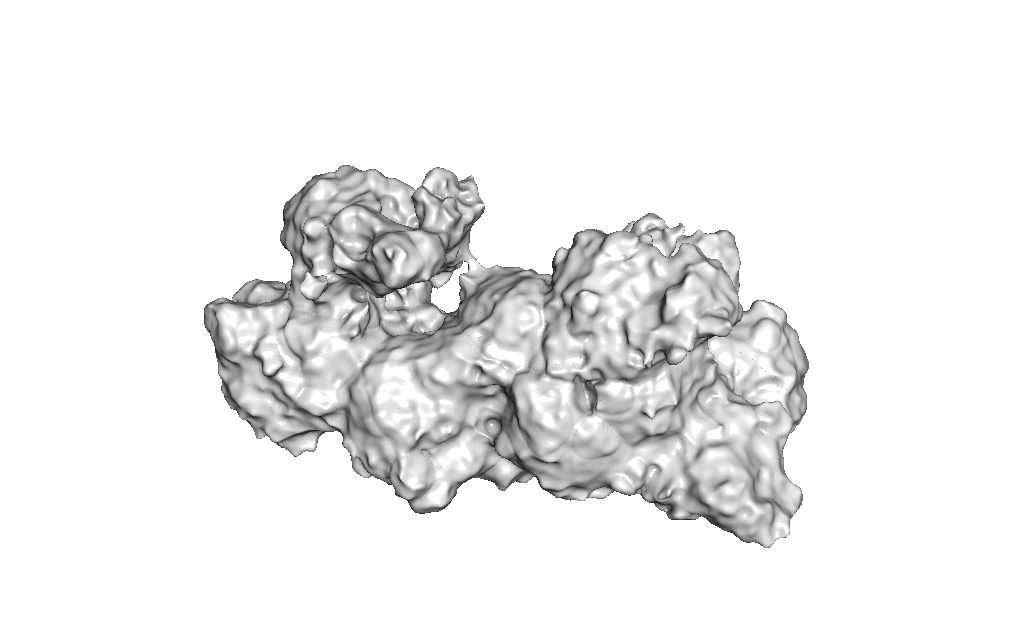}
    \includegraphics[width=0.24\linewidth]{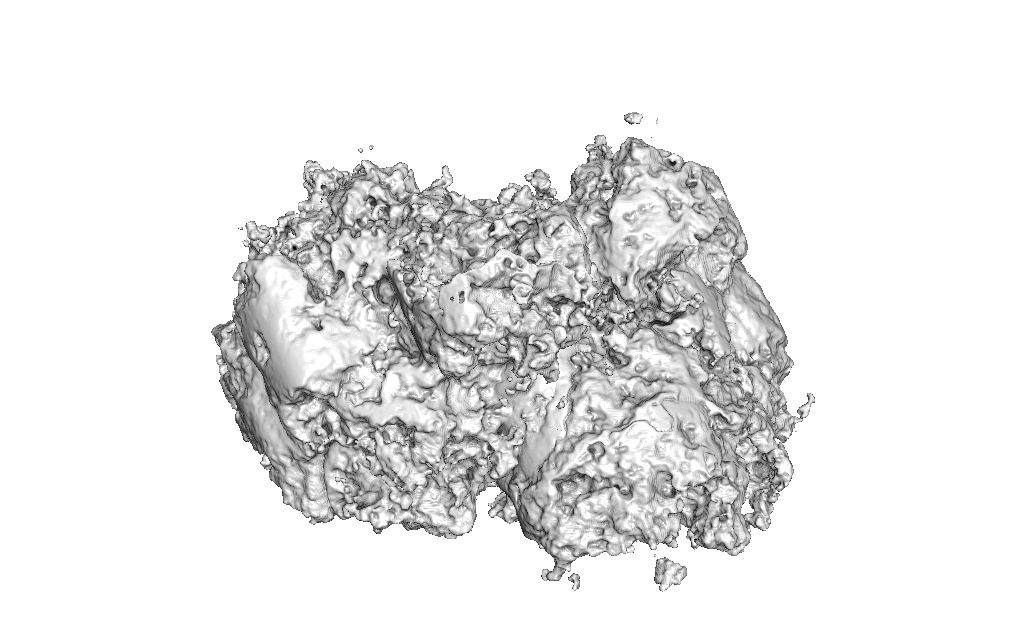}\\
    \includegraphics[width=0.24\linewidth]{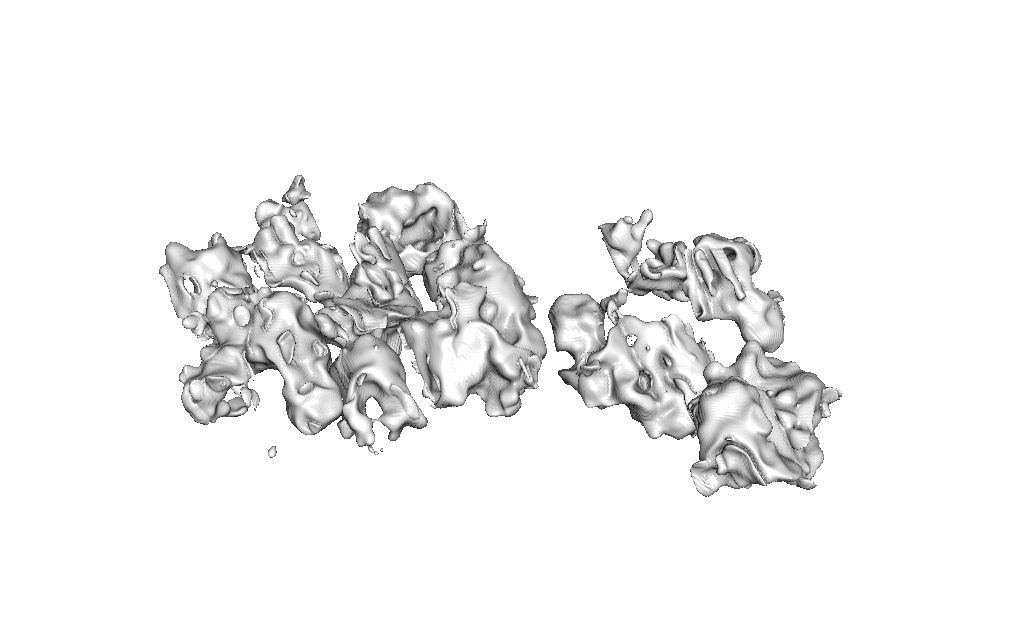}
    \includegraphics[width=0.24\linewidth]{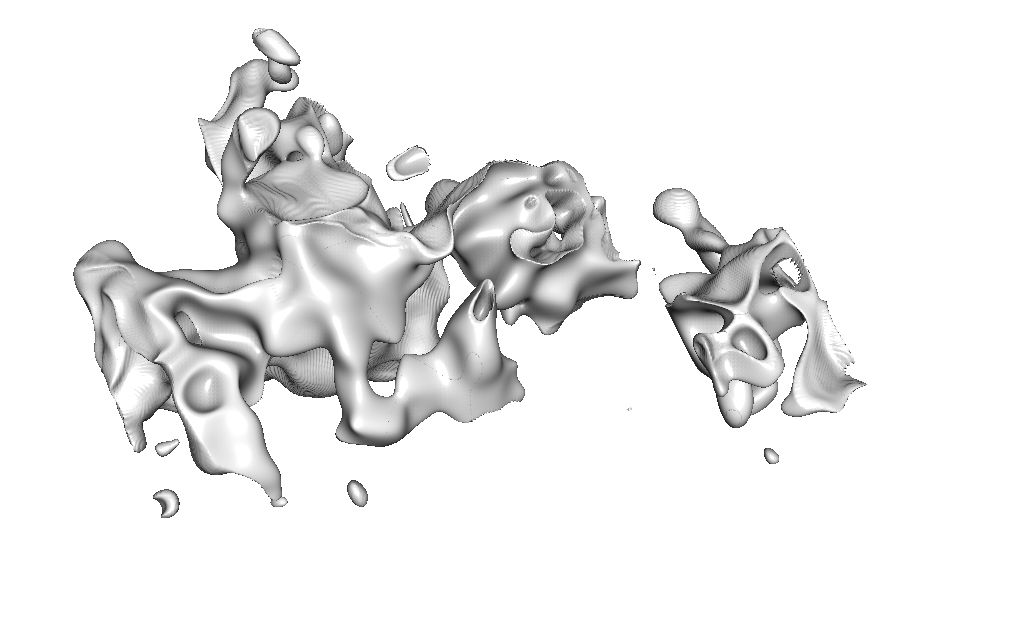}
    \includegraphics[width=0.24\linewidth]{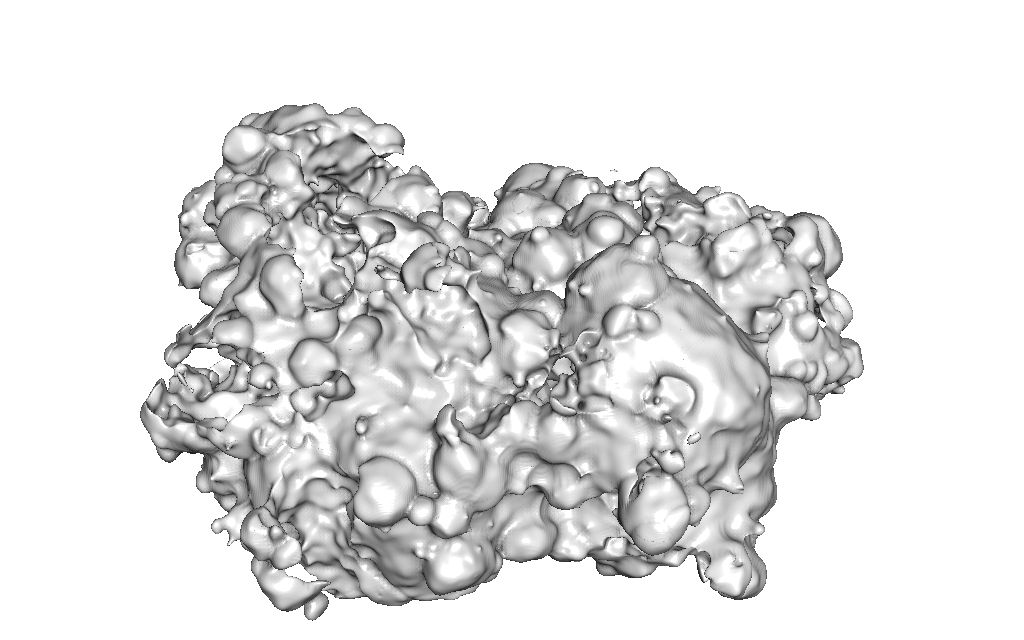}
    \includegraphics[width=0.24\linewidth]{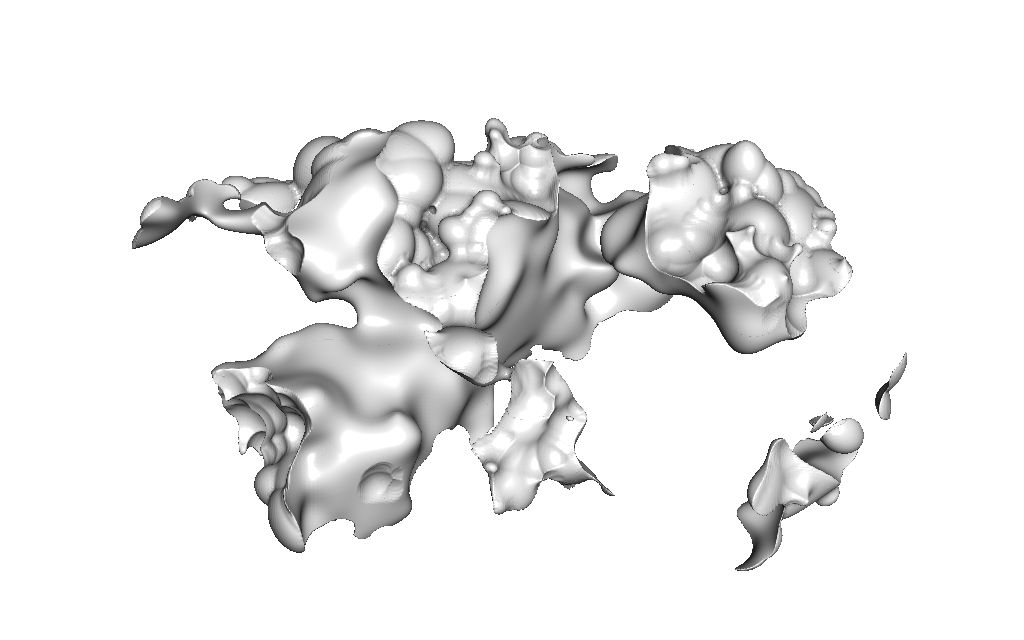}\\
    \includegraphics[width=0.24\linewidth]{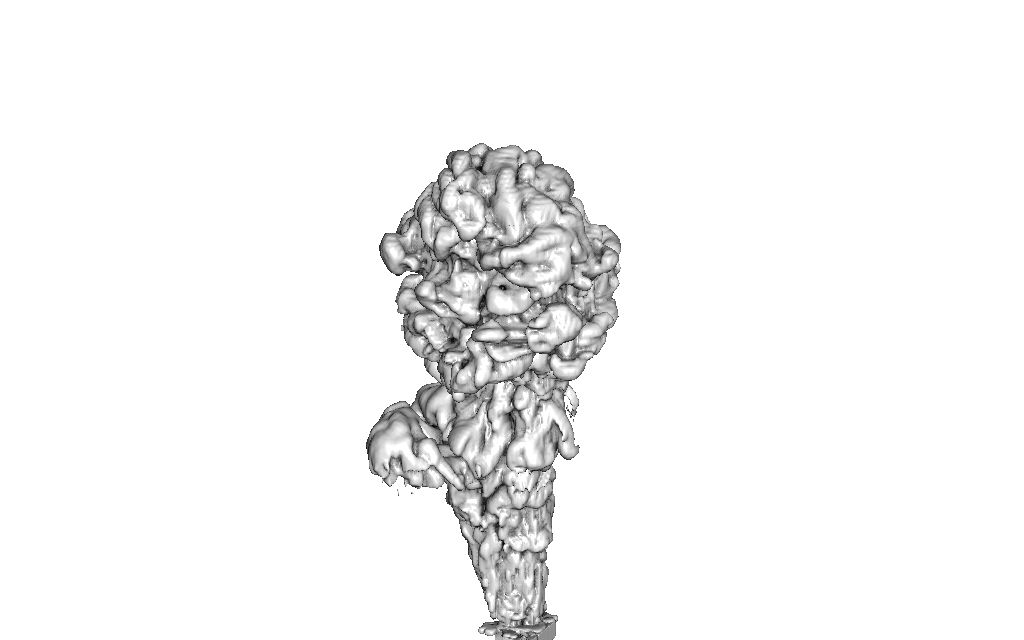}
    \includegraphics[width=0.24\linewidth]{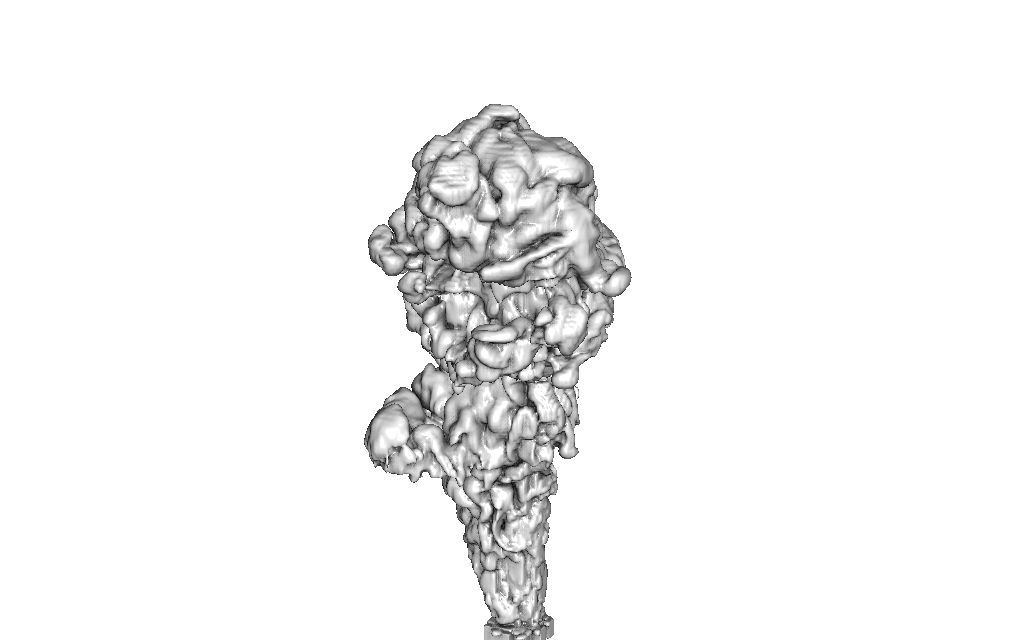}
    \includegraphics[width=0.24\linewidth]{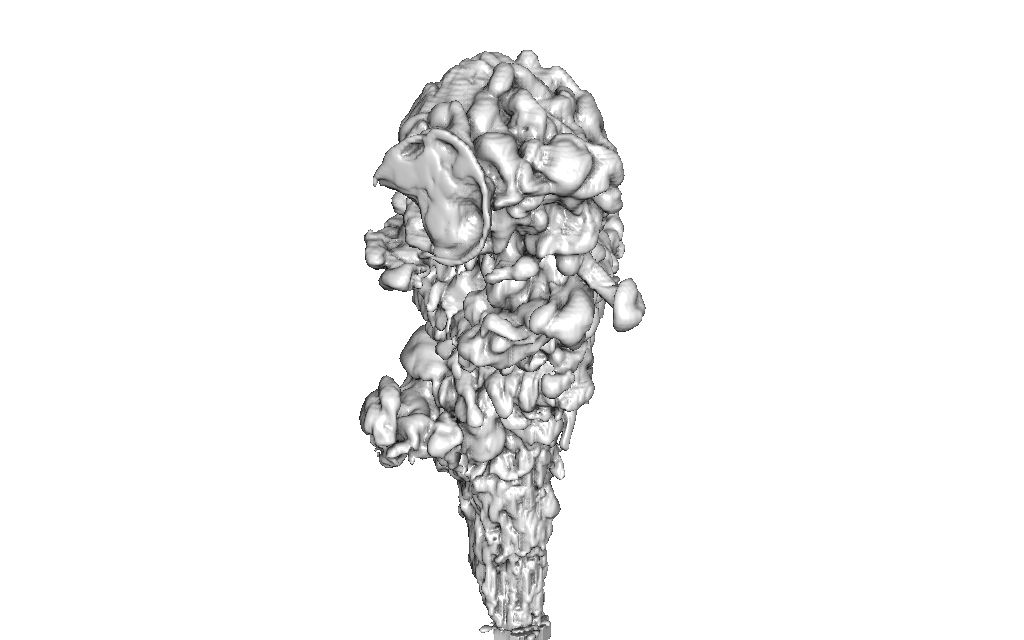} \includegraphics[width=0.24\linewidth]{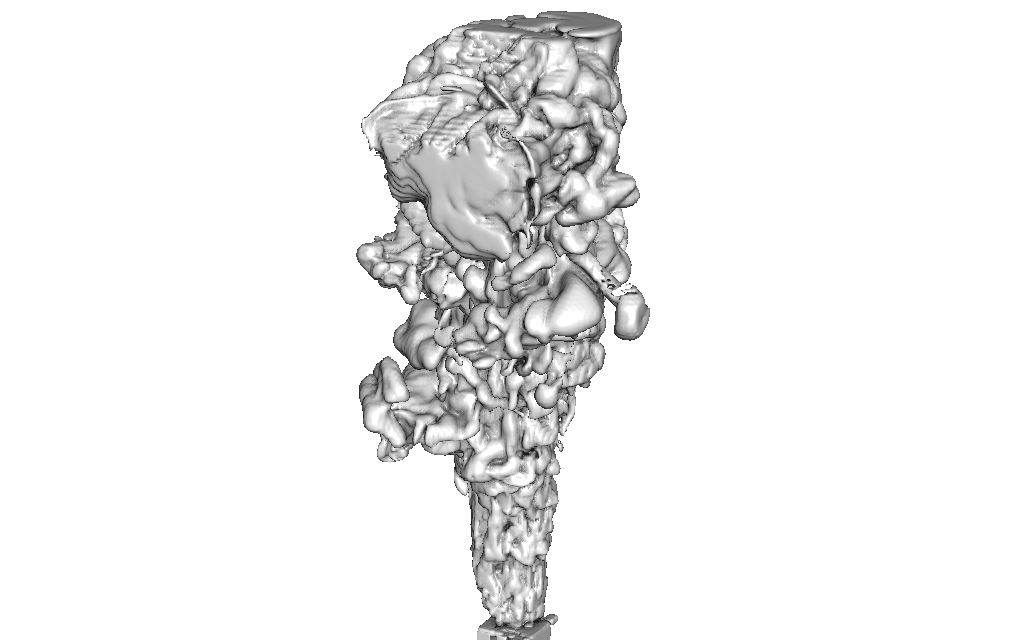}\\
    \includegraphics[width=0.24\linewidth]{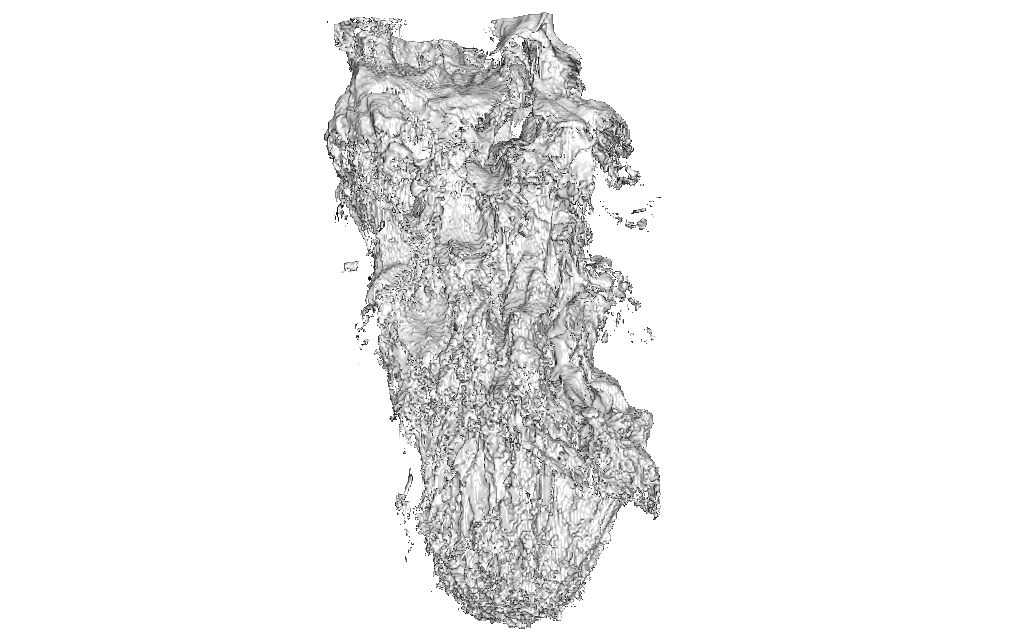}
    \includegraphics[width=0.24\linewidth]{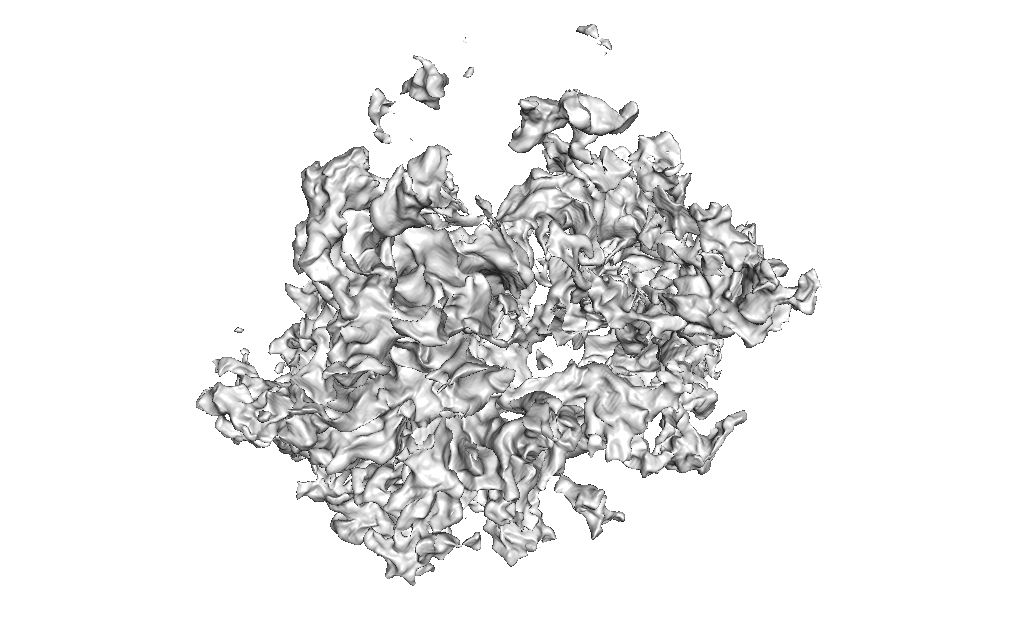}
    \includegraphics[width=0.24\linewidth]{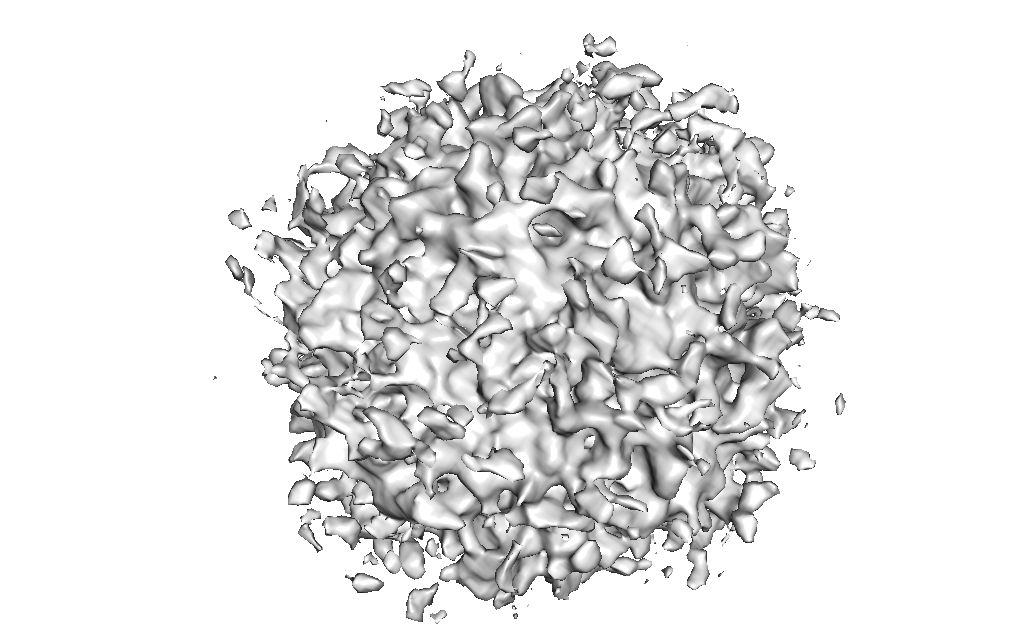}
    \includegraphics[width=0.24\linewidth]{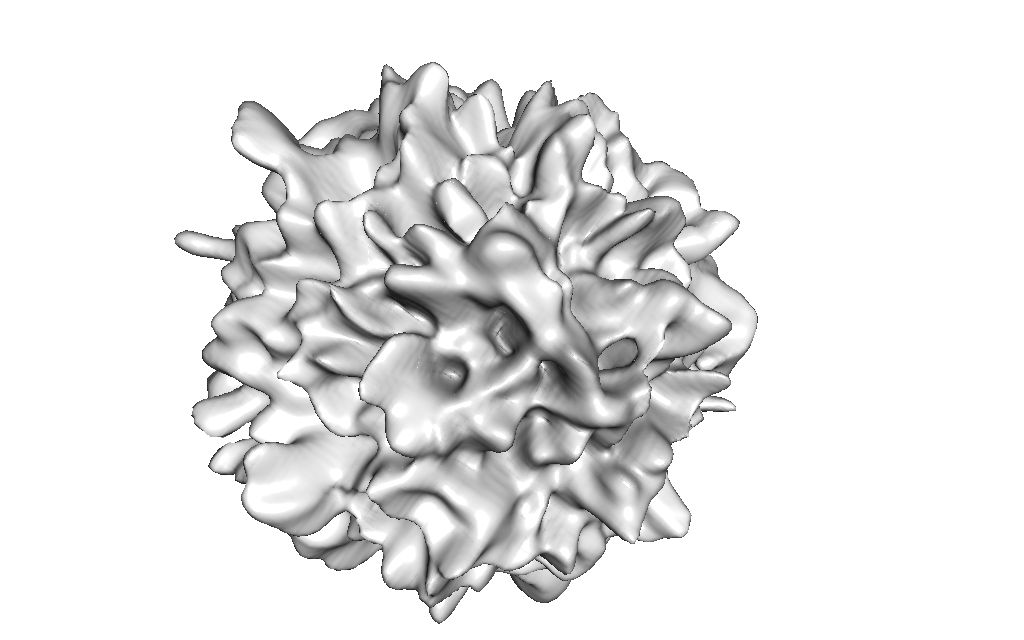}
    \caption{Datasets used for training.}
    \label{fig:TrainingData}
\end{figure}

Our training and validation suite consists of 20 volumes of clouds \cite{Kallweit2017DeepScattering}, smoke plumes simulated with Mantaflow \cite{mantaflow}, and example volumes from OpenVDB \cite{OpenVDB}, see \autoref{fig:TrainingData}.
From these datasets, we rendered 500 sequences of 10 frames each with a low resolution of $128^2$ and random camera movement.
Out of these sequences, 5000 random crops of low resolution size of $32^2$ were taken with the condition of being filled to at least 50\%. These 5000 smaller sequences were then split into training (80\%) and validation (20\%) data.

The ground truth ambient occlusion at each surface point was generated by sampling random directions on the hemisphere and testing if rays along these directions again collide with the isosurface. This gives a much higher visual quality than screen-space ambient occlusion, which test samples against screen-space depth.

We further want to stress that our low-resolution input is directly generated by the raytracer, not a blurred and downscaled version of the original high-resolution image as it is common practice in image and video super-resolution \cite{sajjadi2017enhancenet,chu2018temporally}. Therefore, the input image is very noisy, and poses a challenging task for the network.

To estimate the current frame, the network takes the previous high-resolution prediction $O_{t-1}^\HR$ as input. Since the previous frame is not available for the first frame of a sequence, we evaluated different ways to initialize the first previous high-resolution input:
\begin{enumerate}
    \item All entries set to zero,
    \item Default values: mask=0, normal=$[0,0,1]$, AO=1,
    \item An upscaled version of the current input.
\end{enumerate}
In the literature, the first approach is used most of the time. 
We found that there is hardly any visual difference in the first frame between networks trained with these three options.
For the final networks, we thus used the simplest option 1.

\subsection{Training Characteristics of the Loss Terms}\label{sec:Method:LossDiscussion}



\begin{figure}
    \centering
        \begin{overpic}[width=0.4\linewidth,trim=400 250 400 250, clip]{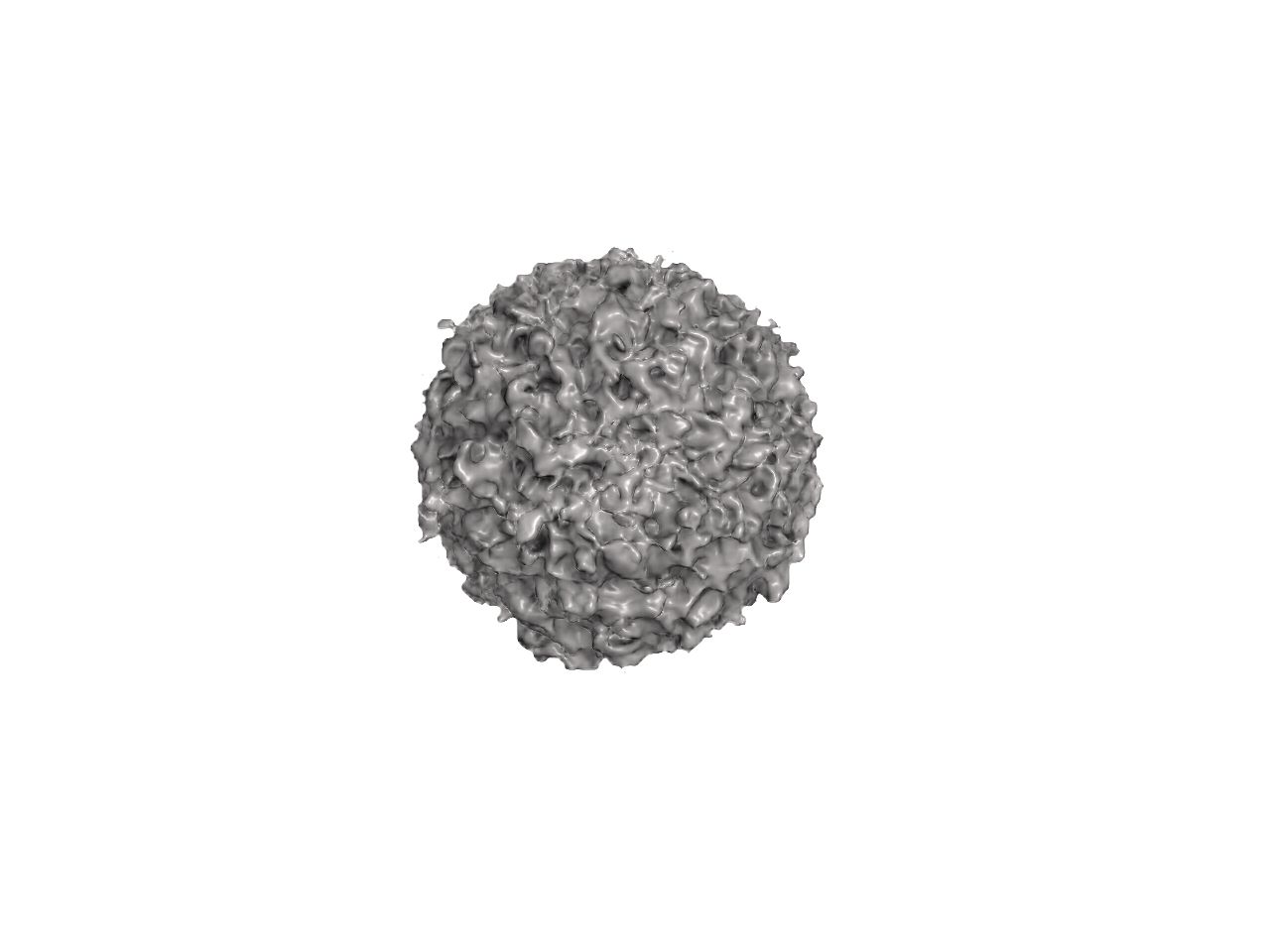}
           \put(5,5){(a)}  
        \end{overpic}
        \begin{overpic}[width=0.4\linewidth,trim=400 250 400 250, clip]{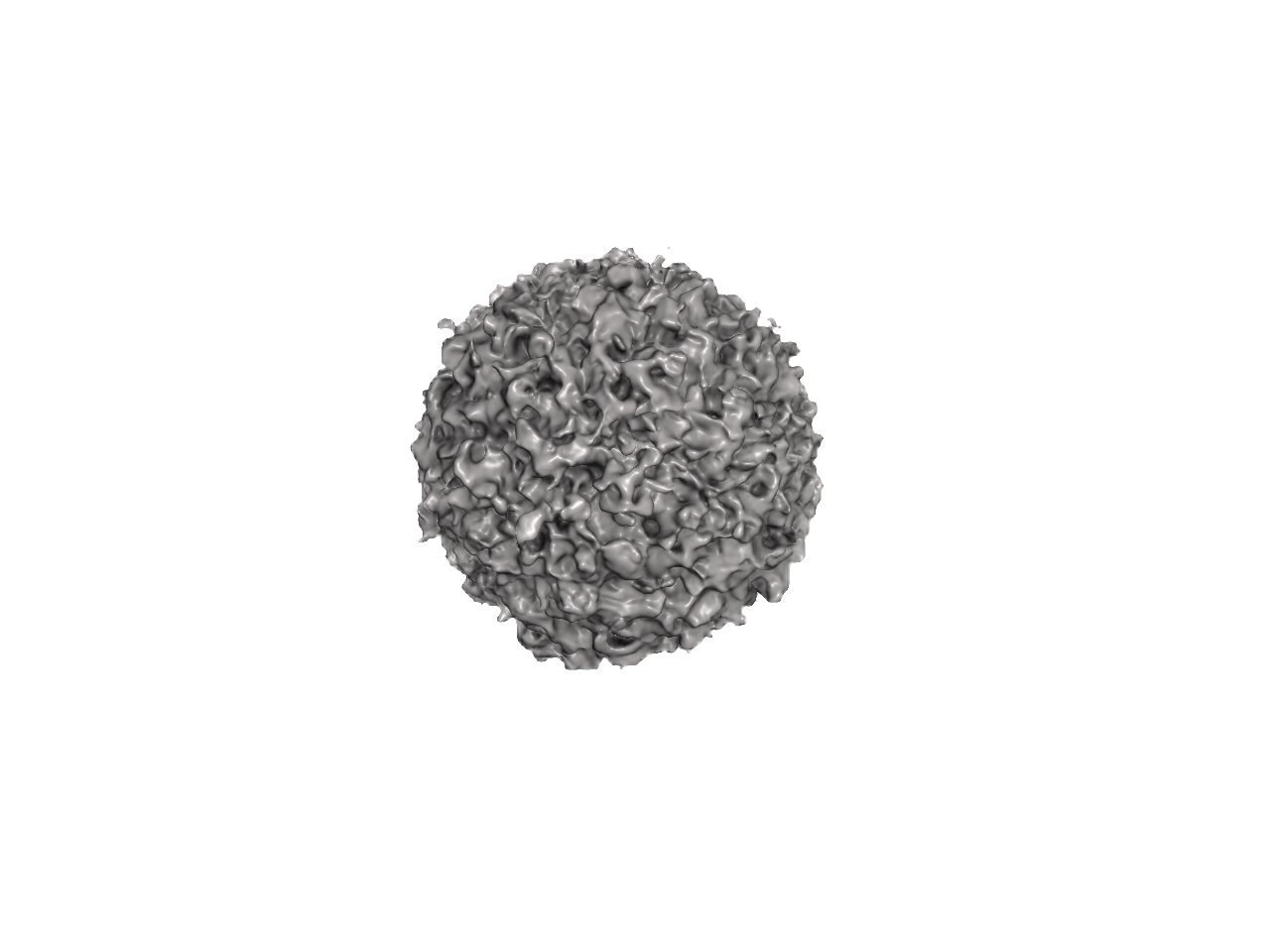}
           \put(5,5){(b)}  
        \end{overpic}
    \caption{Comparison: L1 loss on color (a) vs. on normal (b)}
    \label{fig:L1Comparison}
\end{figure}

Now we give some insights into the effects of the different 
loss term combinations. We found that it is not advisable
to use losses on the color output, and found losses working with the mask, normal and ambient occlusion maps to be preferable. 
Strong weights on the color losses typically degrade the image quality and smooths out results. 
    For a comparison between two networks trained with L1 loss on the color and one with L1 loss on the normal, see \autoref{fig:L1Comparison}. 
    It can be seen that the version with L1 loss on the normal produces much sharper results.
    A comparison between a network with a perceptual loss on the color as dominating term yields the same outcome.
    We think that this is caused by ambiguities in the shading: a darker color can be achieved by a normal more orthogonal to the light, more ambient occlusion or a lower value for the mask. Furthermore, this makes the training depend on the material and light settings, which we aimed to avoid.

The perceptual loss for the normal field $\mathcal{L}_{N,P}$ provides more pronounced details in the output. For the mask and ambient occlusion this is not desirable, as both contain much smoother content in general. Hence, we employ a L1 loss for the mask and ambient occlusion.

The temporal losses $\mathcal{L}_{X,\text{temp}}$ successfully reduce flickering between frames, but can lead to smoothing due to the warping of the previous image. By changing the weighting between temporal losses and e.g. perceptual loss on the normal, 
more focus is put on either sharp details or improved temporal coherence.
Since our adversarial loss includes temporal supervision, we do not include the explicit temporal losses when using $\mathcal{L}_{\text{GAN},G}$.

The adversarial loss generally gives the most details, as can be seen in  \autoref{sec:evaluation:comparison}. Only the GAN loss, however, is typically not sufficient. To overcome plateaus if the discriminator fails to provide gradients or to stabilize the optimization  in the beginning, we combine the GAN loss with a perceptual loss on the normals and an L1 loss on the mask and ambient occlusion with a small weighting factor. 
We also found that the GAN loss improves in quality if the discriminator is provided with the shaded color in addition to masks, normals and depths.
This stands in contrast to the spatial or perceptual loss terms, which did not benefit from the shaded color.


%
%
%


\section{Evaluation}\label{sec:evaluation}

For the evaluation we identified the three best performing networks, two GAN architectures and one dominated by only a L1 loss on the normals, see \autoref{tab:NetworkLosses} for the exact combinations of the losses.
In the following, we evaluate the three architectures in more detail. 

To quantify the accuracy of the super-resolution network, we compute the Peak signal-to-noise ratio (PSNR) of the different networks on the validation data, see also \autoref{tab:NetworkLosses}.
From this table, one can see that the network ``L1-normal''-network performs best.
This is further supported by the visual comparison in the following section.
The PSNR for nearest-neighbor interpolation is $14.282$ and for bilinear interpolation $14.045$.

\subsection{Comparisons with Test Data}\label{sec:evaluation:comparison}
\begin{figure*}[htbp]
    \centering
    \begin{subfigure}[c]{0.24\textwidth}
        \includegraphics[width=\textwidth,trim=150 0 0 0, clip]{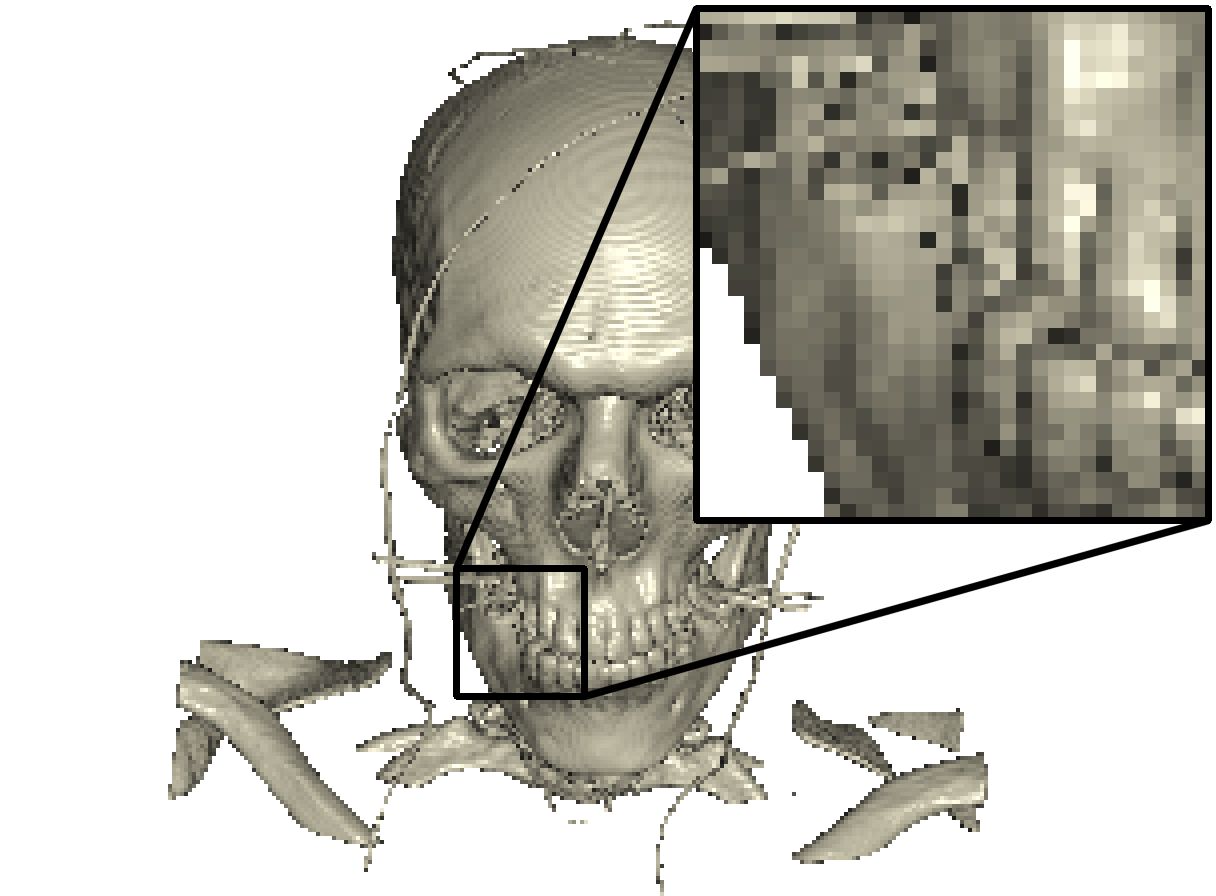}
        \subcaption{Low-resolution input}
    \end{subfigure}
    \begin{subfigure}[c]{0.24\textwidth}
        \includegraphics[width=\textwidth,trim=150 0 0 0, clip]{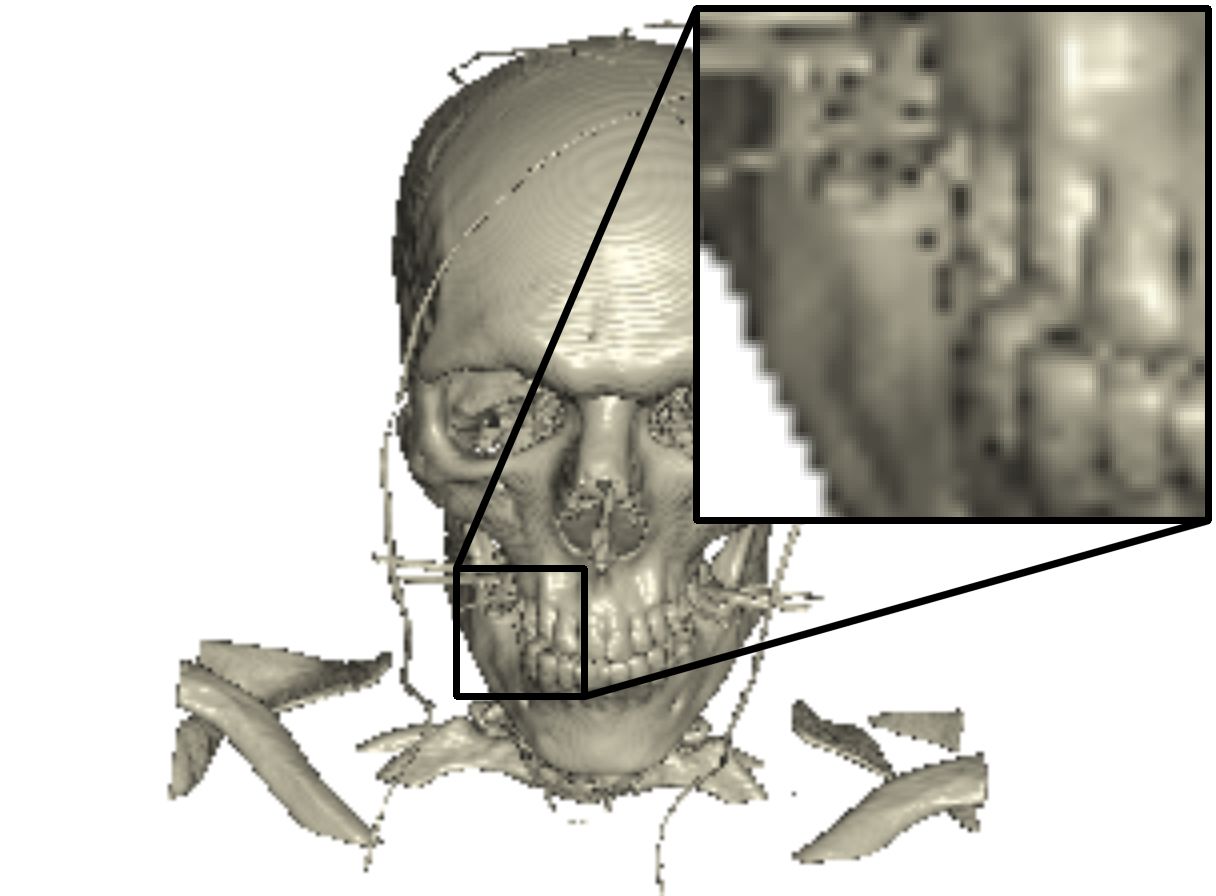}
        \subcaption{Bilinear upscaling}
    \end{subfigure}
    \begin{subfigure}[c]{0.24\textwidth}
        \includegraphics[width=\textwidth,trim=150 0 0 0, clip]{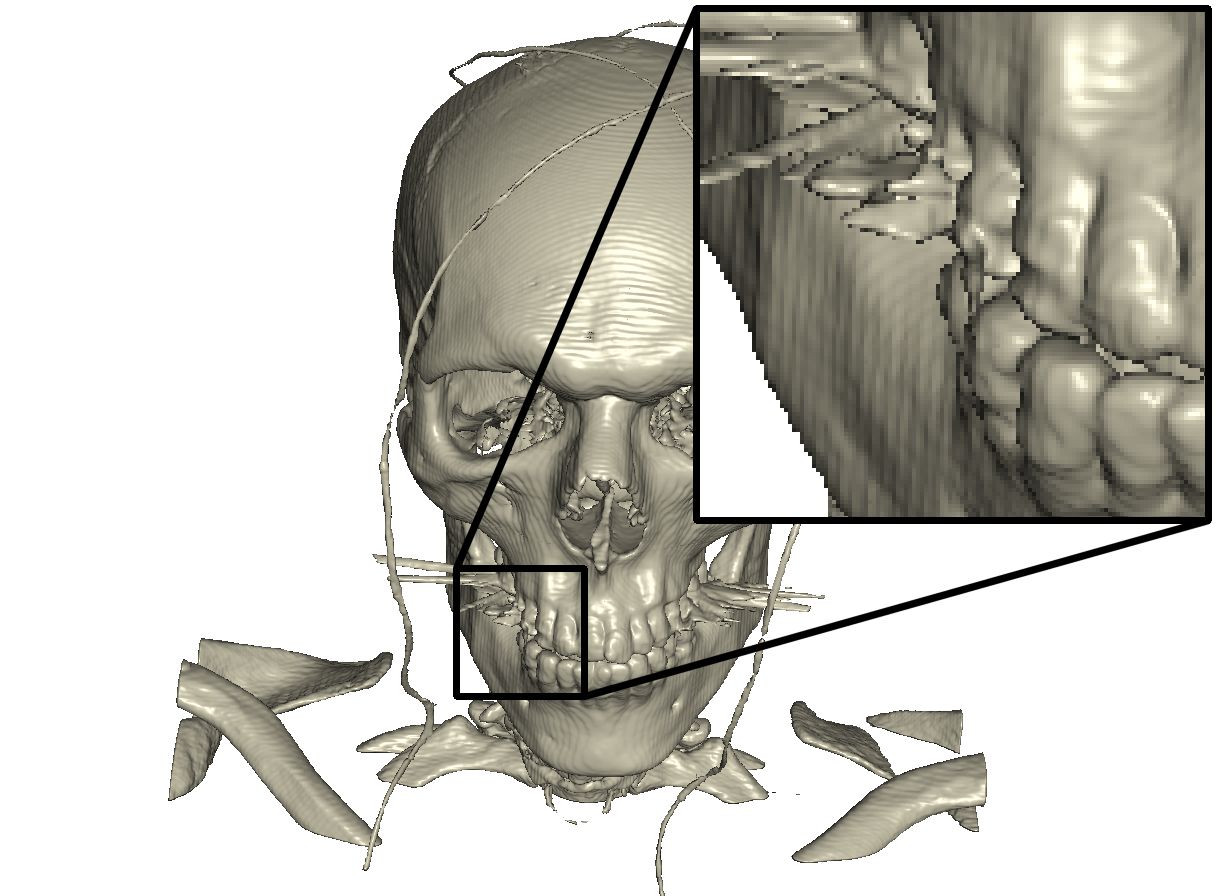}
        \subcaption{Ground truth without AO}
    \end{subfigure}
    \begin{subfigure}[c]{0.24\textwidth}
        \includegraphics[width=\textwidth,trim=150 0 0 0, clip]{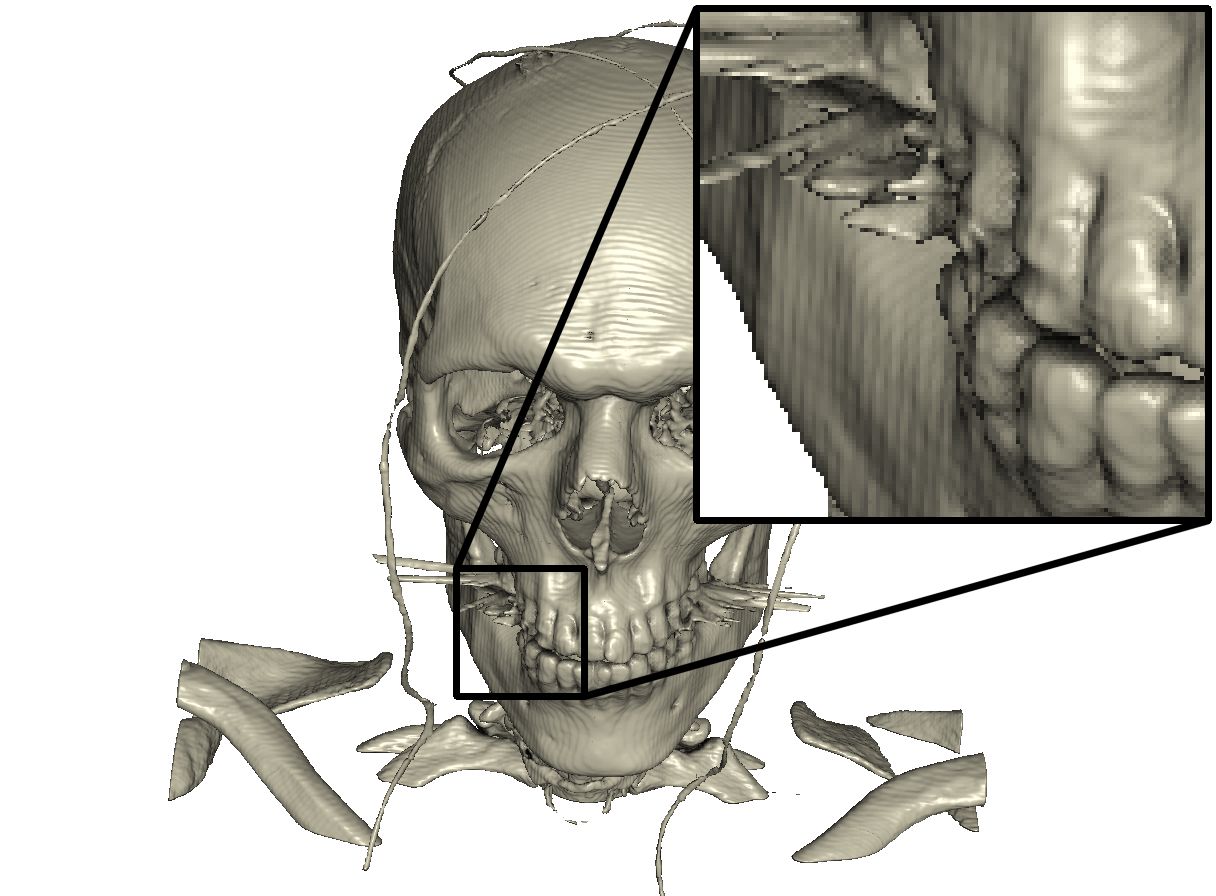}
        \subcaption{Ground truth with AO}
    \end{subfigure}
    \\
    \begin{subfigure}[c]{0.3\textwidth}
        \includegraphics[width=\textwidth]{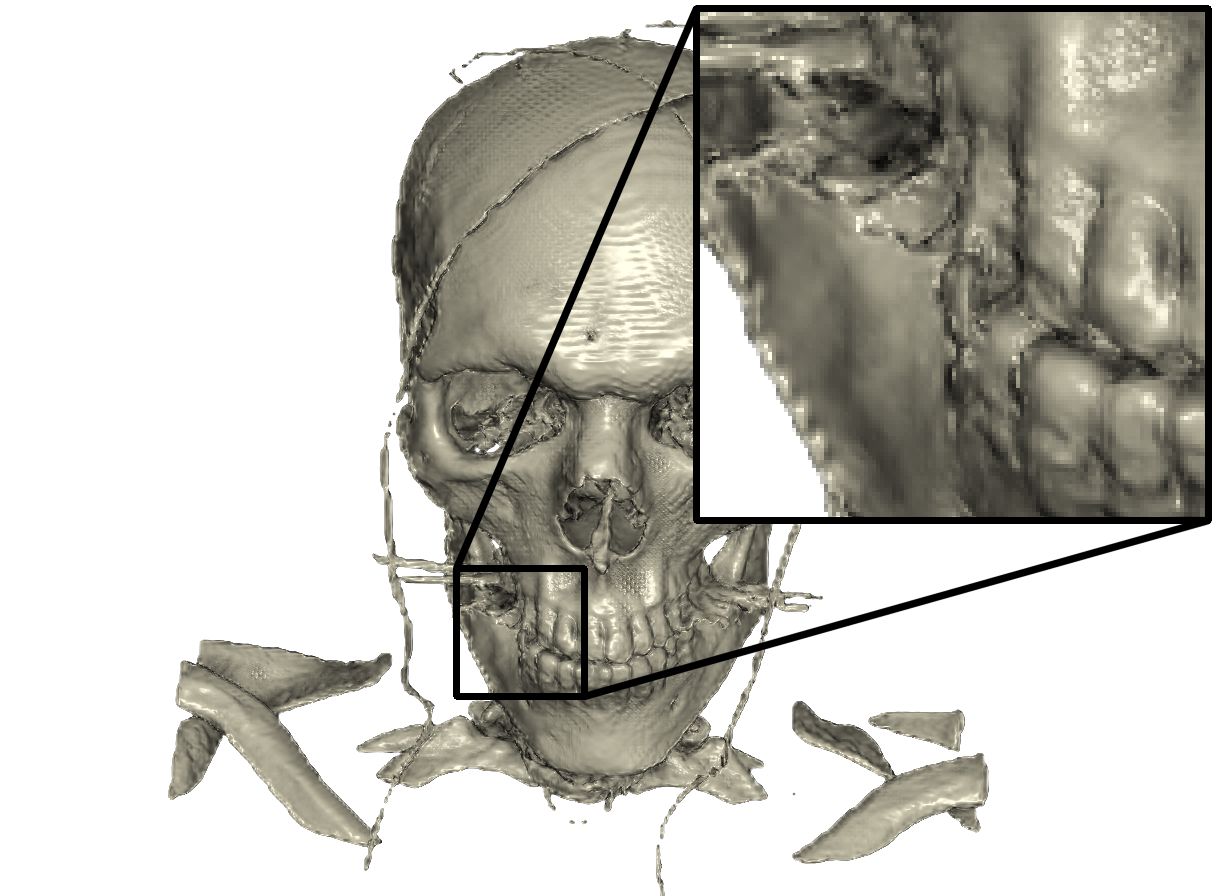}
        \subcaption{GAN 1}
    \end{subfigure}
    \begin{subfigure}[c]{0.3\textwidth}
        \includegraphics[width=\textwidth]{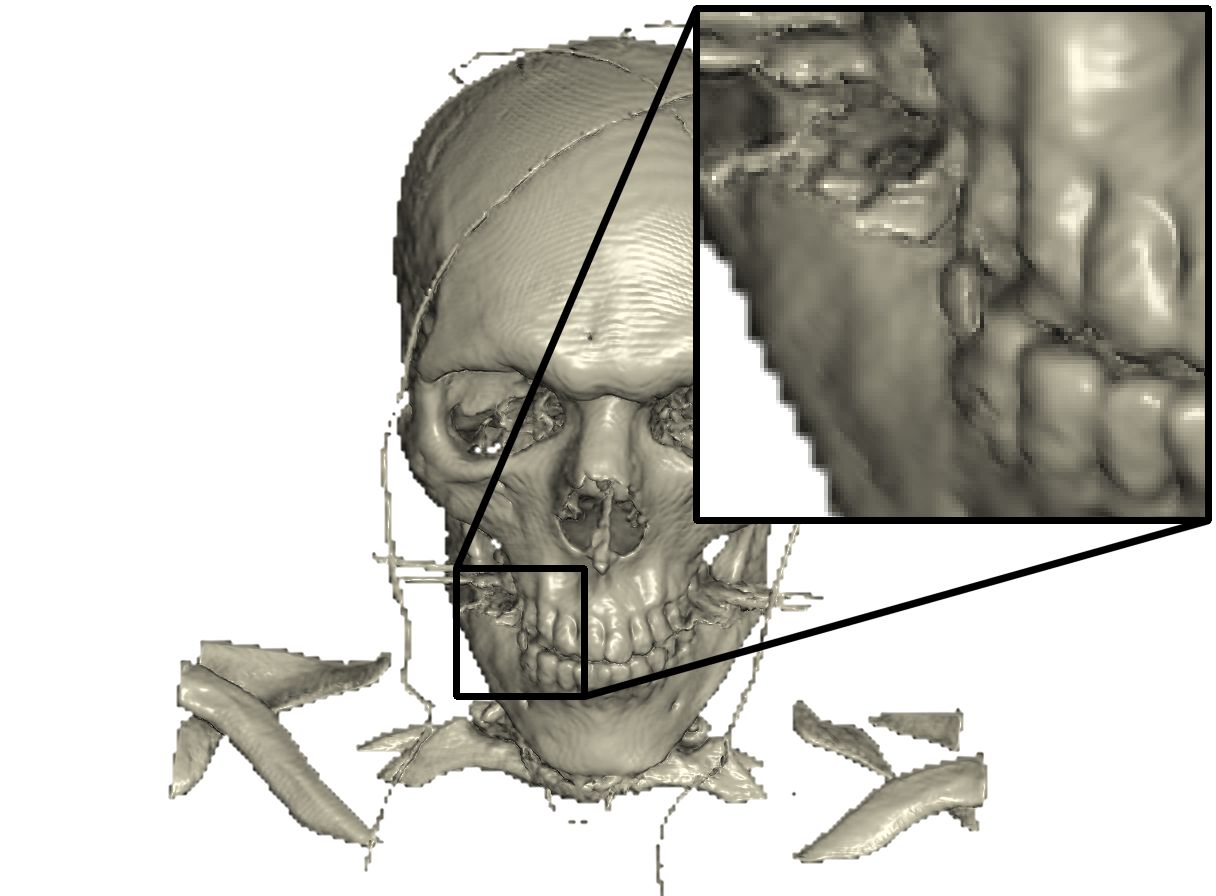}
        \subcaption{GAN 2}
    \end{subfigure}
    \begin{subfigure}[c]{0.3\textwidth}
        \includegraphics[width=\textwidth]{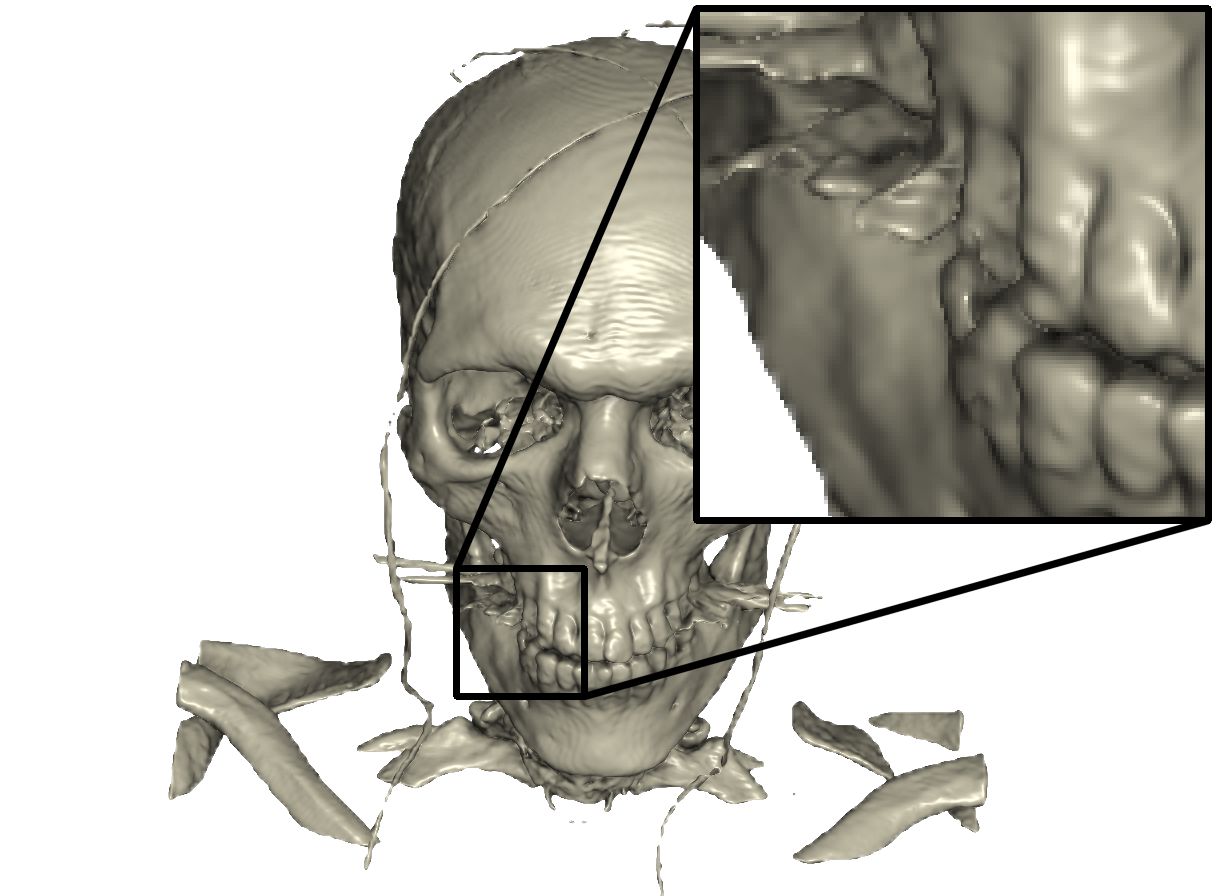}
        \subcaption{L1 loss on normals}
    \end{subfigure}
    \caption{Comparison of different network outputs on a CT scan of the human skull.}
    \label{fig:Comparison:vmhead}
\end{figure*}

\begin{figure*}[htbp]
    \centering
    \begin{subfigure}[c]{0.24\textwidth}
        \includegraphics[width=\textwidth,trim=0 0 100 0, clip]{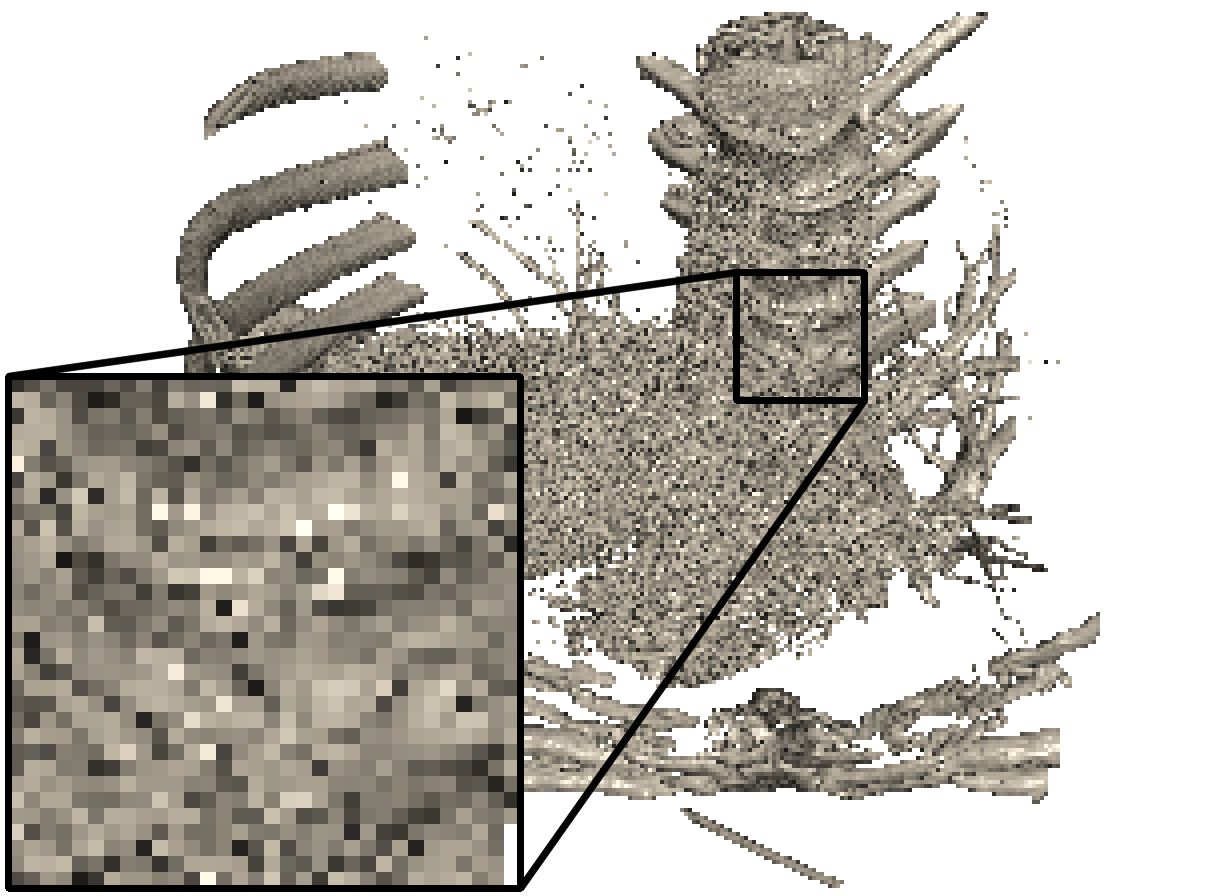}
        \subcaption{Low-resolution input}
    \end{subfigure}
    \begin{subfigure}[c]{0.24\textwidth}
        \includegraphics[width=\textwidth,trim=0 0 100 0, clip]{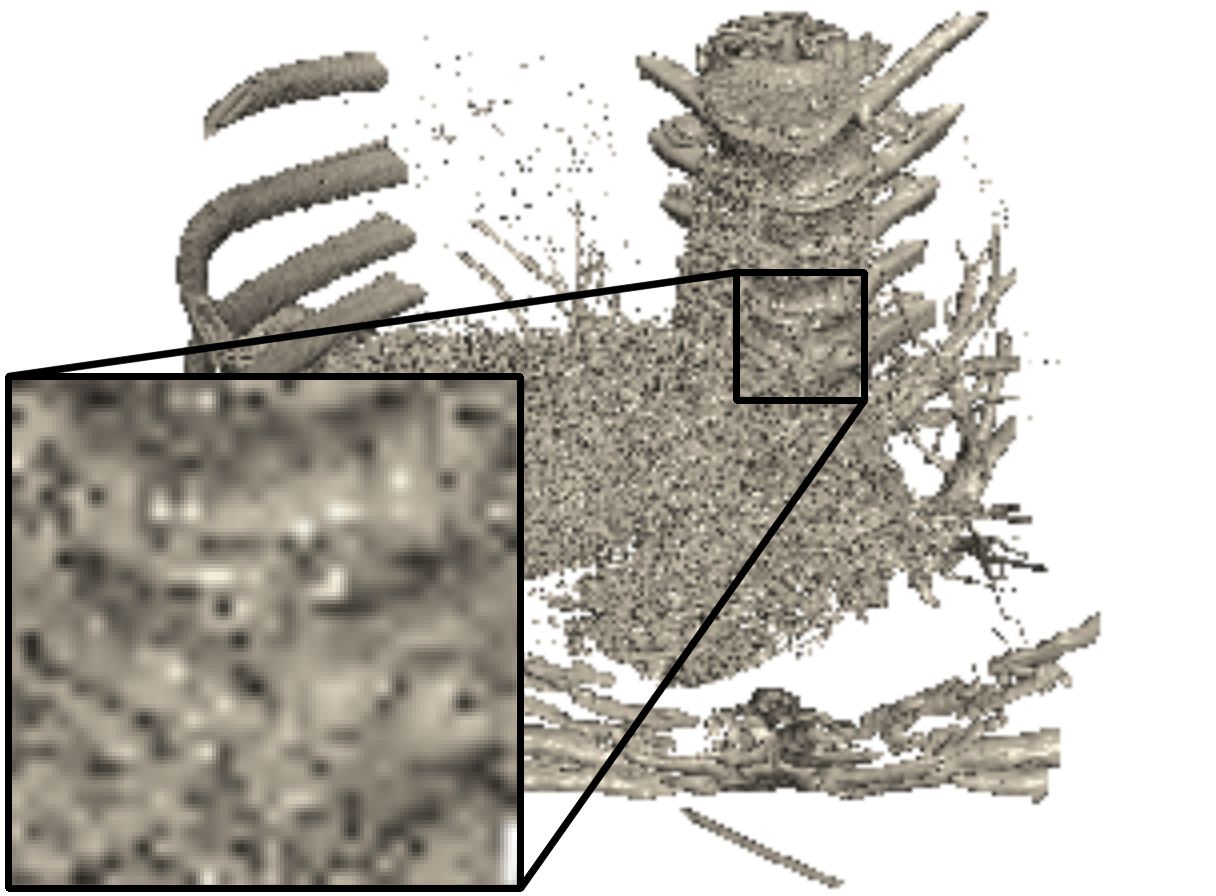}
        \subcaption{Bilinear upscaling}
    \end{subfigure}
    \begin{subfigure}[c]{0.24\textwidth}
        \includegraphics[width=\textwidth,trim=0 0 100 0, clip]{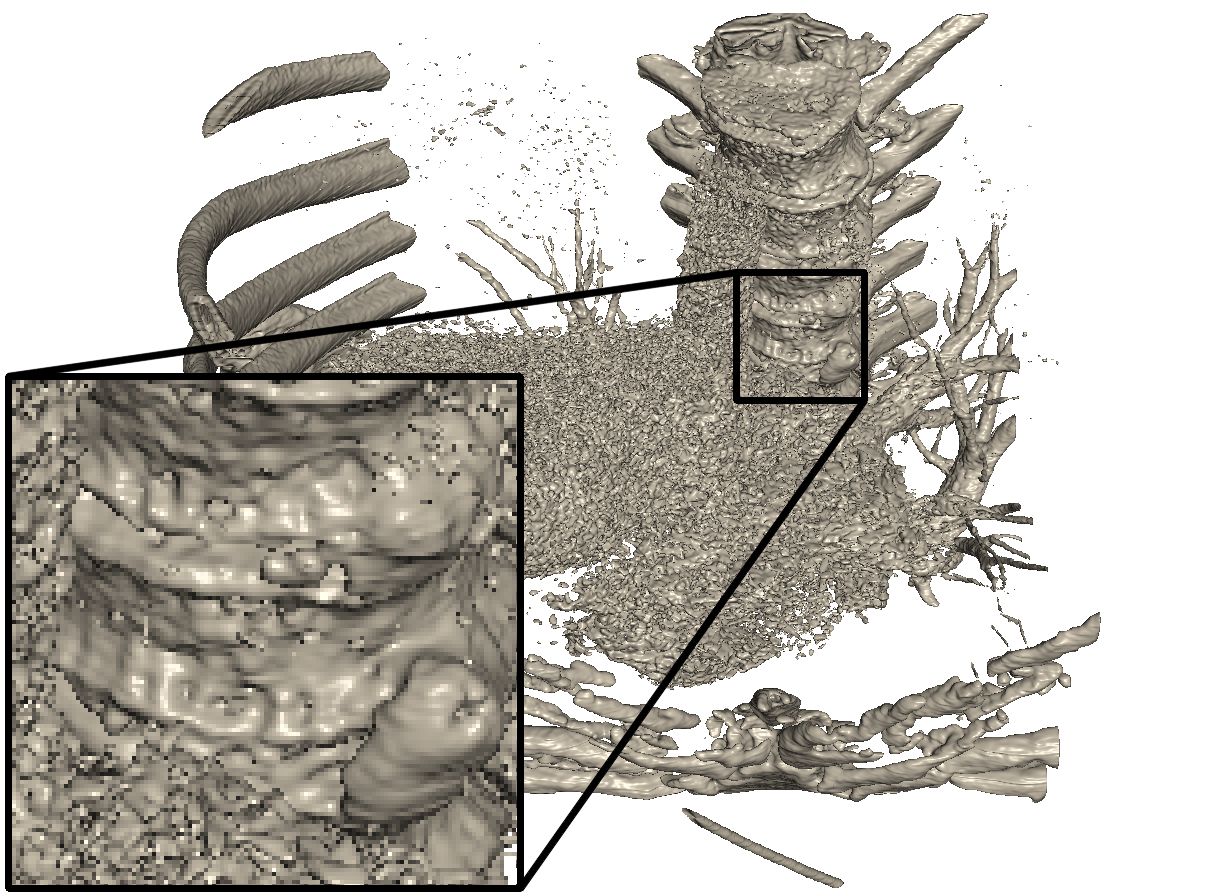}
        \subcaption{Ground truth without AO}
    \end{subfigure}
    \begin{subfigure}[c]{0.24\textwidth}
        \includegraphics[width=\textwidth,trim=0 0 100 0, clip]{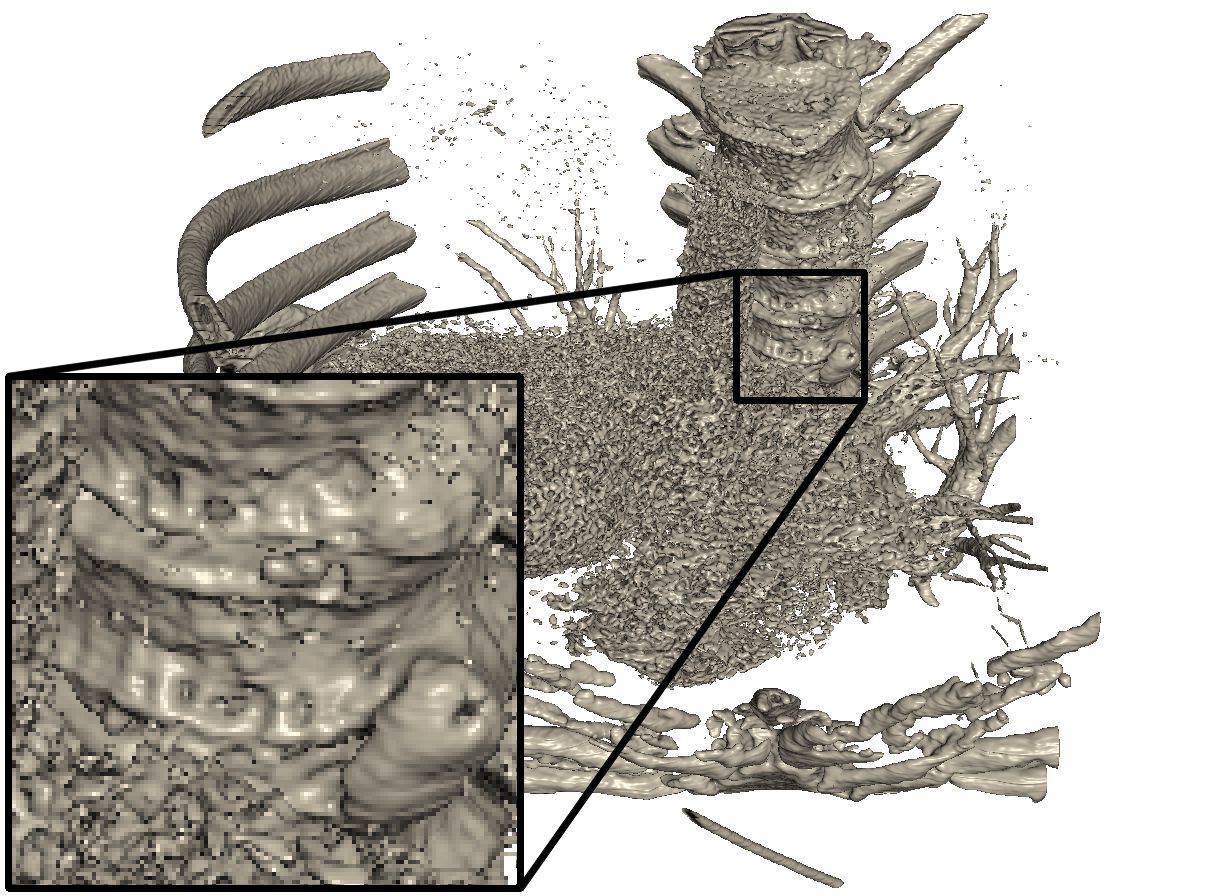}
        \subcaption{Ground truth with AO}
    \end{subfigure}
    \\
    \begin{subfigure}[c]{0.3\textwidth}
        \includegraphics[width=\textwidth]{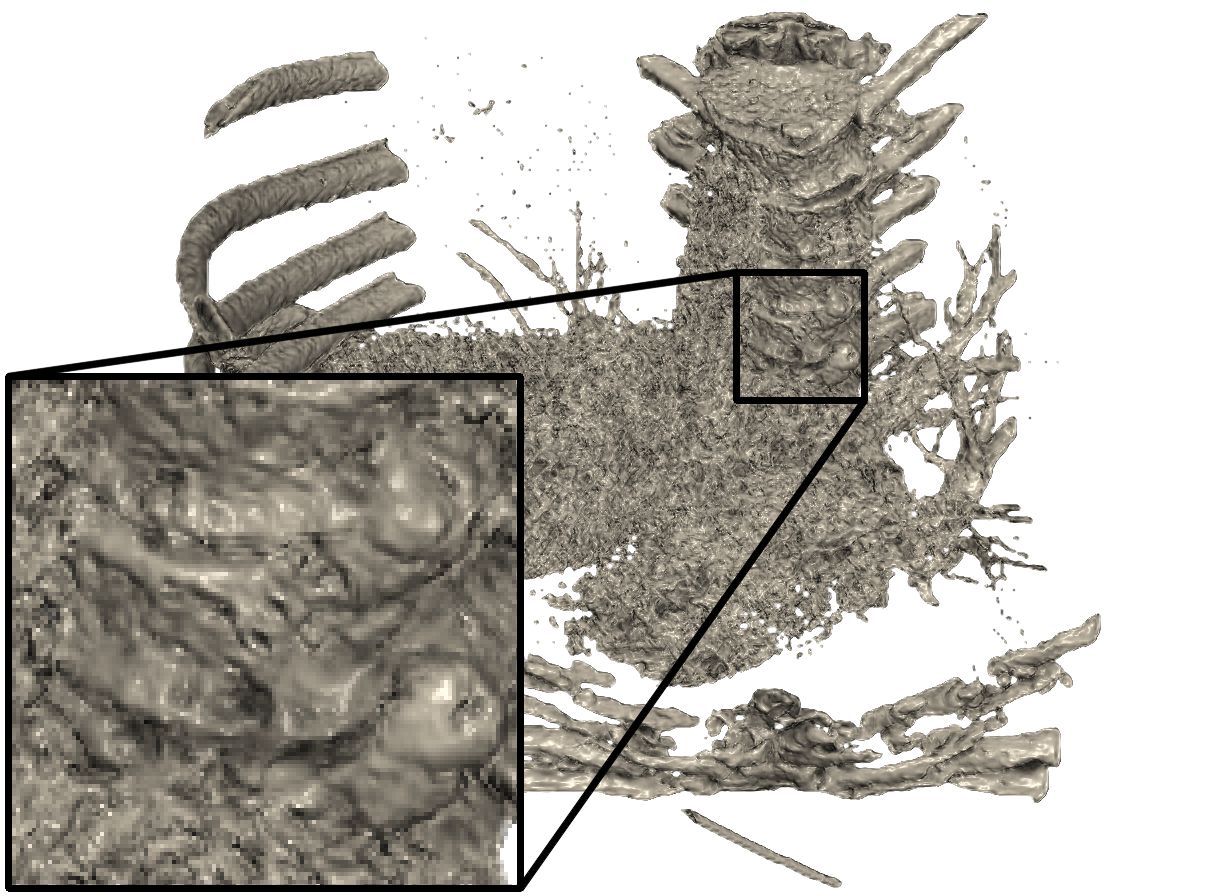}
        \subcaption{GAN 1}
    \end{subfigure}
    \begin{subfigure}[c]{0.3\textwidth}
        \includegraphics[width=\textwidth]{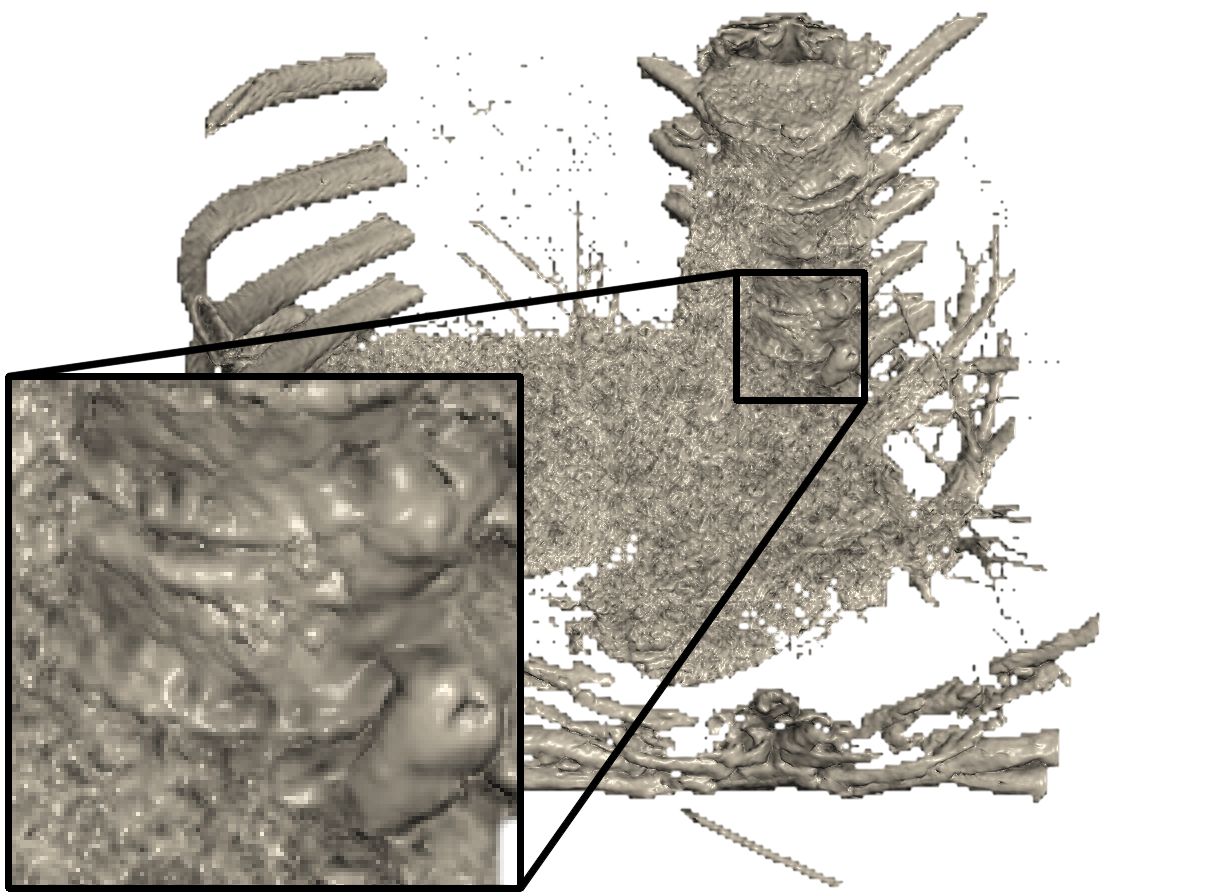}
        \subcaption{GAN 2}
    \end{subfigure}
    \begin{subfigure}[c]{0.3\textwidth}
        \includegraphics[width=\textwidth]{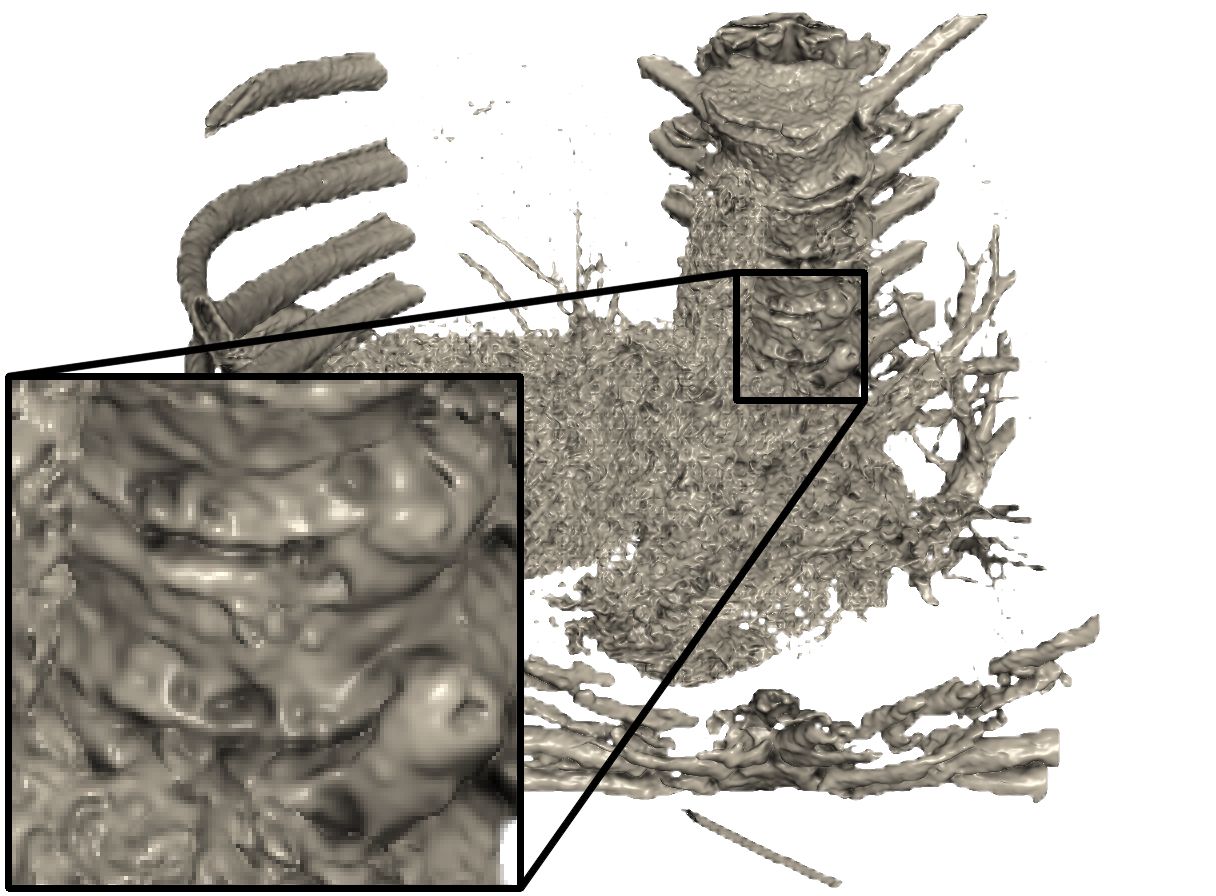}
        \subcaption{L1 loss on normals}
    \end{subfigure}
    \caption{Comparison of different network outputs on a CT scan of the human thorax.}
    \label{fig:Comparison:cleveland}
\end{figure*}

\begin{figure*}[htbp]
    \centering
    \begin{subfigure}[c]{0.24\textwidth}
        \includegraphics[width=\textwidth]{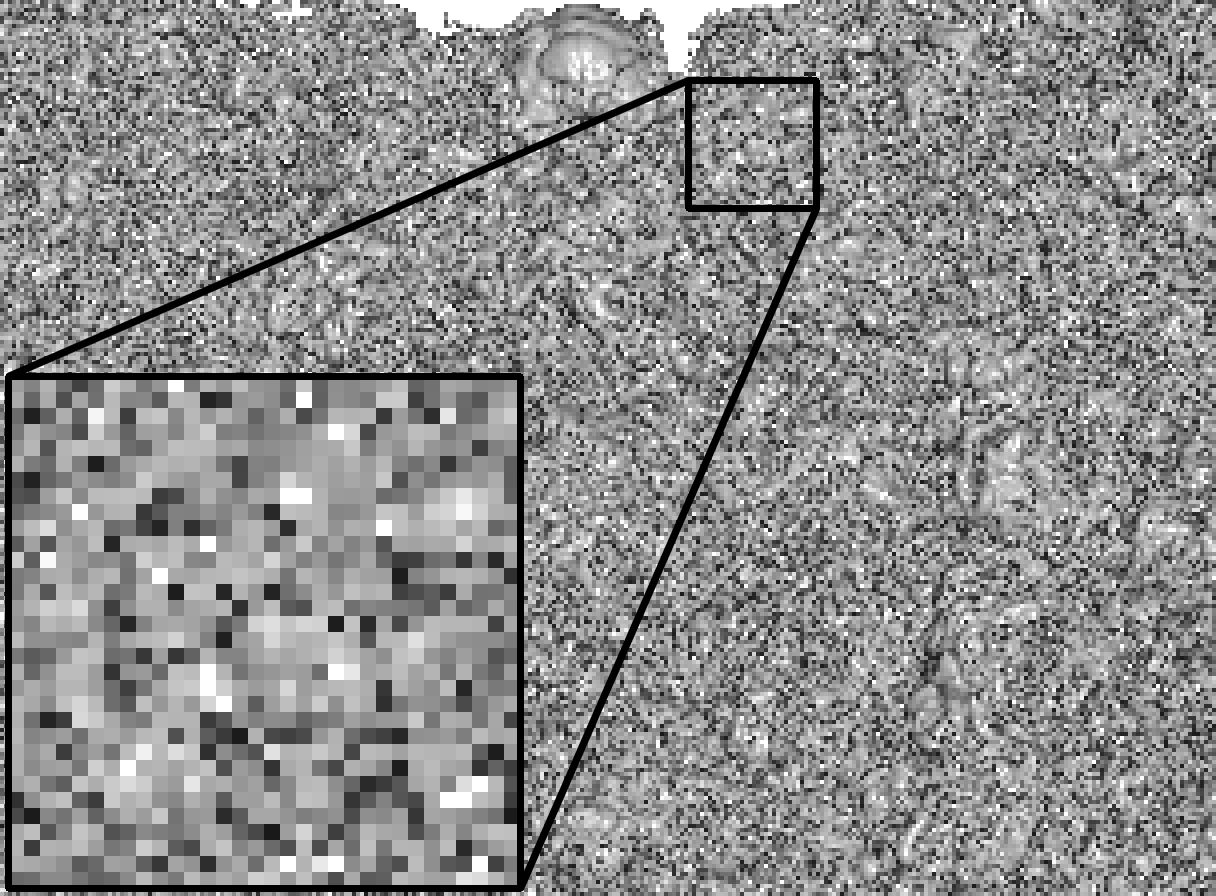}
        \subcaption{Low-resolution input}
    \end{subfigure}
    \begin{subfigure}[c]{0.24\textwidth}
        \includegraphics[width=\textwidth]{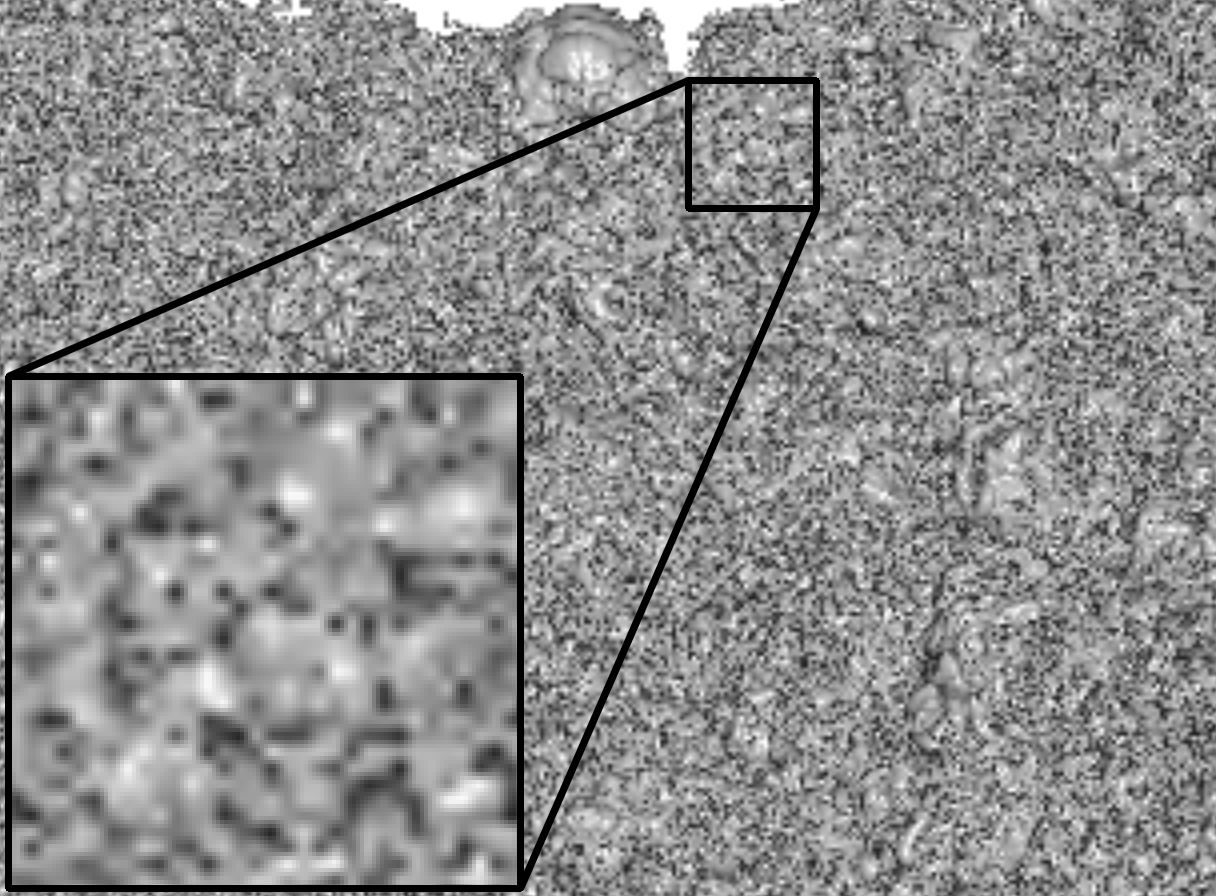}
        \subcaption{Bilinear upscaling}
    \end{subfigure}
    \begin{subfigure}[c]{0.24\textwidth}
        \includegraphics[width=\textwidth]{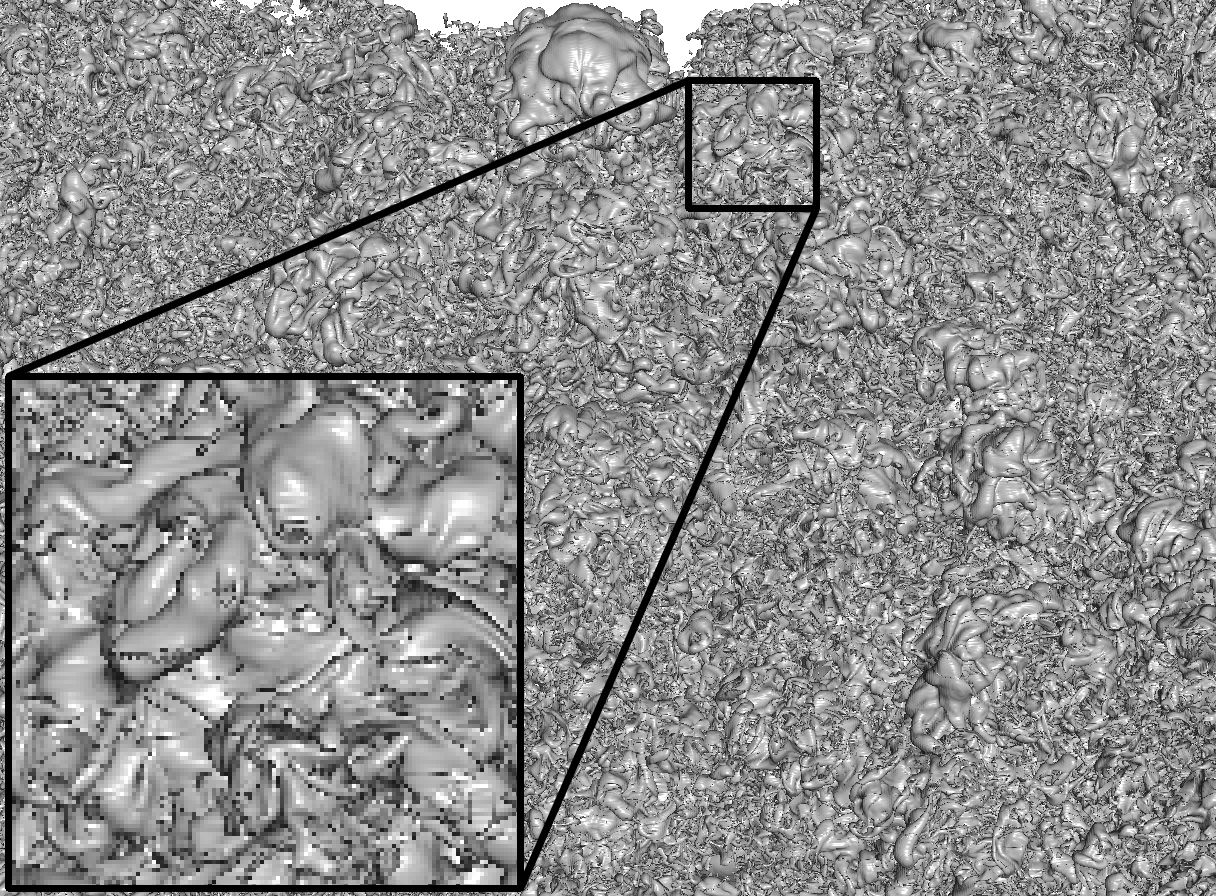}
        \subcaption{Ground truth without AO}
    \end{subfigure}
    \begin{subfigure}[c]{0.24\textwidth}
        \includegraphics[width=\textwidth]{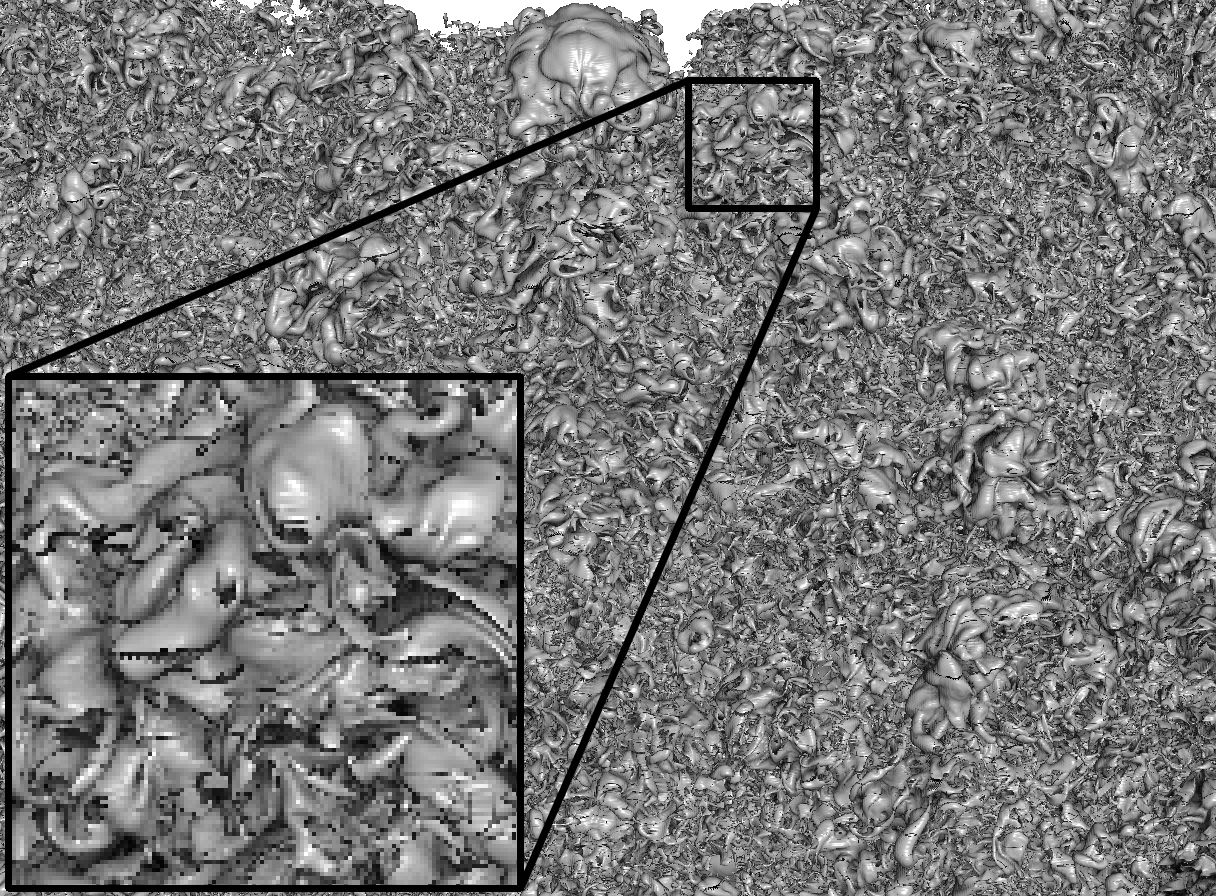}
        \subcaption{Ground truth with AO}
    \end{subfigure}
    \\
    \begin{subfigure}[c]{0.3\textwidth}
        \includegraphics[width=\textwidth]{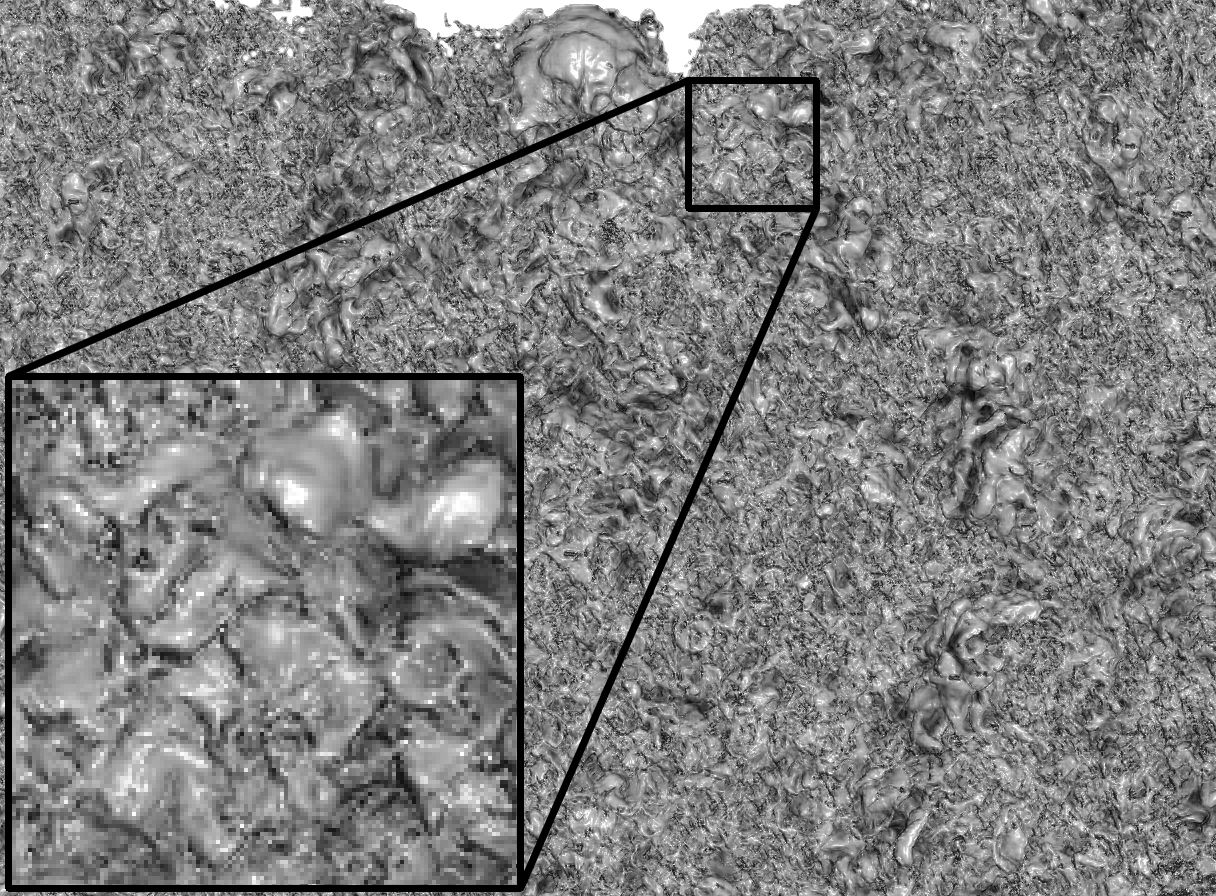}
        \subcaption{GAN 1}
    \end{subfigure}
    \begin{subfigure}[c]{0.3\textwidth}
        \includegraphics[width=\textwidth]{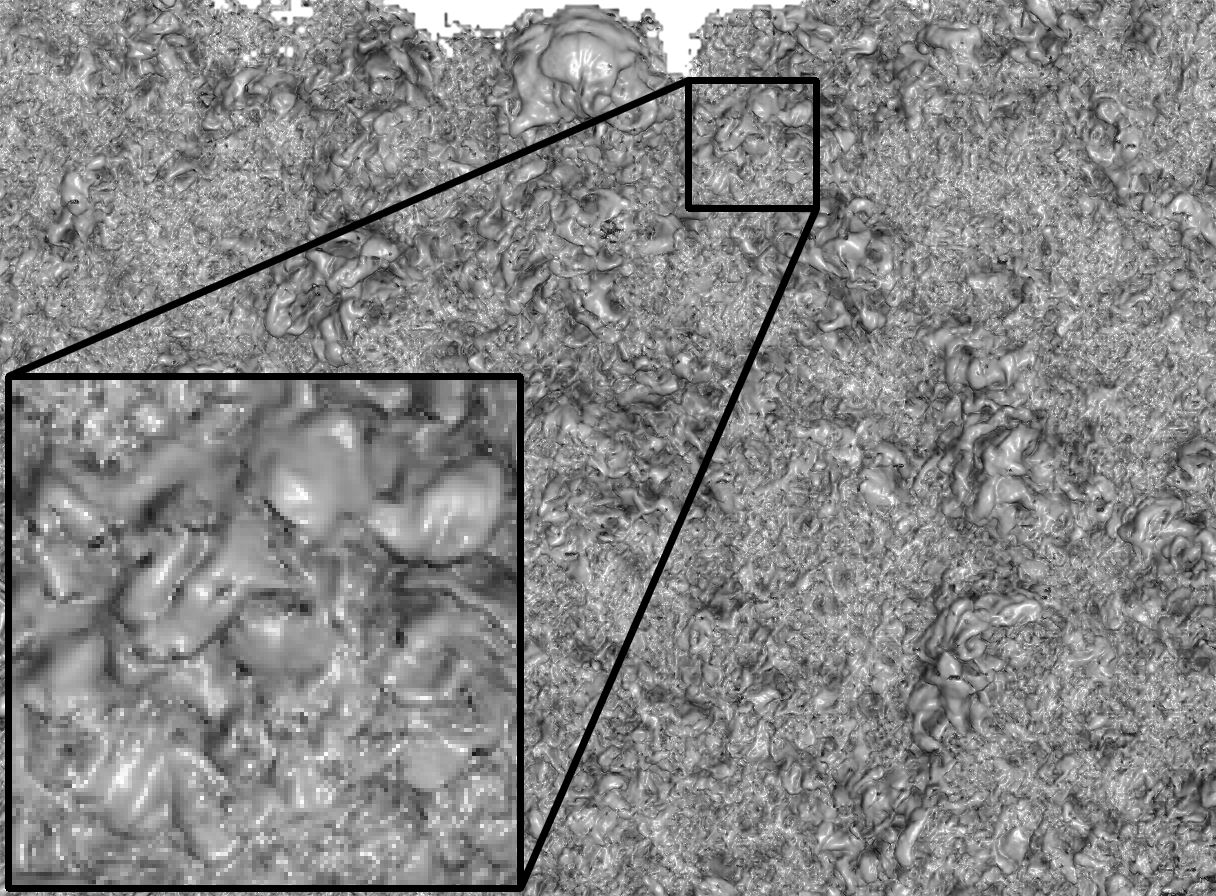}
        \subcaption{GAN 2}
    \end{subfigure}
    \begin{subfigure}[c]{0.3\textwidth}
        \includegraphics[width=\textwidth]{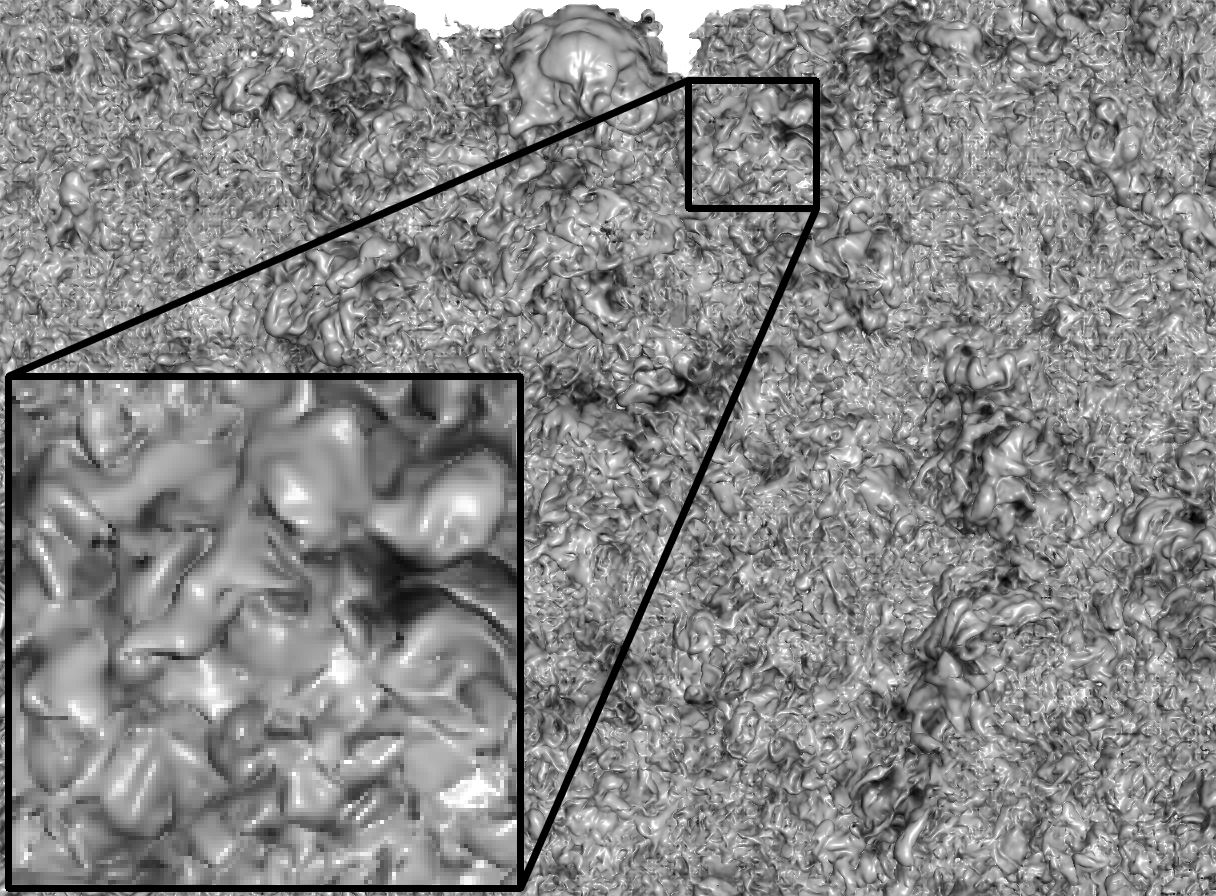}
        \subcaption{L1 loss on normals}
    \end{subfigure}
    \caption{Comparison of different network outputs on an iso-surface of a Richtmyer-Meshkov process.}
    \label{fig:Comparison:rm}
\end{figure*}

To validate how good the trained networks generalize to new data, we compare the three networks on new volumes that were never be shown to the network during training. 
First, a CT scan of a human skull with a resolution of $256^3$ is shown in \autoref{fig:Comparison:vmhead}, followed by a CT scan of a human thorax (\autoref{fig:Comparison:cleveland}) at $256^3$. Third, \autoref{fig:Comparison:rm} shows an isosurface in a $1024^3$ Richtmyer-Meshkov instability.

From these three examples, we found that the network trained with a spatial L1 loss on the normals ($\mathcal{L}_{N,l1}$) gives the most reliable inference results. 
\nils{This ``L1-normal''-network is the most strictly supervised variant among our three versions,
and as such the one that stays closest to image structures that were provided at training time.
Hence, we consider it as the preferred model.} 
In contrast, networks trained with adversarial losses like our two GAN architectures tend to generate 
a larger amount of detailed features. While this can be preferable e.g. for super-resolution of 
live-action movies, this is \nils{typically not desirable for visualizations of isosurfaces.} 

\begin{figure*}[htbp]
    \centering
    \begin{subfigure}[c]{0.24\textwidth}
        \includegraphics[width=\textwidth]{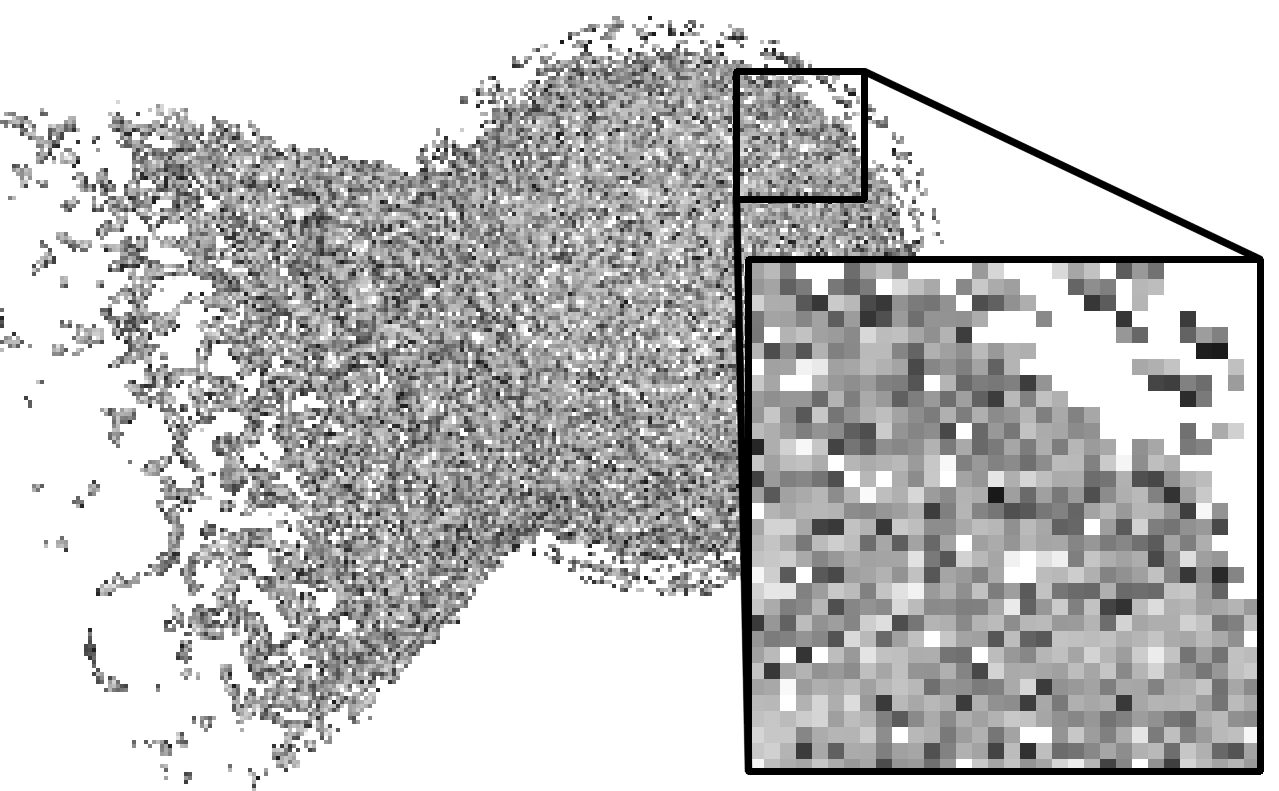}
        \subcaption{Low-resolution input}
    \end{subfigure}
    \begin{subfigure}[c]{0.24\textwidth}
        \includegraphics[width=\textwidth]{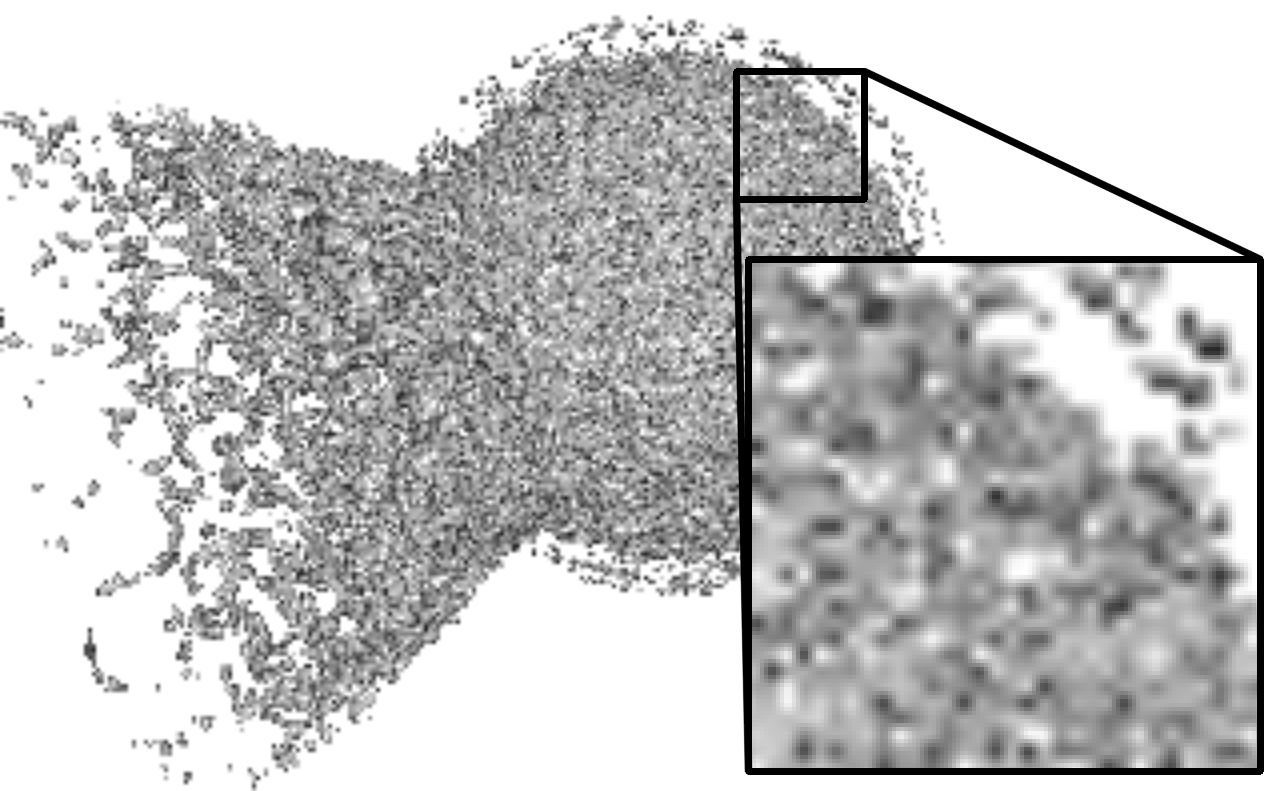}
        \subcaption{Bilinear upscaling}
    \end{subfigure}
    \begin{subfigure}[c]{0.24\textwidth}
        \includegraphics[width=\textwidth]{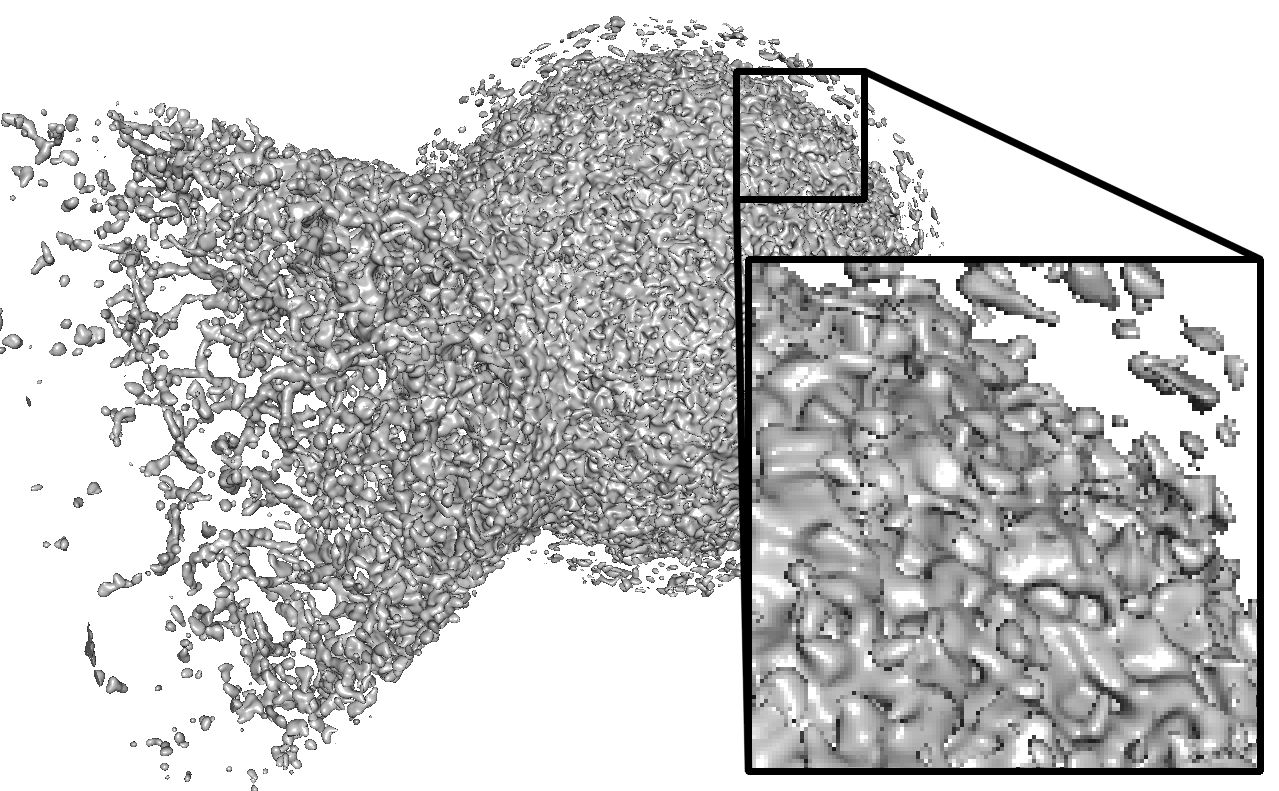}
        \subcaption{Ground truth without AO}
    \end{subfigure}
    \begin{subfigure}[c]{0.24\textwidth}
        \includegraphics[width=\textwidth]{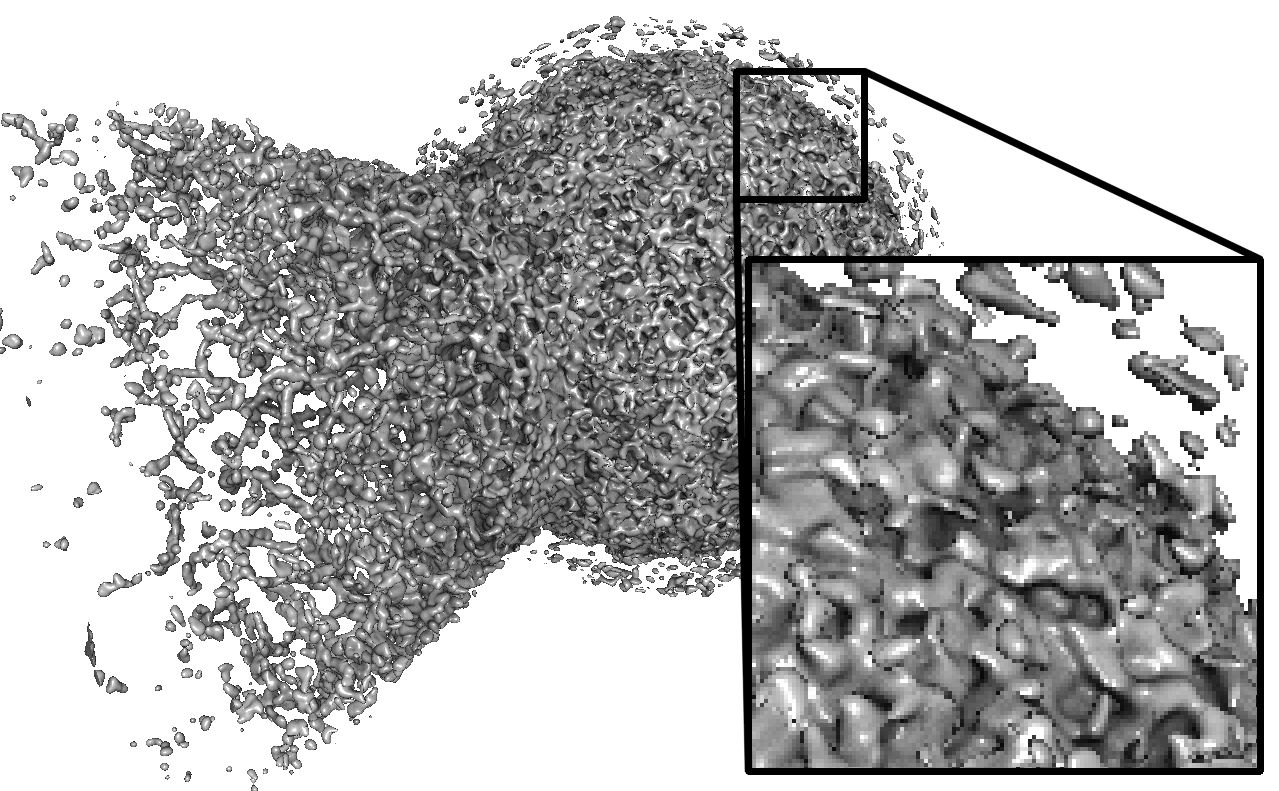}
        \subcaption{Ground truth with AO}
    \end{subfigure}
    \\
    \begin{subfigure}[c]{0.3\textwidth}
        \includegraphics[width=\textwidth]{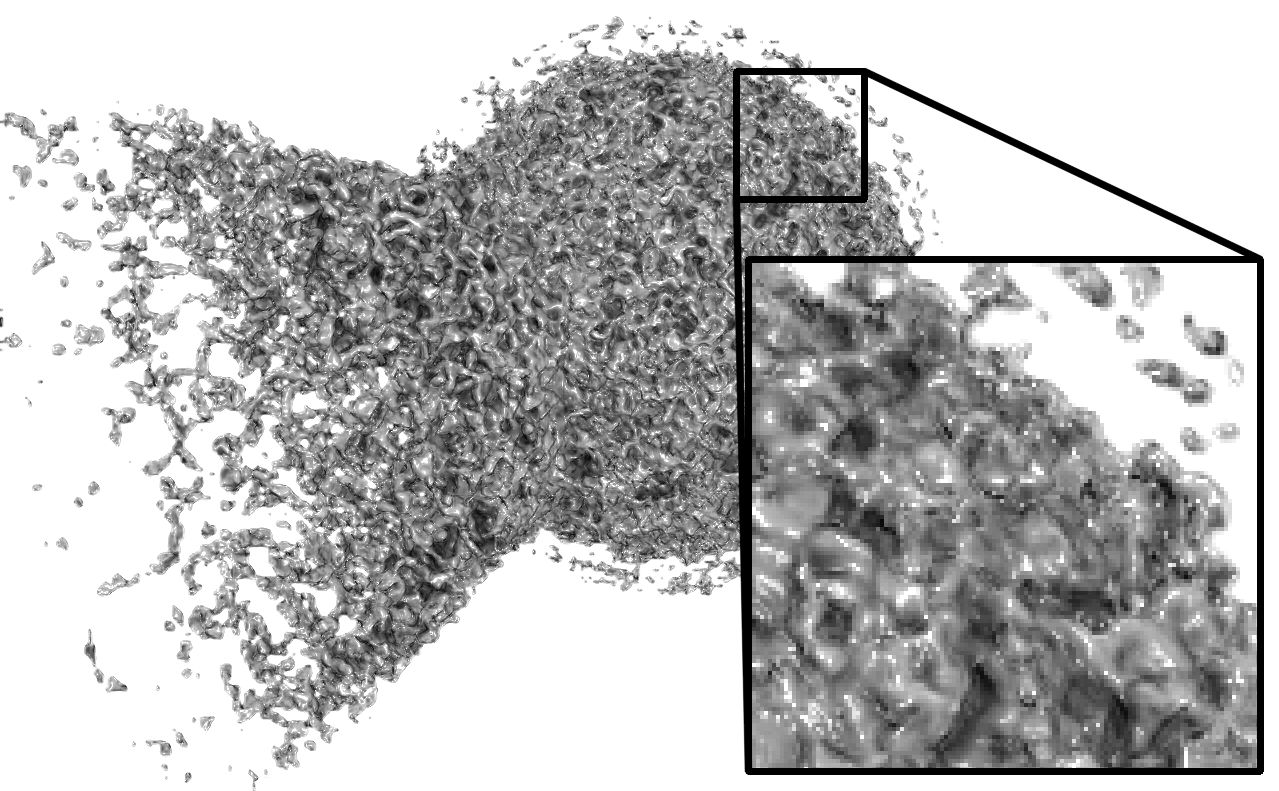}
        \subcaption{GAN 1}
    \end{subfigure}
    \begin{subfigure}[c]{0.3\textwidth}
        \includegraphics[width=\textwidth]{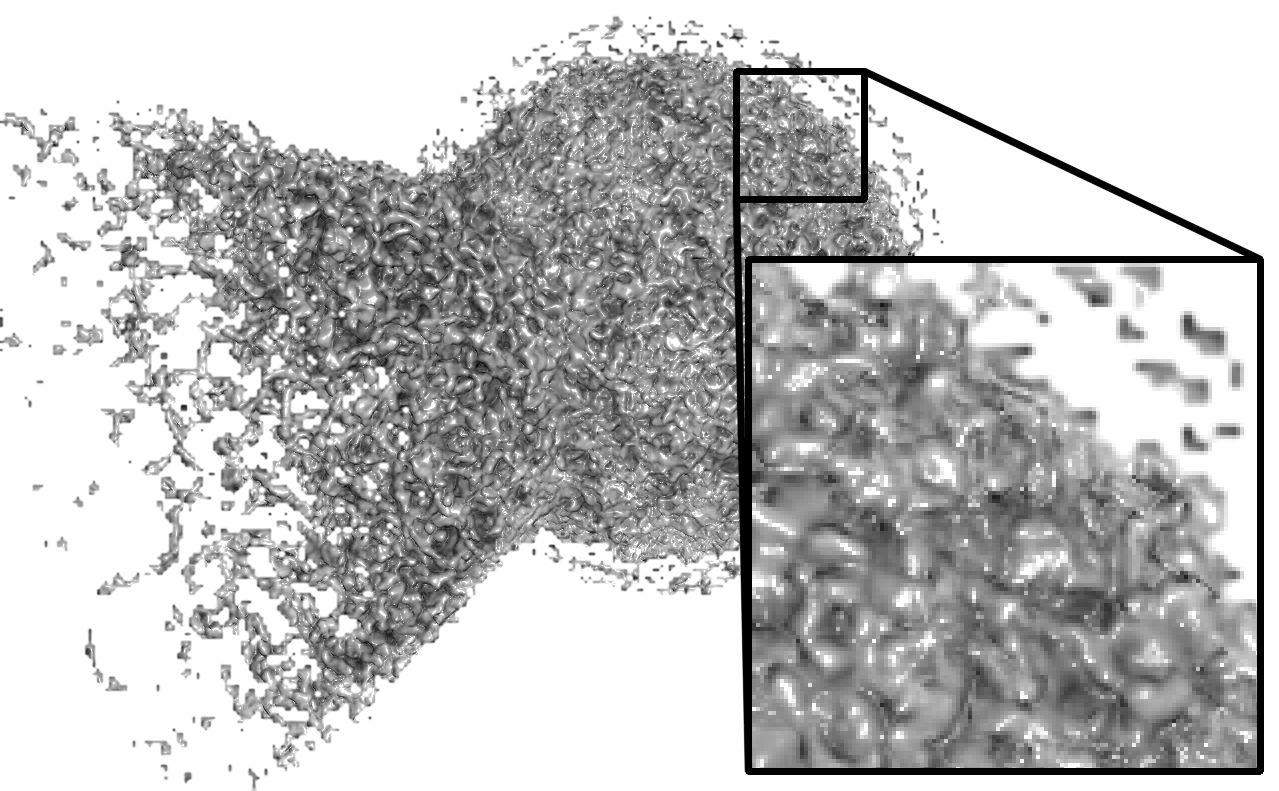}
        \subcaption{GAN 2}
    \end{subfigure}
    \begin{subfigure}[c]{0.3\textwidth}
        \includegraphics[width=\textwidth]{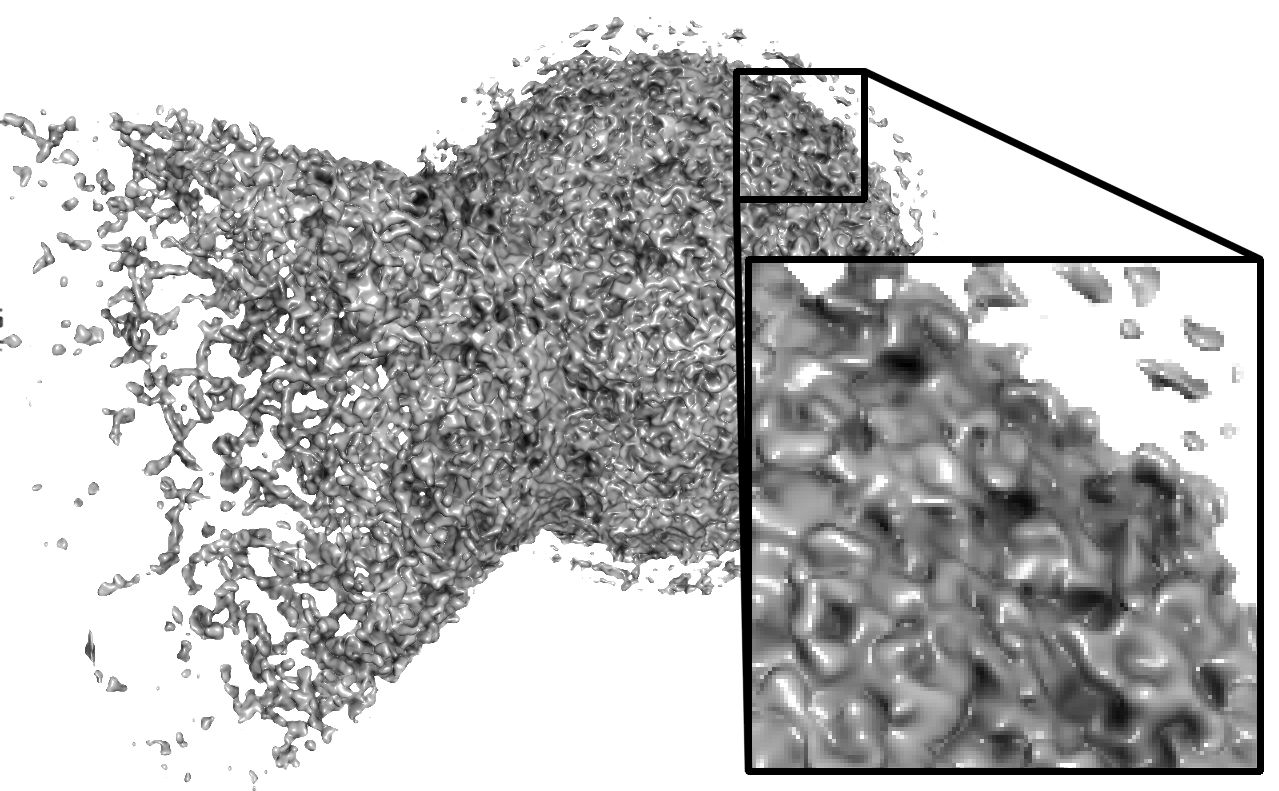}
        \subcaption{L1 loss on normals}
    \end{subfigure}
    \caption{Comparison of different network outputs on an different iso-surface of the Ejecta dataset than the one shown throughout the paper.}
    \label{fig:Comparison:Ejecta1024}
\end{figure*}

The teaser (\autoref{fig:teaser}) demonstrates the ``L1-normal''-network again on the Ejecta dataset, a particle-based fluid simulation resampled to a grid with resolution $1024^3$. 
For a comparison of all three networks on a different isosurface of the same model, see \autoref{fig:Comparison:Ejecta1024}.
The accompanying video also shows results generated with the ``L1-normal''-network for the Ejecta and Richtmyer-Meshkov datasets in motion.

\subsection{Timings}\label{sec:timings}

To evaluate the performance of isosurface super-resolution, we compare it to volumetric ray-casting on the GPU using an empty-space acceleration structure. Rendering times for the isosurfaces shown in the four test data sets are given in \autoref{tab:Timings}, for a view port size of 1920x1080.

The table shows the time to render the ground truth image at full resolution with ambient occlusion, the rendering times for the low-resolution input without ambient occlusion, and the time to perform super-resolution upscaling of the input using the network. As the evaluation time for all three networks is the same, only the average time is reported.
The time to warp the previous image, perform screen-space shading and IO between the renderer and the network are not included.
For rendering the ground truth ambient occlusion, 128 samples were taken. This gives reasonable results, but noise is still visible.

As one can see from \autoref{tab:Timings}, the time to compute ambient occlusion drastically increases the computational cost. 
Because the Ejecta dataset contains less empty blocks that can be skipped during rendering than the Richtmyer-Meshkov dataset, the computation time for the first hit (high-resolution image without AO) is twice as high that for the Richtmyer-Meshkov dataset.
As expected, the time to evaluate the super-resolution network stays constant for all four datasets as it only depends on the screen size.

As an example, the total time to render the input and perform the super-resolution on the Richtmyer-Meshkov dataset is $0.014s+0.072s=0.086s$.
Hence the network approximately takes the same time as rendering the full resolution without ambient occlusion ($0.088s$), but produces a smooth ambient occlusion map in addition.
Once ambient occlusion is included in the high-resolution rendering, the rendering time drastically increases to $14.5s$, hence the super-resolution outperforms the high-resolution renderer by two orders of magnitudes.

The isosurface renderer is implemented with Nvidia's GVDB library~\cite{gvdb}, an optimized GPU raytracer written in CUDA. The super-resolution network uses Pytorch.
The timings were performed on a workstation running Windows 10, equipped with an Intel Xeon W-2123, 3.60Ghz, 8 logical cores, 64GB RAM, and a Nvidia RTX Titan GPU.

\begin{table}[tb]
    \centering
    \begin{tabular}{r|p{1.2cm}|p{1.3cm}|p{1.2cm}|p{1.3cm}}
        Dataset & High-res\newline (no AO) & High-res\newline (with AO) & Low-res & Super-res \\\hline
        Skull $256^3$ & 0.057 & 4.2 & 0.0077 & 0.071 \\
        Thorax $256^3$ & 0.069 & 9.1 & 0.010 & 0.071 \\
        R.-M. $1024^3$ & 0.088 & 14.5 & 0.014 & 0.072 \\
        Ejecta $1024^3$ & 0.163 & 18.6 & 0.031 & 0.072 
    \end{tabular}
    \vspace{-6pt}
    \caption{Timings in seconds for rendering an isosurface in FullHD (1920x1080) resolution, averaged over 10 frames.}
    \vspace{-6pt}
    \label{tab:Timings}
\end{table}

\section{Discussion}

Our results demonstrate that deep learning-based methods have great potential for upscaling tasks beyond classical image-based approaches. The trained network seems to infer well the geometric properties of isosurfaces in volumetric scalar fields. We believe this result is of theoretical interest on its own, and has potential to spawn further research, e.g. towards the upscaling of the volumetric parameter fields themselves. Furthermore, we immediately see a number of real-world scenarios where the proposed super-resolution isosurface rendering technique can be applied.

\subsection{Use Cases}

An interesting use case is remote visualization. As designers and engineers in today’s supercomputing environments are working with increasingly complex models and simulations, remote visualization services are becoming an indispensable part of these environment. 
Recent advances in GPU technology, virtualization software, and remote protocols including image compression support high quality and responsiveness of remote visualization systems. However, in practice the bandwidth of the communication channel across which rendered images are transmitted often limits the streaming performance. Thus, the degree of interactivity often falls below what a user expects. To weaken this limitation, modern remote visualization systems perform sophisticated image-processing operations, like frame-to-frame change identification, temporal change encoding as well as image compression for bandwidth reduction.

We see one application of deep learning-based super-resolution along this streaming pipeline. During interaction, compressed low-resolution images can be streamed to the client-side, and decompressed and upscaled using trained networks. We are confident that during interaction the reconstruction error does not lead to the suppression of relevant surface features, and for a selected view a full resolution version can then be streamed.

In the imagined scenario, the networks need to be specialized on certain data types, such as images of certain physical simulations or visualization output from systems for specific use-cases like terrain rendering. We believe that custom networks need to be developed, which take into account specific properties of the data that is remotely visualized, as well as application-specific display parameters like color tables and feature-enhancements. Even though our method can so far only be used to upscale renderings of isosurfaces in volumetric scalar fields, we are convinced that the basic methodology can also be used for other types of visualizations. In particular the application to direct volume rendering using transparency and color will be an extremely challenging yet rewarding research direction.

As a second use case we see in-situ visualization. Since often the data from supercomputer simulations can only be saved at every $n$-th time step due to memory bandwidth limitations, in-between frames need to be extrapolated from these given key frames. In such scenarios, it needs to be investigated whether the network can infer in-between isosurface images or even volumetric scalar fields to perform temporal super-resolution. 

Our third use case is focus$+$context volume visualization. Images of isosurfaces often include a barrage of 3D information like shape, appearance and topology of complex structures, quickly overwhelming the user. Thus, often the user prefers to focus on particular areas in the data while preserving context information at a coarser, less detailed scale. In this situation, our proposed network can be used to reconstruct the image from a sparse set of samples in the context region, merged with an accurate rendering of the surface in the focus region. To achieve this, we will investigate the adaptations to let the network infer the surface from a sparse set of samples, and smoothly embed the image of the focus region.

\subsection{Conclusion and Future Work}

We have investigated a first deep learning technique
for isosurface super-resolution with ambient occlusion. Our network yields detailed high-resolution images
of isosurfaces at a performance that is two orders of magnitudes faster than that of an optimized ray-caster at full resolution. Our recurrent network architecture with temporally coherent adversarial training make it possible to retrieve detailed images from highly noisy low-resolution input renderings.

Despite these improvements in runtime and quality, our method only represents a very first step towards the use of deep learning methods for scientific visualization, and there are numerous promising and interesting avenues for future work.
Among others, it will be important to 
analyze how sparse the input data can be so that a network can still infer on the geometry of the underlying structures.
In addition, additional rendering effects such as 
soft shadows could be included in future versions of our approach.
While we have focused on isosurface super-resolution networks in the current work, the extension to support transparency and multiple-scattering effects is not straight forward and needs to be investigate in the future. 

\acknowledgments{
This work is supported by the ERC Starting Grant {\em realFlow} (StG-2015-637014).
}

\clearpage
\bibliographystyle{abbrv-doi}

\bibliography{ms}

\end{document}